\documentclass[11pt,a4paper]{article}

\usepackage{graphicx}
\usepackage{float}
\usepackage{afterpage}
\usepackage{epsfig,cite}
\usepackage{amssymb}
\usepackage{amsmath}
\usepackage{dsfont}
\usepackage{multirow}
\usepackage{epstopdf}
\usepackage{url,hyperref}
\usepackage{listings}

\usepackage{url}
\usepackage{hyperref}

\newcommand{\be}{\begin{equation}}
\newcommand{\ee}{\end{equation}}
\newcommand{\bea}{\begin{eqnarray}}
\newcommand{\eea}{\end{eqnarray}}
\newcommand{\bi}{\begin{itemize}}
\newcommand{\ei}{\end{itemize}}
\newcommand{\ben}{\begin{enumerate}}
\newcommand{\een}{\end{enumerate}}

\newcommand{\lc}{\left[}
\newcommand{\rc}{\right]}
\newcommand{\lp}{\left(}
\newcommand{\rp}{\right)}

\def\frac#1#2{{{#1}\over {#2}}}
\def\gsim{\mathrel{\rlap{\lower4pt\hbox{\hskip1pt$\sim$}}
    \raise1pt\hbox{$>$}}}         
\def\lsim{\mathrel{\rlap{\lower4pt\hbox{\hskip1pt$\sim$}}
    \raise1pt\hbox{$<$}}}         

\newcommand{\draft}[1]{}

\def\beq{\begin{equation}}  
\def\eeq{\end{equation}}  

\def \n0{N_j^{(0)}}

\def\lapprox{\lower .7ex\hbox{$\;\stackrel{\textstyle <}{\sim}\;$}}
\def\gapprox{\lower .7ex\hbox{$\;\stackrel{\textstyle >}{\sim}\;$}}

\begin{document}
\begin{figure}[h]
\epsfig{width=0.45\textwidth,figure=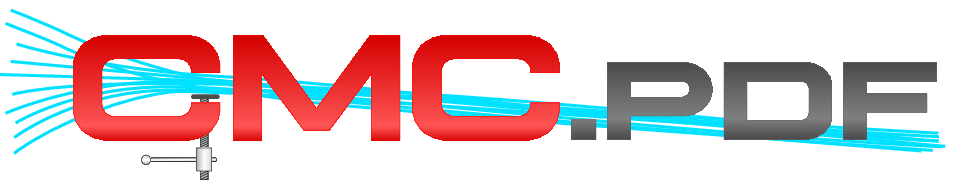}
\end{figure}
\begin{flushright}
TIF-UNIMI-2015-2\\
OUTP-15-01P\\
IPPP/15/22\\
DCPT/15/44\\
\end{flushright}
\vspace{.1cm}

\begin{center}
  {\Large \bf A compression algorithm for the combination of
  PDF sets}
\vspace{.7cm}

Stefano~Carrazza$^{1}$, Jos\'e~I.~Latorre$^2$,  Juan~Rojo$^{3}$ and Graeme Watt$^{4}$

\vspace{.3cm}
{\it
~$^1$ Dipartimento di Fisica, Universit\`a di Milano and
INFN, Sezione di Milano,\\ Via Celoria 16, I-20133 Milano, Italy\\
~$^2$ Departament d'Estructura i Constituents de la Mat\`eria, 
Universitat de Barcelona,\\ Diagonal 647, E-08028 Barcelona, Spain\\
~$^3$ Rudolf Peierls Centre for Theoretical Physics, 1 Keble Road,\\ University of Oxford, Oxford OX1 3NP, UK\\
~$^4$ Institute for Particle Physics Phenomenology, Durham University, Durham DH1 3LE, UK\\}
\end{center}

\vspace{0.1cm}

\begin{center}
{\bf \large Abstract}

\end{center}
The current PDF4LHC recommendation to estimate uncertainties due to parton distribution functions (PDFs)
in theoretical predictions for LHC processes involves the combination
of separate predictions computed using PDF sets from different groups, each of which comprises
a relatively large number of either Hessian eigenvectors or Monte Carlo (MC) replicas.
While many fixed-order and parton shower programs allow the
evaluation of PDF uncertainties for a single PDF set at no additional CPU cost,
this feature is not universal,
and moreover the {\it a posteriori} combination of the predictions
using at least three different PDF sets
is still required.
In this work, we present a strategy
for the statistical combination of individual PDF sets,
based on the MC representation of Hessian sets,
followed by
a compression
algorithm for the reduction of the number of MC replicas.
We illustrate our strategy with
the combination and compression
of the recent NNPDF3.0, CT14 and MMHT14 NNLO PDF sets.
The resulting Compressed Monte Carlo PDF (CMC-PDF) sets
are validated at the level of parton luminosities and 
LHC inclusive cross-sections and differential distributions.
We determine that
around 100 replicas provide an adequate representation
of the probability distribution for the
original combined PDF set, suitable for
general applications to LHC phenomenology.

\clearpage

\tableofcontents

\clearpage

\section{Introduction}
\label{sec:introduction}

Parton distribution functions (PDFs) are an essential ingredient for
LHC phenomenology~\cite{Forte:2010dt,Forte:2013wc,Perez:2012um,Ball:2012wy,Watt:2011kp,DeRoeck:2011na,Rojo:2015acz}.
They are one of the limiting theory factors for the extraction
of Higgs couplings from LHC data~\cite{Heinemeyer:2013tqa},
they reduce the reach of many BSM searches, particularly in the
high-mass region~\cite{AbelleiraFernandez:2012cc,Borschensky:2014cia,Kramer:2012bx},
and they are the dominant source of systematic uncertainty in precision
electroweak measurements such as the $W$ mass at the
LHC~\cite{Bozzi:2011ww,ATL-PHYS-PUB-2014-015,Bozzi:2015hha}.
A crucial question is therefore how to estimate the total PDF
uncertainty that affects the various processes listed above.

While modern PDF sets~\cite{CooperSarkar:2011aa,Ball:2014uwa,Alekhin:2013nda,Gao:2013xoa,Harland-Lang:2014zoa,Accardi:2011fa,Jimenez-Delgado:2014twa,Dulat:2015mca} provide their own estimates of
the associated PDF error, using a single
set might not lead to a robust enough estimate of the total
uncertainty arising from our imperfect knowledge of the PDFs in LHC
computations. For instance, different global PDF sets,
based on similar input
datasets and theory assumptions, while in reasonable agreement, can still
differ for some PDF flavours and $(x,Q^2)$ regions by a non-negligible
amount~\cite{Ball:2012wy,Alekhin:2010dd,Watt:2011kp}.
These differences are likely to arise from the different
fitting methodologies or from sources of theoretical uncertainty that are not
yet accounted for, such as missing higher orders
or parametric uncertainties.  For these reasons, while an improved
understanding of the origin of these differences
is achieved, from the practical point of view it is necessary to combine
different PDF sets to obtain a more reliable estimate of the
total PDF uncertainty in LHC applications.

That was the motivation underlying the original 2010 recommendation
from the PDF4LHC Working Group to compute the total PDF uncertainty
in LHC processes~\cite{Botje:2011sn,Alekhin:2011sk}.
The prescription was to take the envelope and midpoint of the
three global sets available at the time (CTEQ6.6~\cite{Nadolsky:2008zw}, MSTW08~\cite{Martin:2009iq} and
NNPDF2.0~\cite{Ball:2010de}),
each at their default value of $\alpha_s(M_Z)$, and where
each set included the combined PDF+$\alpha_s$ uncertainty using
the corresponding prescription~\cite{Lai:2010nw,Martin:2009bu,Demartin:2010er}.
This prescription has been updated~\cite{PDF4LHCrecom} to the most
recent sets from each group, and currently these
are CT14~\cite{Dulat:2015mca},
MMHT14~\cite{Harland-Lang:2014zoa} and NNPDF3.0~\cite{Ball:2014uwa}.
More recently, PDF4LHC has simplified the prescription for
the combination of PDF+$\alpha_s$ uncertainties: the current
recommendation~\cite{alphasprop} is now to take the three
global sets at a common
value of $\alpha_s(M_Z)=0.118$, close enough to the
most recent PDG average~\cite{Agashe:2014kda},
and then add in quadrature the additional uncertainty due to $\alpha_s$.
This procedure has been shown to be exact within the
Gaussian approximation in the case of CT~\cite{Lai:2010nw}, and
close enough to the exact prescription
for practical applications in the cases of MMHT and
NNPDF~\cite{Martin:2009bu,Demartin:2010er}.

One criticism that has been raised to this PDF4LHC recommendation is
that defining the total PDF uncertainty by the envelope of the
predictions from different sets does not have a well defined
statistical interpretation.
However,
as originally proposed by Forte in~\cite{Forte:2010dt}, and developed in some more detail later by Forte and Watt in~\cite{Watt:2012tq,Forte:2013wc,Watt:2013},
it is possible to modify the PDF4LHC prescription
to give the combination of PDF sets a robust statistical meaning as follows.
The first step consists in transforming the Hessian PDF sets into
Monte Carlo PDF sets using the Watt-Thorne method~\cite{Watt:2012tq}.
Then one can consider that each of the replicas from each set is a different
instance of a common probability distribution, thus the combination
of the different sets can be achieved by simply adding together their
Monte Carlo replicas.
Assuming that each PDF set that enters the
combination has the same {\it a priori} probability,
the same number of replicas should be chosen from each set.
The predictions from this combined Monte Carlo PDF set, which now
clearly have a well defined statistical meaning, turn out to be in reasonable
agreement to those of the original envelope and midpoint method proposed by
PDF4LHC.
However, the resulting PDF uncertainties will generally
be slightly smaller, since the envelope method gives more
weight to the outliers than the MC combination method.

In general, any method for the combination
of PDF sets from different groups presents
practical difficulties at the
implementation level.
The first one is purely computational: theoretical predictions
have to be computed from all the eigenvectors/replicas of the various
PDF sets, which in total require the same calculation to be redone
around $\mathcal{O}\lp 200\rp$ times for the
PDF4LHC envelope or around $\mathcal{O}\lp 900\rp$ times
for the Monte Carlo combination,
a very CPU-intensive task.
Fortunately, some of the most widely-used Monte Carlo
event generators, such as {\sc\small MadGraph5\_aMC@NLO}~\cite{Alwall:2014hca,Frederix:2011ss} or {\sc\small
  POWHEG}~\cite{Alioli:2010xd}, and
NNLO codes like {\sc\small FEWZ}~\cite{Gavin:2012sy},
now allow computation of PDF uncertainties at no extra cost.
However, this is not the case for all the theory tools used for the
LHC experiments, and even when this feature is available,
in the case of the envelope method
the {\it a posteriori} combination of the
results obtained with the three sets still needs to be performed, which
can be quite cumbersome (as well as error-prone)
especially in the case of exclusive calculations that require very large
event files.

The above discussion provides the motivation to
develop new strategies for the
combination of individual PDF sets, and the subsequent
reduction to a small number of eigenvectors or replicas.
One possible approach in this direction, the Meta-PDFs method,
has been proposed in~\cite{Gao:2013bia}.
The basic idea is to fit a common meta-parameterization
to the PDFs from different groups at some common scale $Q_0$, and then
use the Monte Carlo combination of the different
input sets to define the 68\% confidence-level intervals
of these fit parameters.
A Meta-PDF set combining MSTW08~\cite{Martin:2009iq},
CT10~\cite{Gao:2013xoa} and NNPDF2.3~\cite{Ball:2012cx} at
NNLO was produced in~\cite{Gao:2013bia} based on $N_{\rm eig}=50$ asymmetric eigenvectors.
In addition, using the dataset
diagonalization method proposed in~\cite{Pumplin:2009bb}, it is possible to further
reduce the number of eigenvectors in the Meta-PDF sets for specific
physical applications, such as for Higgs production processes.

The main limitation of the Meta-PDF method is the possible
dependence on the choice of input
meta-parametrization.
Indeed,  the statement that the common
parameterization
that is used to refit all PDF sets is flexible enough
depends on which input sets enter in the combination, thus
it needs to be checked and
adjusted every time the procedure is repeated.
In addition, at least for NNPDF, the Meta-PDF parameterization is bound to
be insufficient, particularly in extrapolation regions like large-$x$,
which are crucial for New Physics searches.

Recently, an alternative Hessian reduction approach,
the {\tt MC2H} method, has been developed~\cite{Carrazza:2015aoa}.
This method adopts the MC replicas themselves as expansion basis,
thus avoiding the need to choose a specific functional form.
It uses Singular Value Decomposition methods with
Principal Component Analysis to construct a representation
of the PDF covariance matrix as a linear combination of MC replicas.
The main advantage of the {\tt MC2H} method is that the construction is
exact, meaning that the accuracy of the new Hessian representation
is only limited by machine precision.
In practice, eigenvectors which carry little information are discarded,
but even so with $N_{\rm eig}=100$ eigenvectors central values and covariances
of the prior combination can be reproduced with $\mathcal{O}\lp 0.1\%\rp$
accuracy or better.

However, a central limitation of any Hessian reduction method
is the impossibility of reproducing non-Gaussian features present
in the input combination.
It should be noted that
even in the case where all the input sets in the combination
are approximately Gaussian, their combination in general will be non-Gaussian.
This is particularly relevant in extrapolation regions where PDF
uncertainties are large and the underlying probability distributions
for the PDFs are far from
Gaussian.
Failing to reproduce non-gaussianities implies that the assumption
of equal prior likelihood of the individual sets that enter
the combination is artificially modified: for instance, if
two sets peak at some value and another one at some other value
(so we have a double hump structure), a Gaussian reduction
effectively will be adding more weight to the second set
as compared to the first two.
To overcome this limitation is
the main motivation for this work,
where we  propose an alternative reduction
strategy based on the compression of the original Monte Carlo
combined set into a smaller subset of replicas, which however
reproduces the main statistical features of the
input distribution.

The starting point of our method is, as
in the case of the Meta-PDF and {\tt MC2H} methods,
the Monte Carlo combination
of individual PDF sets, and then a compression
algorithm follows in order to select a reduced number of replicas
while reproducing the basic statistical properties of the original
probability distribution, such as means, variances, correlations
and higher moments.
This compression is based on the Genetic Algorithms (GA) exploration
of the space of minima of suitably defined error functions, a similar
strategy as that used for the neural network training in
the NNPDF fits~\cite{tau,DelDebbio:2004qj}.
The resulting Compressed Monte Carlo PDFs, or CMC-PDFs
for short, are then validated for a wide variety of
LHC observables, both at the level of inclusive cross-sections,
differential distributions, and correlations, finding that using around
$N_{\rm rep}=100$ replicas are enough to reproduce the original results
for all the processes we have considered.

Another important application of the compression algorithm
is to native Monte Carlo PDF sets.
For instance, in the NNPDF framework, a large number of replicas,
around $N_{\rm rep}=1000$, are required to reproduce fine
details of the underlying probability distribution such as small
correlations.
Therefore, we can apply the same compression algorithm also to
native MC PDF sets,
and end up with a much smaller
number of replicas conveying the same information as the original
probability distribution.
Therefore, in this work we will also present results of this compression
of the NNPDF3.0 NLO $N_{\rm rep}=1000$ set.
Note that despite the availability of the compressed sets, PDF sets
with $N_{\rm rep}=1000$ replicas are still needed for other
applications, for instance for Bayesian reweighting~\cite{Ball:2010gb,Ball:2011gg}.

The outline of this paper is as follows.
First of all in Sect.~\ref{sec:combination} we review the
Monte Carlo method for the combination of individual PDF sets, and we present results
for the combination of the NNPDF3.0, MMHT14 and CT14 NNLO, both at the
level of PDFs and for selected benchmark LHC cross-sections.
Then in Sect.~\ref{sec:compression} we describe
the compression
algorithm used to reduce the number of replicas of a MC
PDF set.
Following this, in Sect.~\ref{sec:results}
we present our main results for the CMC-PDFs, and
validate our approach for the PDF central values, variances and correlations,
together with selected parton luminosities.
We also validate the compression of native MC sets, in particular using
NNPDF3.0 NLO with $N_{\rm rep}=1000$ replicas.
Then in Sect.~\ref{sec:lhcpheno}
we perform the validation of the CMC-PDFs at the level of LHC
cross-sections and differential distributions.
Finally, in Sect.~\ref{sec:conclusions} we summarize and discuss the delivery
of our results, both for the CMC-PDFs to be made available in
{\sc\small LHAPDF6}~\cite{Buckley:2014ana} and for the compression code,
which is also made publicly available~\cite{compressor}.
Appendix~\ref{sec:compressioncode} contains a concise user
manual for the compression code, which allows construction of
CMC-PDFs starting from an
arbitrary input combination
of PDF sets.

The detailed comparison of the CMC-PDFs with those of the
Meta-PDF and {\tt MC2H} methods will be presented in the upcoming
PDF4LHC report with the recommendations about PDF usage at Run II.

\section{Combining PDF sets using the Monte Carlo method}
\label{sec:combination}

In this section we review the Monte Carlo method for combination of
different PDF sets, and we provide results for the combination
of the recent NNPDF3.0, CT14 and MMHT14 NNLO PDF sets.
We then compare this combined PDF set with the predictions from
the three individual sets for a number of 
benchmark LHC inclusive cross-sections and their correlations.

\subsection{Combination strategy}
\label{sec:combination_strategy}

Our starting point is the same
as that originally suggested by Forte in Ref.~\cite{Forte:2010dt}.  First of all
we decide which sets enter the combination, then transform the
Hessian sets into a Monte Carlo representation using the Watt-Thorne
method~\cite{Watt:2012tq} and
finally combine the desired number of replicas from each set
to construct the joint probability distribution of the combination.
This strategy was already used in~\cite{Watt:2012tq,Forte:2013wc,Watt:2013}
to compare
the predictions of the Monte Carlo combination of PDF
sets with those of the original PDF4LHC
envelope recommendation~\cite{Alekhin:2011sk,Botje:2011sn}.

Let us recall that a Monte Carlo representation for a Hessian set
can be constructed~\cite{Watt:2012tq} by generating
a multi-Gaussian distribution in the space of fit parameters, with mean
value corresponding to the best fit result, and with width
determined by the Hessian matrix.
This is most efficiently done
in the basis where the Hessian matrix is diagonal,
and in this case Monte Carlo replicas can be generated using
\be
\label{eq:conversion}
F^k = F(q_0) + \frac{1}{2} \sum_{j=1}^{N_{\rm eig}} \lc F(q_j^+) - F(q_j^-) \rc\,R_j^k \, ,
\quad k=1,\dots,N_{\rm rep} \, ,
\ee
where $q_0$ and $q_j^\pm$ are, respectively, the best-fit and the asymmetric $j$-th eigenvector
PDF member, and $R_j^k$ are univariate Gaussian random numbers.
For most practical applications, $N_{\rm rep}=100$ are
enough to provide an accurate representation of
the original Hessian set~\cite{Watt:2012tq}.
In this work we use the {\small\sc LHAPDF6}~\cite{Buckley:2014ana}
implementation\footnote{ Note that Eq.~(6.5) of Ref.~\cite{Watt:2012tq}
  and the current {\small\sc LHAPDF 6.1.5} code
  contain a mistake which
  has been corrected in the {\sc\small Mercurial} repository and the
  correction will be included 
  in the upcoming {\small\sc LHAPDF 6.1.6} release; see
  Eq.~(22) of Ref.~\cite{Buckley:2014ana}. }
of Eq.~(\ref{eq:conversion}).
In particular, we use the {\small\sc LHAPDF6} program {\tt examples/hessian2replicas.cc}
to convert an entire Hessian set into its corresponding MC
representation.
In Eq.~(\ref{eq:conversion}) the quantity $F$ represents
the value of a particular PDF at $(x,Q)$ and flavours corresponding
to the original {\small\sc LHAPDF6} grids.

Once Hessian PDF sets have been converted into their Monte Carlo
representations, one needs to
decide how many replicas $N_{\rm rep}^{(i)}$ of each PDF set $i$ will be
included in the combination.
The combined probability distribution is simply $P = \sum_{i=1}^n w_i\,P_i$,
where $P_i$ ($i=1,\ldots,n$) are the probability distributions for each of
the $n$ individual PDF sets and the weights
$w_i=N_{\rm rep}^{(i)}/\widetilde{N}_{\rm rep}$ ($i=1,\ldots,n$), where
$\sum_{i=1}^n w_i = 1$ and $\widetilde{N}_{\rm rep}=\sum_{i=1}^n N_{\rm rep}^{(i)}$
is the total number of replicas.
The simplest case, corresponding to an equal degree
of belief in the predictions from each of the PDF sets in the
combination, is to use the same number of replicas,
say $N_{\rm rep}^{(i)}=300$, from each set.
This approach is justified in the case of fits based on
a similar global dataset and comparable theory inputs,
as will be the case in this work.
Choosing the correct value of $N_{\rm rep}^{(i)}$ for sets based
on a reduced dataset, or with very different theory inputs,
is a more complex problem which is not discussed here.
Note that taking the average over a large number of Monte Carlo replicas
generated using Eq.~(\ref{eq:conversion}) will recover the best-fit PDF
member $F(q_0)$ only up to statistical fluctuations.

Using this Monte Carlo combination method, we have produced a combined set with
$\widetilde{N}_{\rm rep}=900$ replicas from adding together
$N_{\rm rep}^{(i)}=300$ replicas of the NNPDF3.0, CT14 and MMHT14
NNLO sets.
Study of the properties of the prior with respect $\widetilde{N}_{\rm rep}$
show that at least 900 replicas are required to eliminate
the statistical fluctuations from Eq.~(\ref{eq:conversion}) down
to an acceptable level.
For the three groups we use a common
value of $\alpha_s(M_Z)=0.118$.
One requirement for the validation of
this procedure is that
the combination of the same number of instances of $n$ different
probability
distributions should have mean
$
\mu \approx \frac{1}{n}\sum_{i=1}^n \mu_i
$
and variance
$
\sigma^2 \approx \sum_{i=1}^n \lp \mu_i^2 + \sigma^2_i \rp/n - \mu^2.
$
The equality only holds when the three input
distributions are Gaussian, which in the case
of NNPDF is approximately true in the
experimental data region.

\begin{figure}[h]
  \centering 
  \includegraphics[scale=0.38]{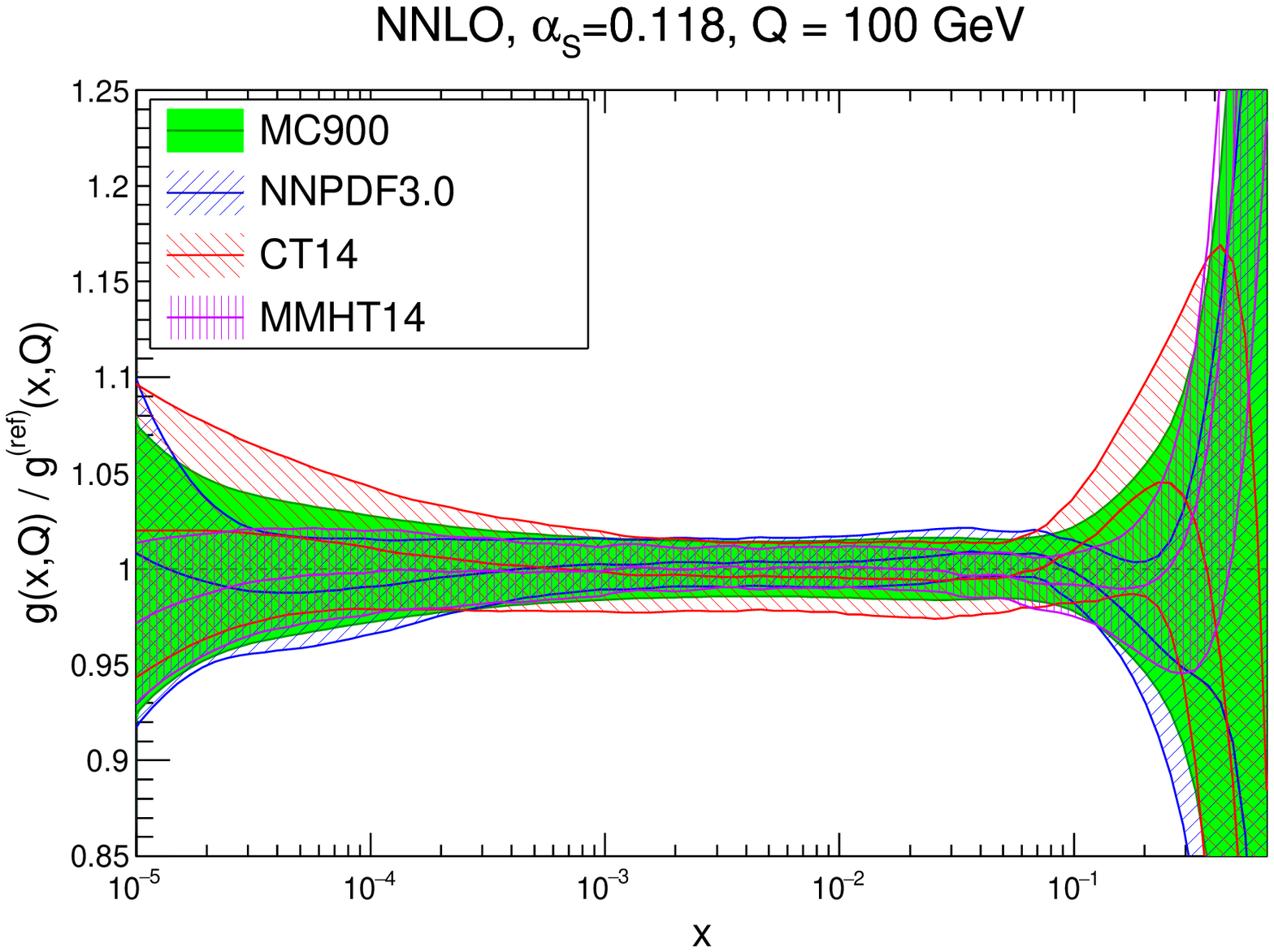}
  \includegraphics[scale=0.38]{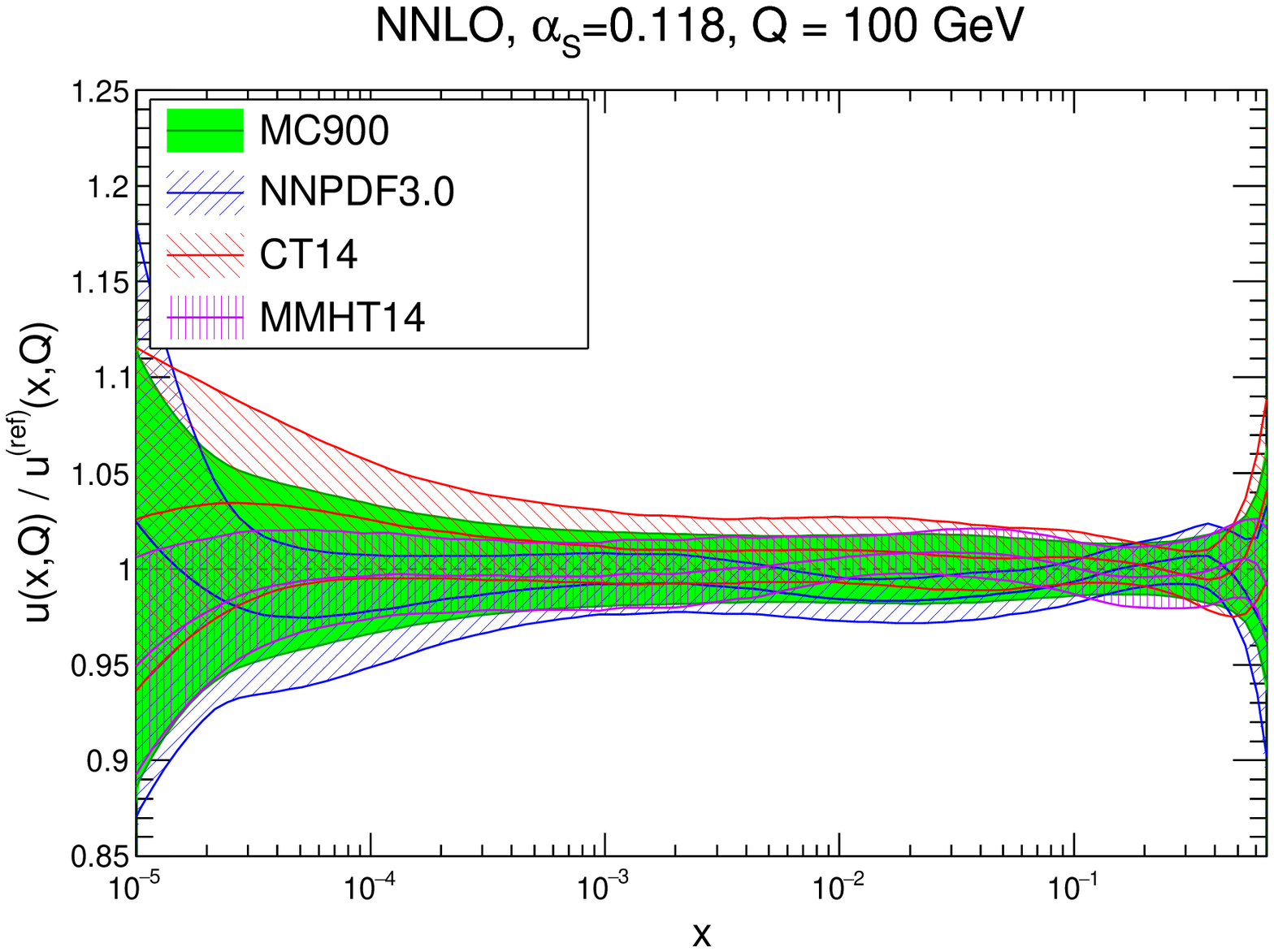}
  \includegraphics[scale=0.38]{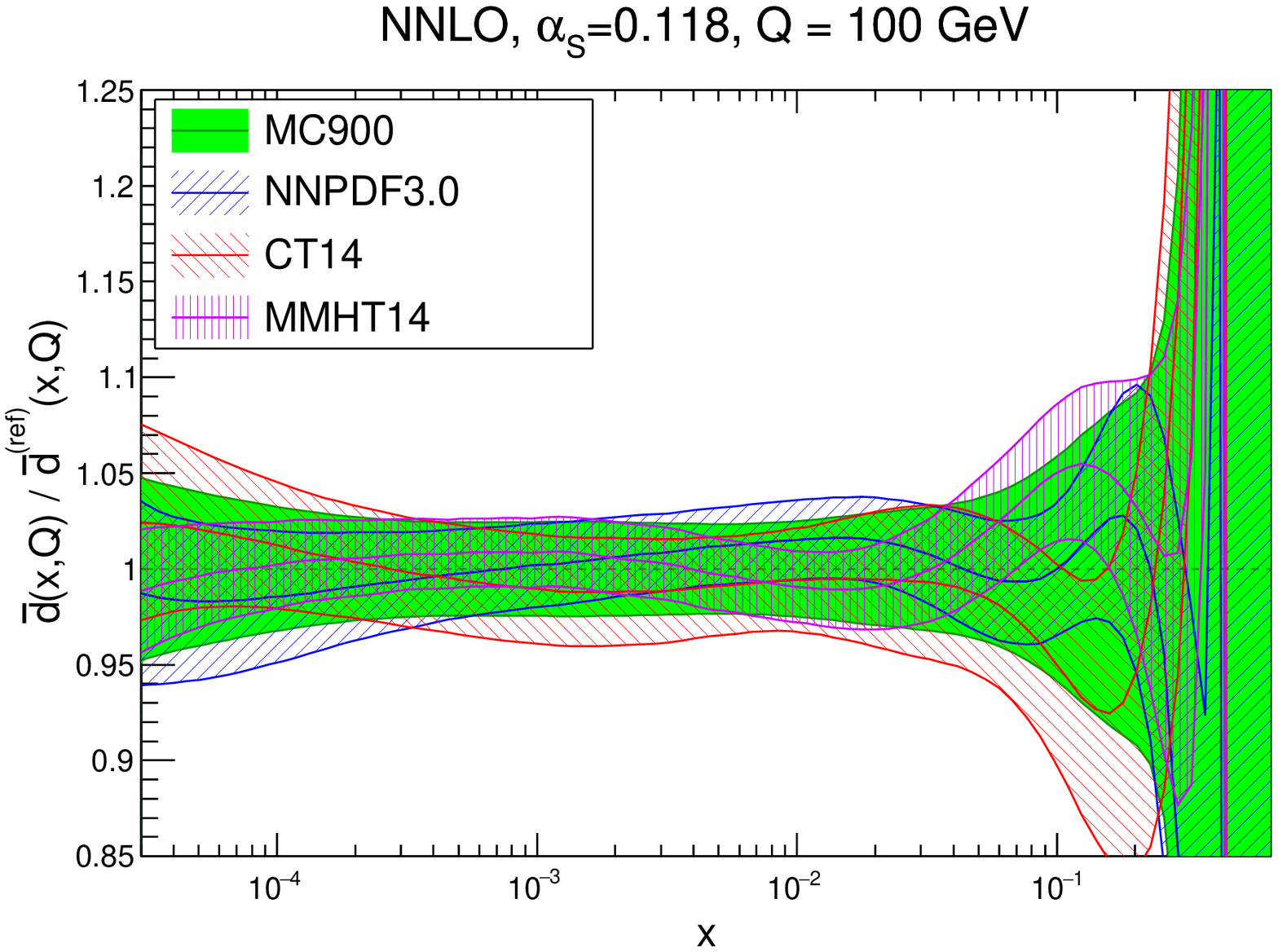}
  \includegraphics[scale=0.38]{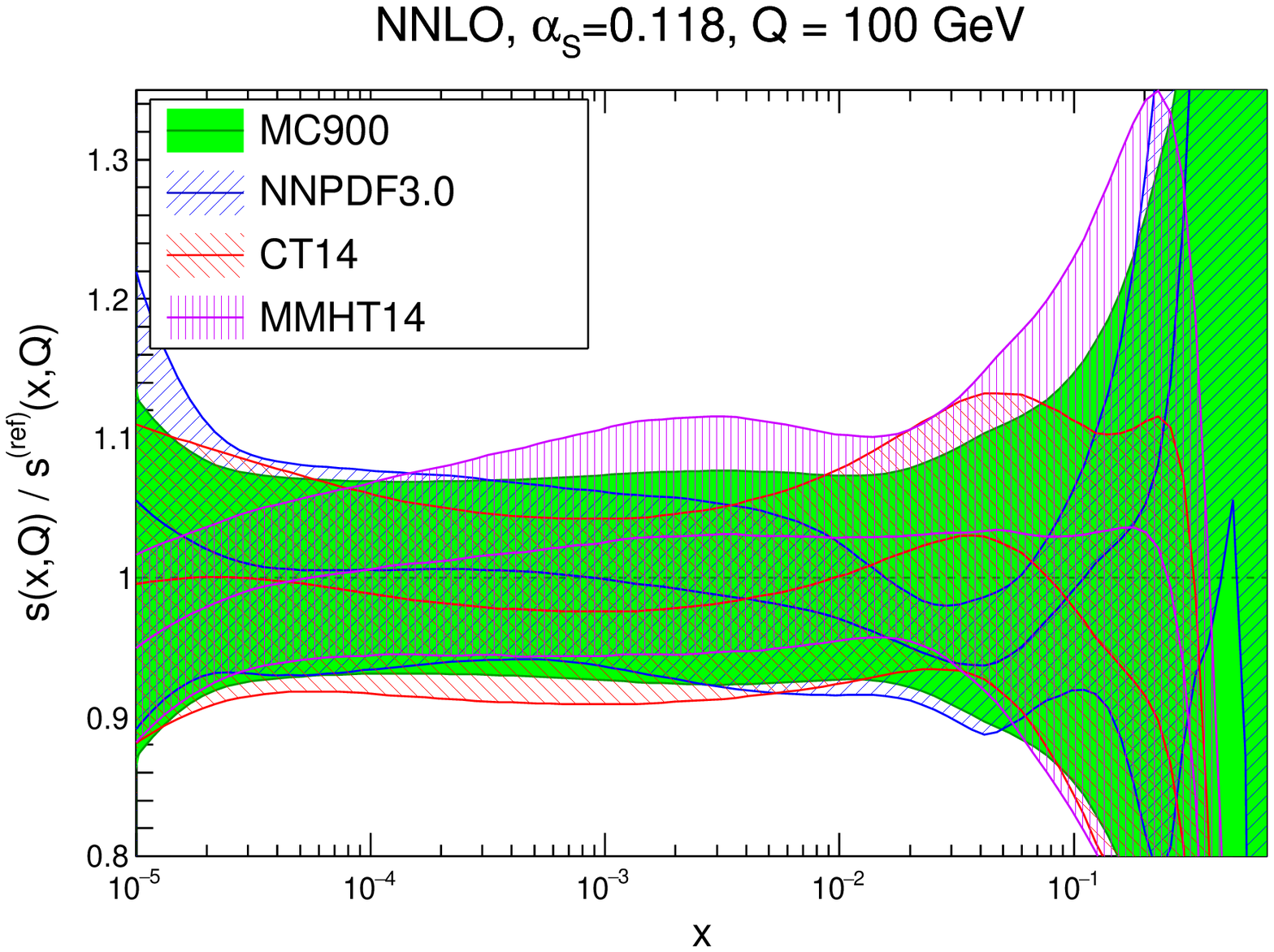}
  \caption{\small Comparison of the individual
    NNPDF3.0, CT14 and MMHT14 NNLO sets with
    the corresponding Monte Carlo combination MC900.
    The comparison is performed at a typical LHC scale of
    $Q=100$~GeV, and the PDFs are normalized to the central value
    of the combined set MC900.
  }
  \label{fig:PDFcomp}
\end{figure}

In this MC 
combination strategy, which is a common ingredient of the CMC-PDF, Meta-PDF,
and MC2H methods,
the theoretical inputs from each PDF group, like the method
of solution of the DGLAP evolution equations,
or the values of the heavy-quark masses, are not modified.
Given that the current MC combination is
 based on PDF sets with different choices of the heavy quark masses $m_c$ and
 $m_b$, and different heavy quark schemes, for applications which depend
 sizably on the values of the heavy quark masses and/or
 of the PDFs  close to the heavy quark thresholds, one should use
 the individual PDF sets rather than their combination.
    This might however change in future combinations if these are based on
    PDF sets with common settings for the treatment of heavy quarks.

    While the starting point is common,
    the differences between the three reduction methods arises
    in the strategies adopted to decrease the number of error PDF sets
    in the combination, which is achieved by compressing the
    MC representation (CMC-PDFs) or by constructing a Hessian representation,
    based either on a meta-parametrization (Meta-PDFs) or
    in a linear expansion over the MC replicas themselves (MC2H).
    In the  Meta-PDF approach~\cite{Gao:2013bia},
     common theory settings are used to evolve upwards
the meta-parameterization starting from $Q_0=8$~GeV using
{\sc\small HOPPET}~\cite{Salam:2008qg}, while CMC-PDF and MC2H maintain
the original theory settings of each individual PDF set.
It has been concluded, following a careful benchmarking between
the two groups, that both options provide an adequate
enough representation of the MC prior for $Q > m_b$, and in any case
the current combined PDFs should not be used for $Q \lsim m_b$.

In Fig.~\ref{fig:PDFcomp} we show
the comparison of the individual PDF sets,
NNPDF3.0, CT14 and MMHT14, with
their Monte Carlo combination with
$\widetilde{N}_{\rm rep}=900$.
In the following, we will denote by MC900 this prior combination.
The comparison is performed at a typical LHC scale of
$Q=100$~GeV, and the PDFs are normalized to the central value
of the combined set.
As can be seen there is reasonable agreement between the three
individual sets, and the resulting combined set is a good
measure of their common overlap.
Note that at large-$x$ differences between the three sets
are rather marked, and we expect the resulting combined
probability distribution to be rather non-Gaussian.

\begin{figure}[t]
  \centering 
  \includegraphics[scale=0.39]{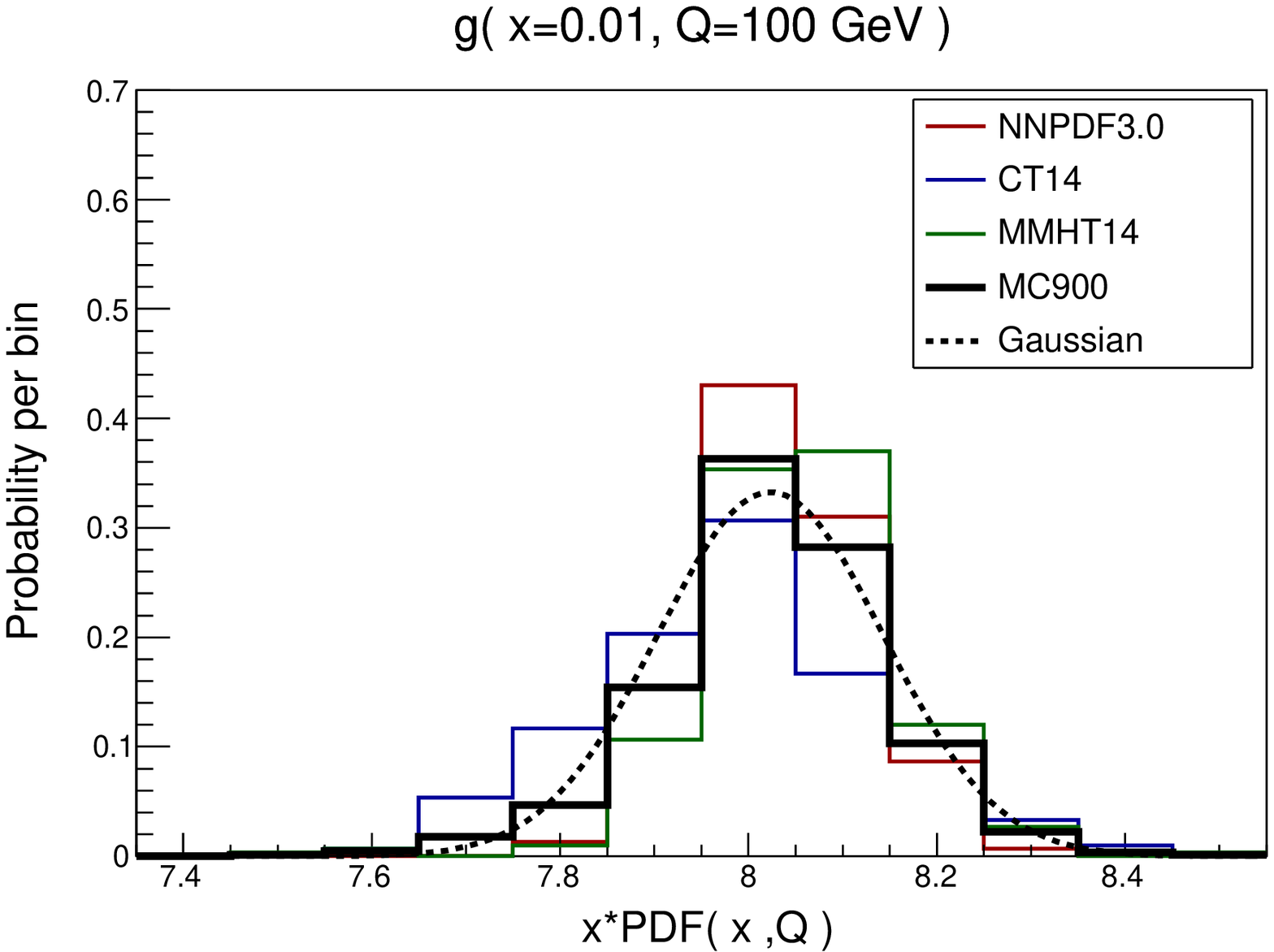}
  \includegraphics[scale=0.39]{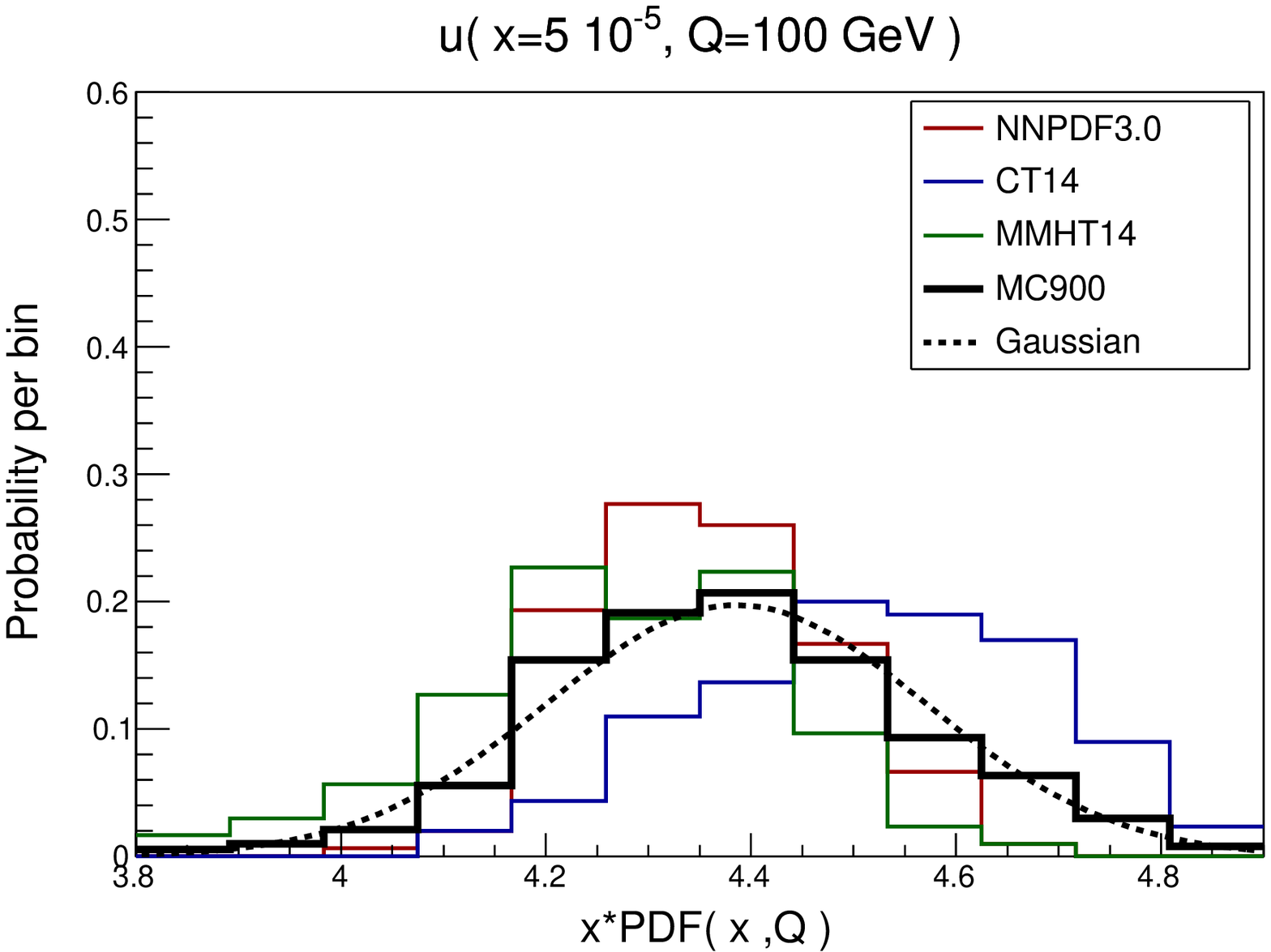}
  \includegraphics[scale=0.39]{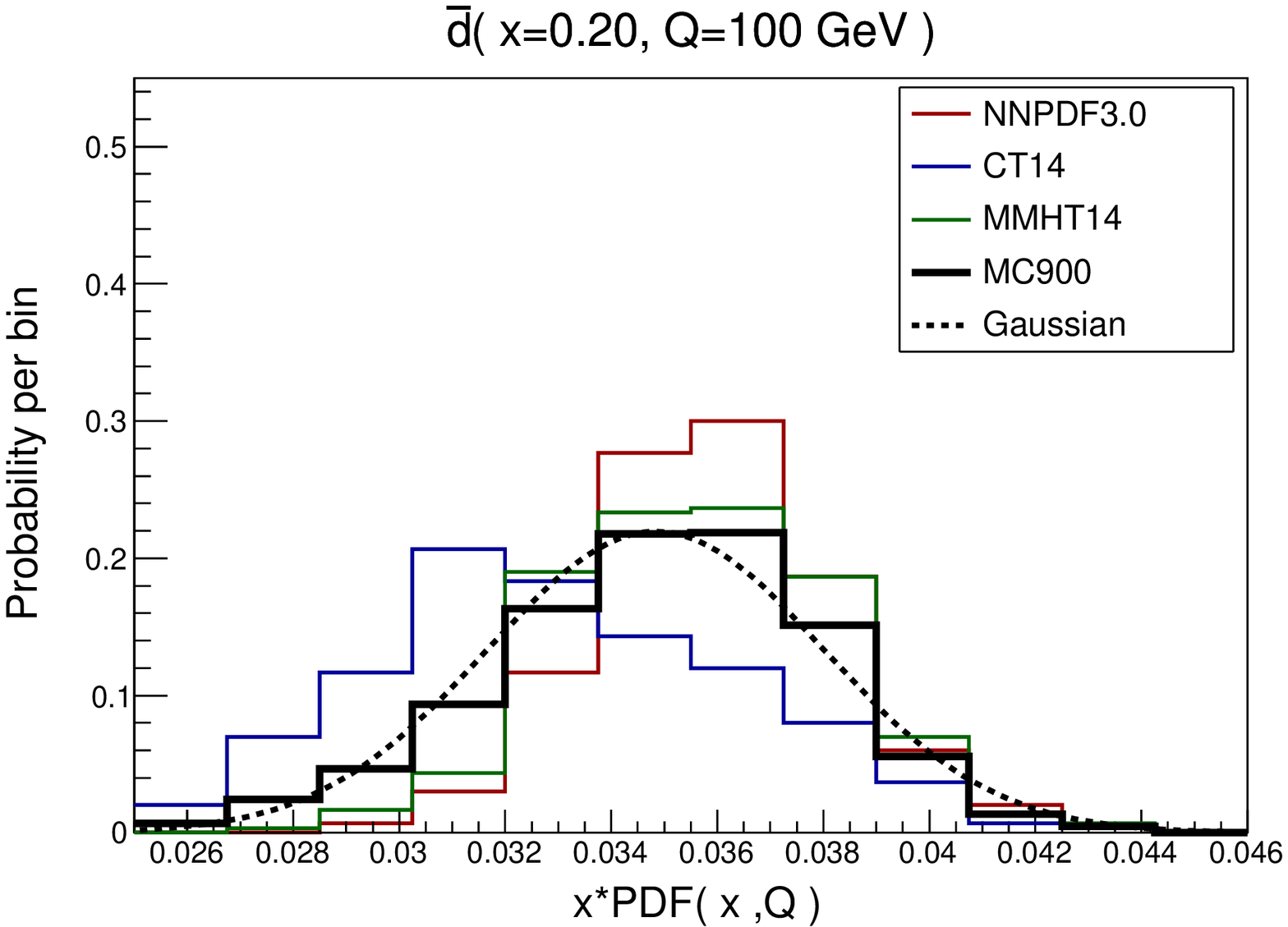}
  \includegraphics[scale=0.39]{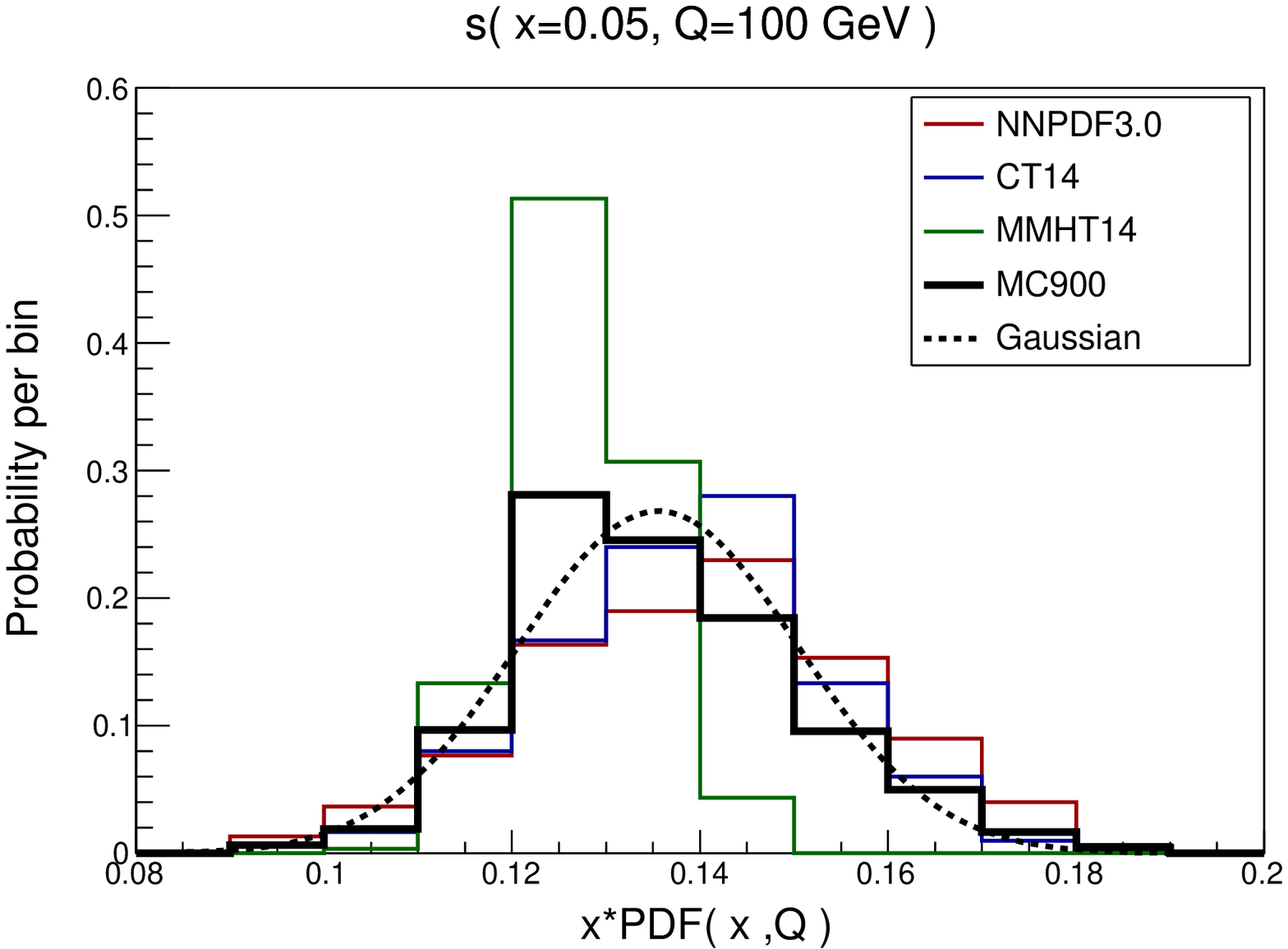}
  \caption{\small Histograms representing the probability distribution of
    Monte Carlo replicas for both the individual PDF sets and for the
    combined set, for different flavours and values of $(x,Q)$.
    From top to bottom and from left to right we show the
    gluon at $x=0.01$, the up quark at $x=5\cdot10^{-5}$,
    the down antiquark for $x=0.5$ and the strange PDF
    for $x=0.05$.
    All PDFs have been evaluated at $Q=100$ GeV.
    A  Gaussian distribution computed with from the
    mean and variance of the MC900 prior  
    is also shown. 
  }
  \label{fig:PDFcomp_histo}
\end{figure}

In Fig.~\ref{fig:PDFcomp_histo} we show the histograms representing the distribution of
Monte Carlo replicas in the individual PDF sets and in the
combined set, for different flavours and values of $(x,Q)$.
From top to bottom and from left to right we show the
gluon at $x=0.01$ (relevant for Higgs production
in gluon fusion), the up quark at $x=5\cdot10^{-5}$ (at the lower
edge
of the region covered by HERA data),
the down antiquark for $x=0.2$ (relevant for high-mass searches)
and the strange PDF
for $x=0.05$ (accessible at the LHC through $W$+charm production).
All PDFs have been evaluated at $Q=100$ GeV.

The histograms for the MC900 prior  allow us to determine
in each case
how close the combined distribution is to a normal distribution,
by comparison with a
Gaussian  computed using the same mean
and variance of the MC900 set.
From this comparison
in Fig.~\ref{fig:PDFcomp_histo}, we  see that while in some cases the
underlying distribution of the MC900 PDFs is reasonably Gaussian,
like for $g(x=0.01)$ and $u(x=5\cdot 10^{-5})$, in others,
for $\bar{d}(x=0.2)$ and $s(x=0.05)$,
the Gaussian
approximation is not satisfactory.
Deviations from a Gaussian distribution are in general more important
for PDFs in extrapolation regions with limited experimental information.

Concerning the treatment of the PDF+$\alpha_s$ uncertainties, the updated PDF4LHC
recommendation~\cite{alphasprop} proposes a simplified prescription
based on the addition in quadrature of the
separated $\delta\sigma^{\rm PDF}$ and $\delta\sigma^{\alpha_s}$ uncertainties, based on the realization
that this always gives approximately the same answer as
more sophisticated methods, and in some procedures exactly the same answer.
In the case of the Monte Carlo combination, this prescription can be implemented by simply constructing
the central values of the MC900 prior with, say, $\alpha_s(M_Z)=0.1165$
and $\alpha_s(M_Z)=0.1195$ as the mean of the
central values of the NNPDF3.0, MMHT14 and CT14 sets,
each with the corresponding value of $\alpha_s(M_Z)$.
Half of the spread of the predictions computed with the central values of the CMC-PDFs
with $\alpha_s(M_Z)=0.1165$ and $\alpha_s(M_Z)=0.1195$ defines then the one-sigma $\delta\sigma^{\alpha_s}$
uncertainty.
This assumes $\alpha_s(M_Z)=0.118\pm 0.0015$ as an external input, but a different
value of $\delta\alpha_s$ can be implemented by a simple rescaling.
Note also that for the MC900 sets with $\alpha_s(M_Z)\ne 0.118$, only the central values are required.


\subsection{PDF dependence of benchmark LHC cross-sections }
\label{sec:bench}

As stated in the introduction, the goal of this work is to compress the
MC900 prior
by roughly an order of magnitude, from the starting
$\widetilde{N}_{\rm rep}=900$ to at least $N_{\rm rep}\simeq 100$,
and to validate
the results of this compression for a number of
LHC observables.
In Sect.~\ref{sec:results} we will show the results of applying the compression
strategy of Sect.~\ref{sec:compression} to the combined MC set.
But first let us explore how the predictions from
MC900 prior
compare with the individual PDF sets
for a variety of LHC cross-sections.
We also compare the correlations between physical observables
for the individual PDF sets to their combination.

In the following we consider a number of NNLO inclusive cross-sections:
Higgs production in gluon fusion,
computed
using {\sc\small ggHiggs}~\cite{Ball:2013bra},
top-quark pair production, using
{\sc\small top++}~\cite{Czakon:2011xx},
and inclusive $W$ and $Z$ production, using
{\sc\small VRAP}~\cite{Anastasiou:2003ds}.
In all cases we use the default settings in each of these codes,
since our goal is to study similarities and differences
between the predictions of each of the PDF sets, for
fixed theory settings.

\begin{figure}[t]
  \centering 
  \includegraphics[scale=0.39]{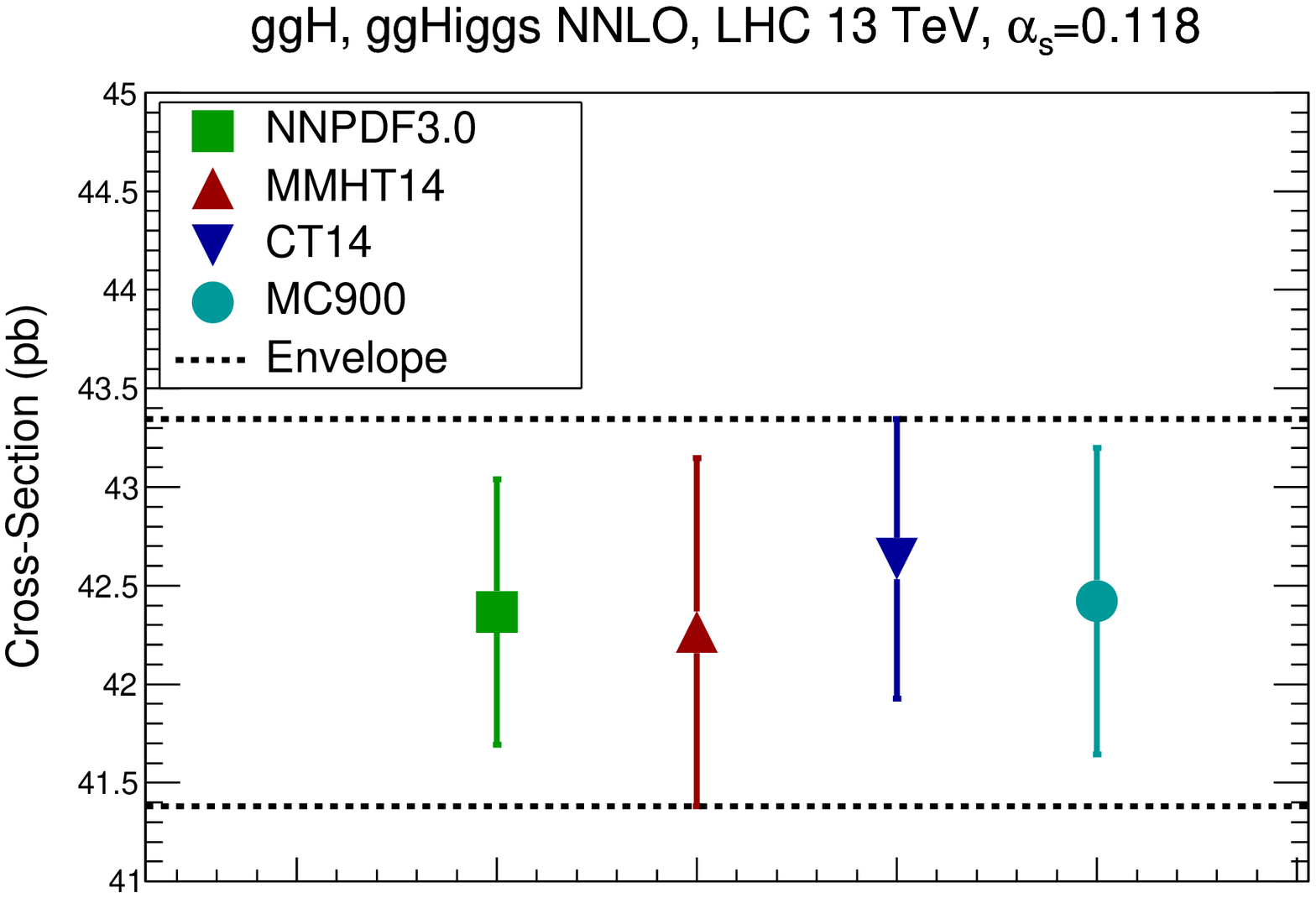}
  \includegraphics[scale=0.39]{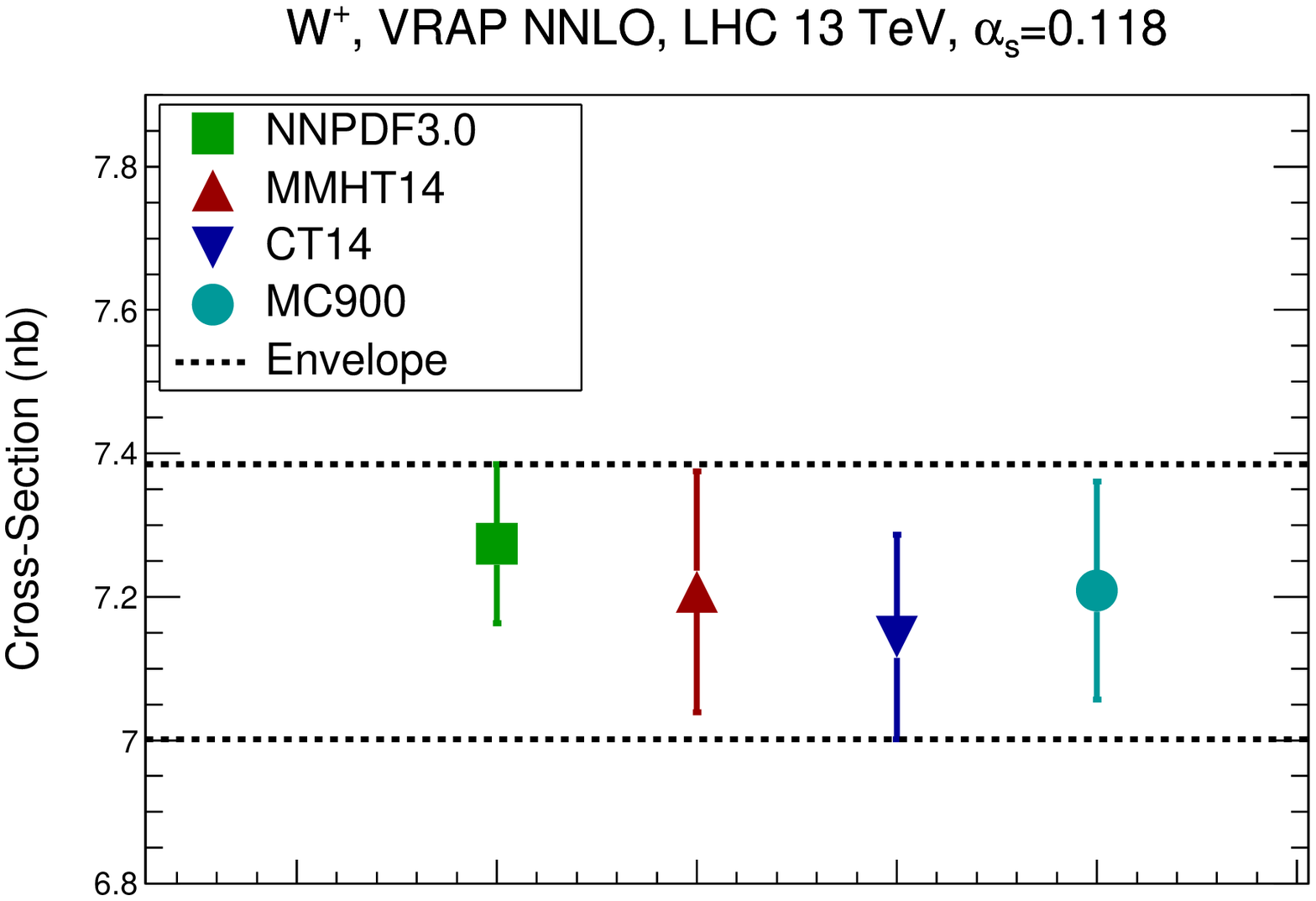}
  \includegraphics[scale=0.39]{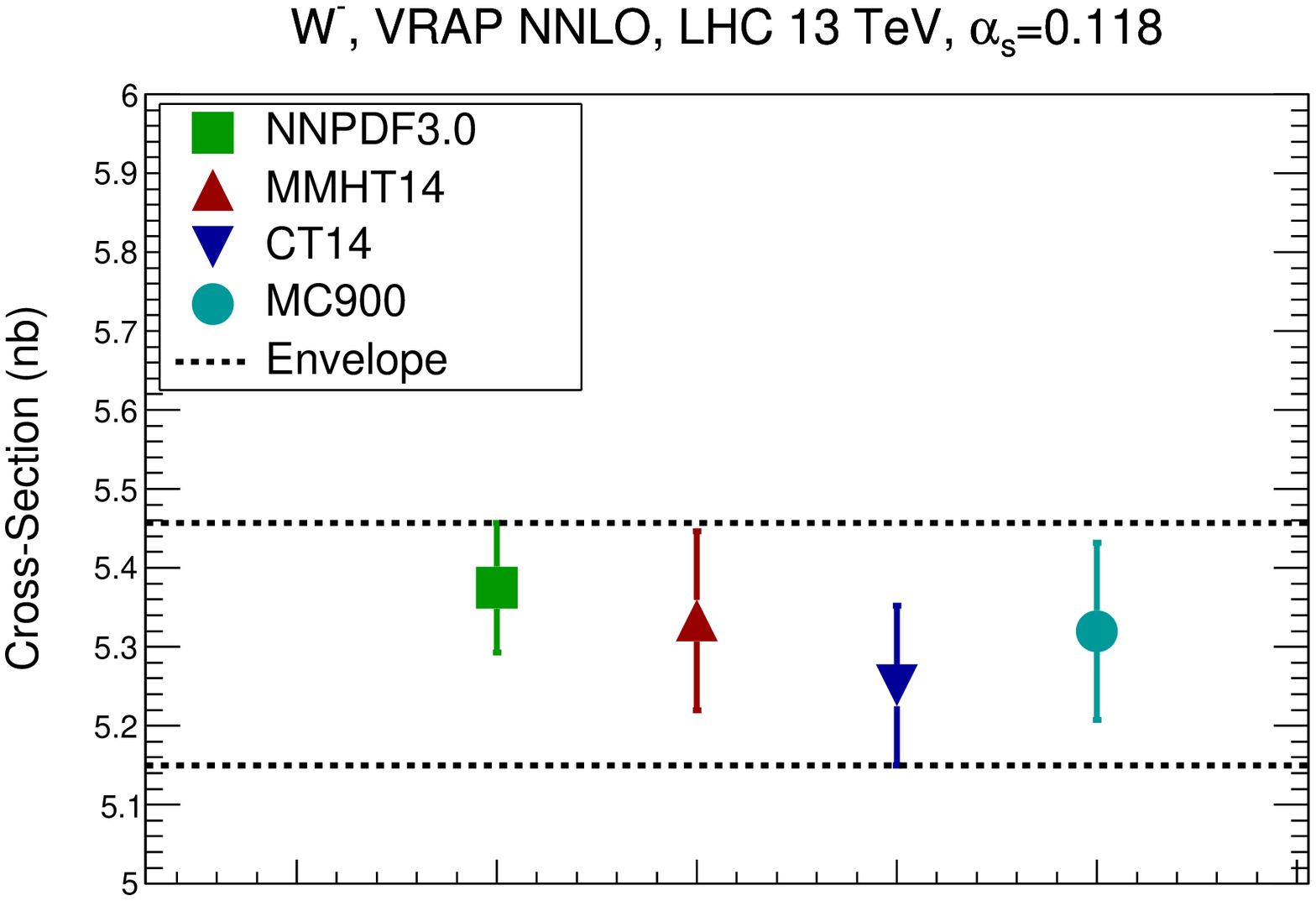}
  \includegraphics[scale=0.39]{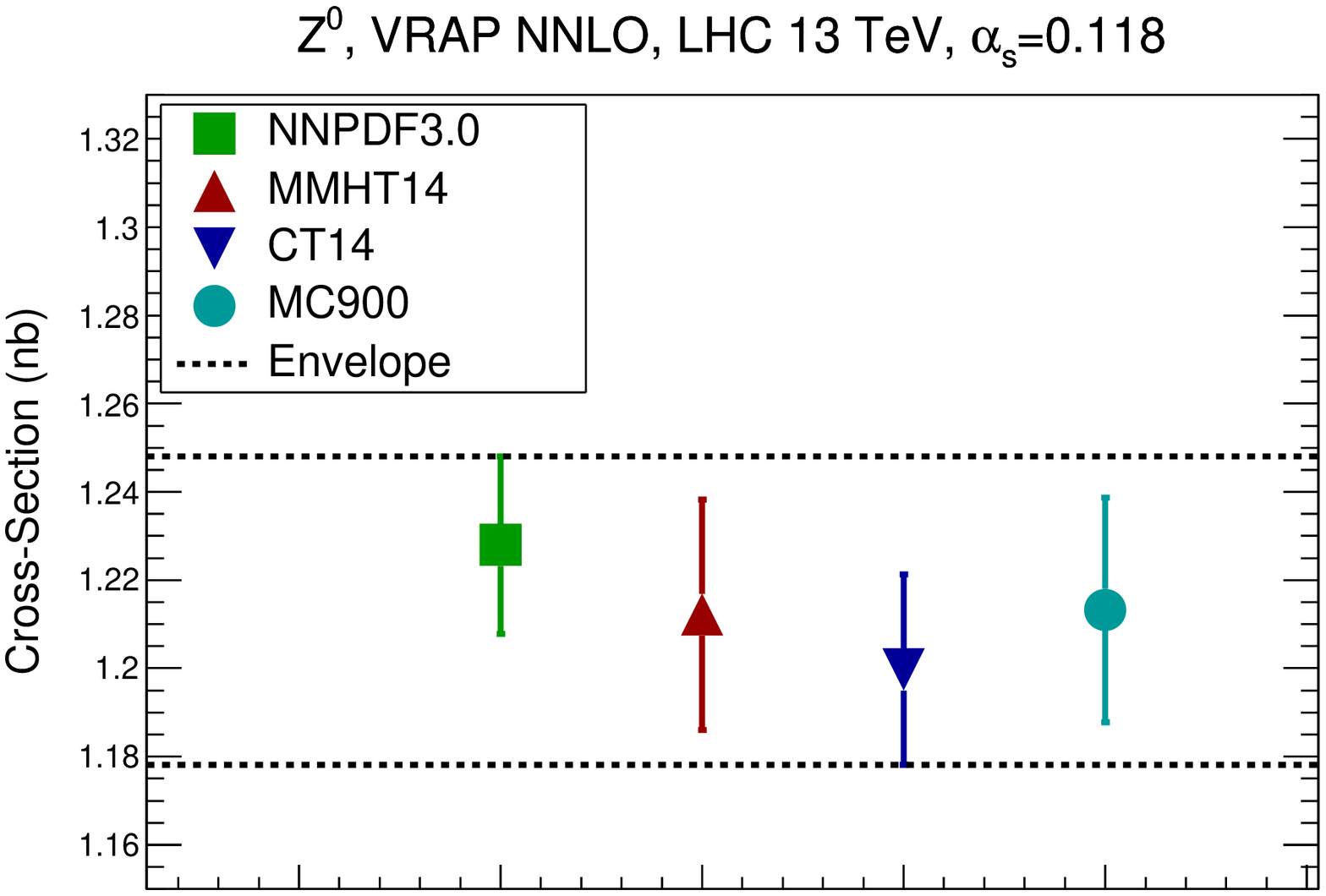}
  \includegraphics[scale=0.39]{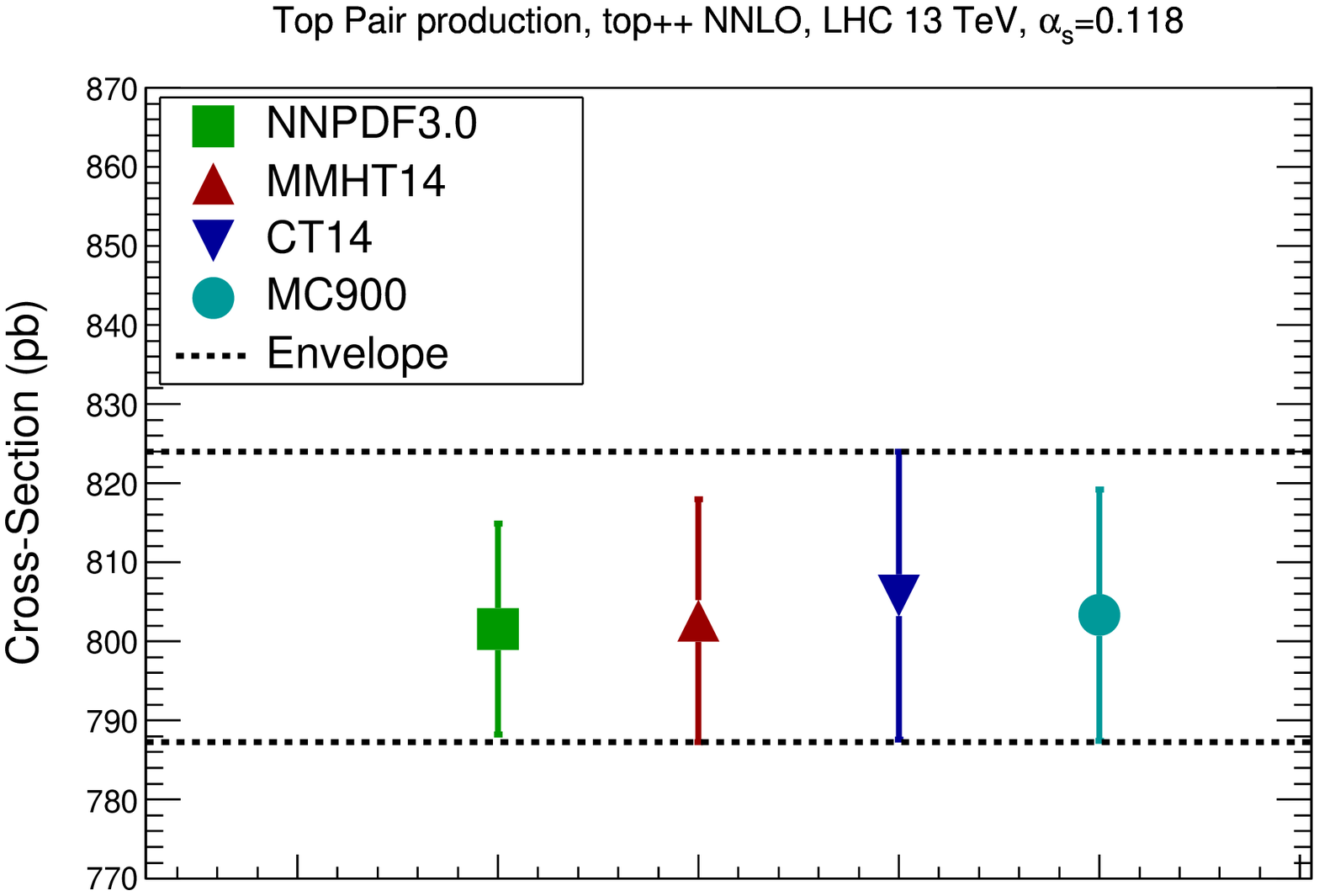}
  \caption{\small Comparison of the predictions from the
    NNPDF3.0, MMHT14 and CT14 NNLO sets, with those of their
    Monte Carlo combination MC900, for
    a number of inclusive benchmark LHC cross-sections.
    For illustration, we also indicate the envelope of the
    predictions of the three different PDF sets, which would determine
    the total PDF uncertainty in the current PDF4LHC recommendation.
    From top to bottom and from left to right: Higgs production
    in gluon fusion, $W^+$, $W^-$
    and $Z$ production, and
    top quark pair production.
    All processes have been computed at the LHC
    with a centre-of-mass energy of 13 TeV.
  }
  \label{fig:xsecbenchI}
\end{figure}

The results for these inclusive
cross-sections are shown in Fig.~\ref{fig:xsecbenchI}.
We also show with dashed lines the envelope of the one-sigma
range obtained from the three individual sets, which
would correspond to the total PDF uncertainty for this process
if obtained following the present PDF4LHC recommendation.
We see that in general the two methods, the MC combination
and the envelope, give similar results, the former leading to a smaller
estimate of the total PDF uncertainty since the envelope assigns
more weight to outliers than what would be required on a statistical
basis.

\begin{figure}[t]
  \centering 
  \includegraphics[scale=0.36]{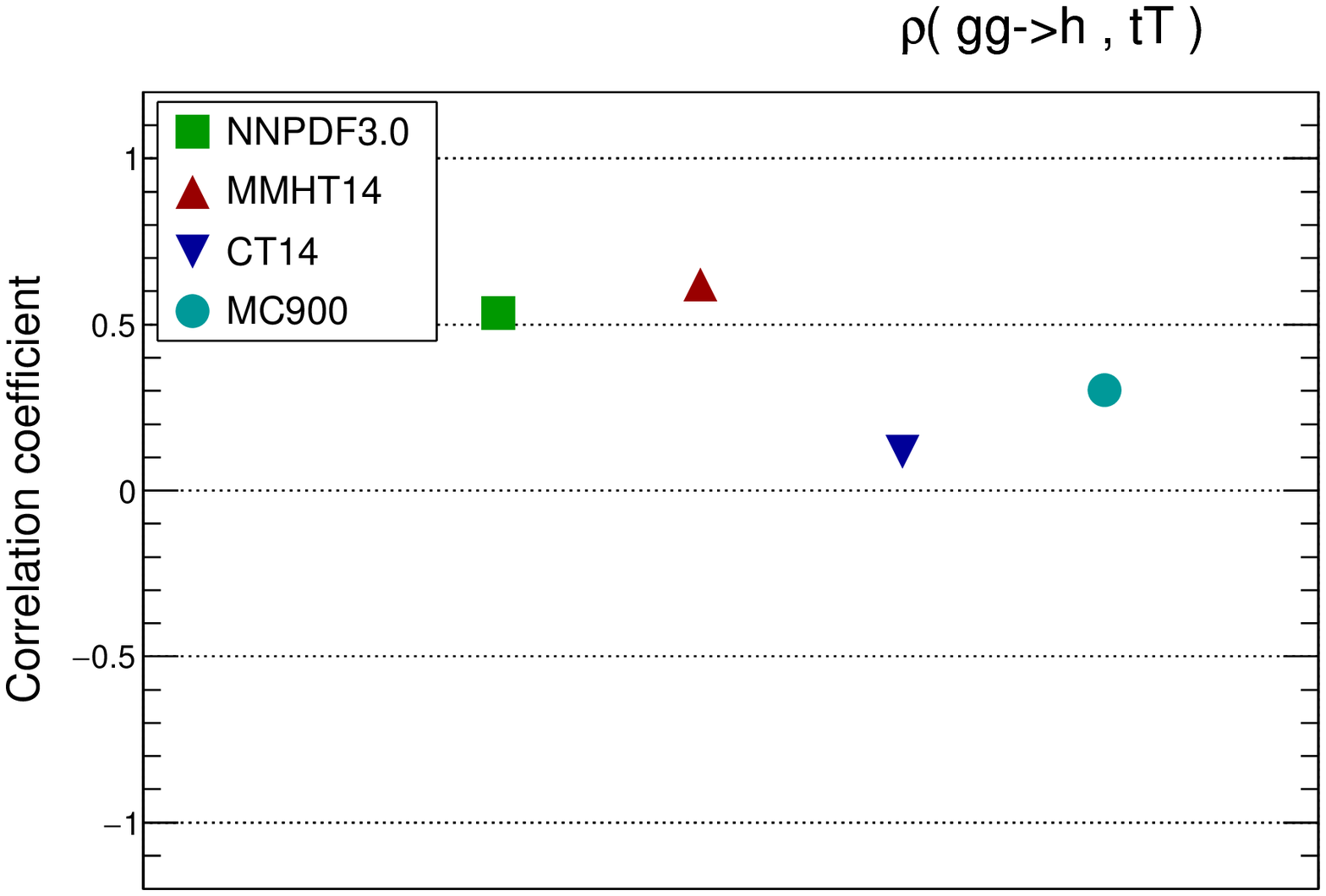}
  \includegraphics[scale=0.36]{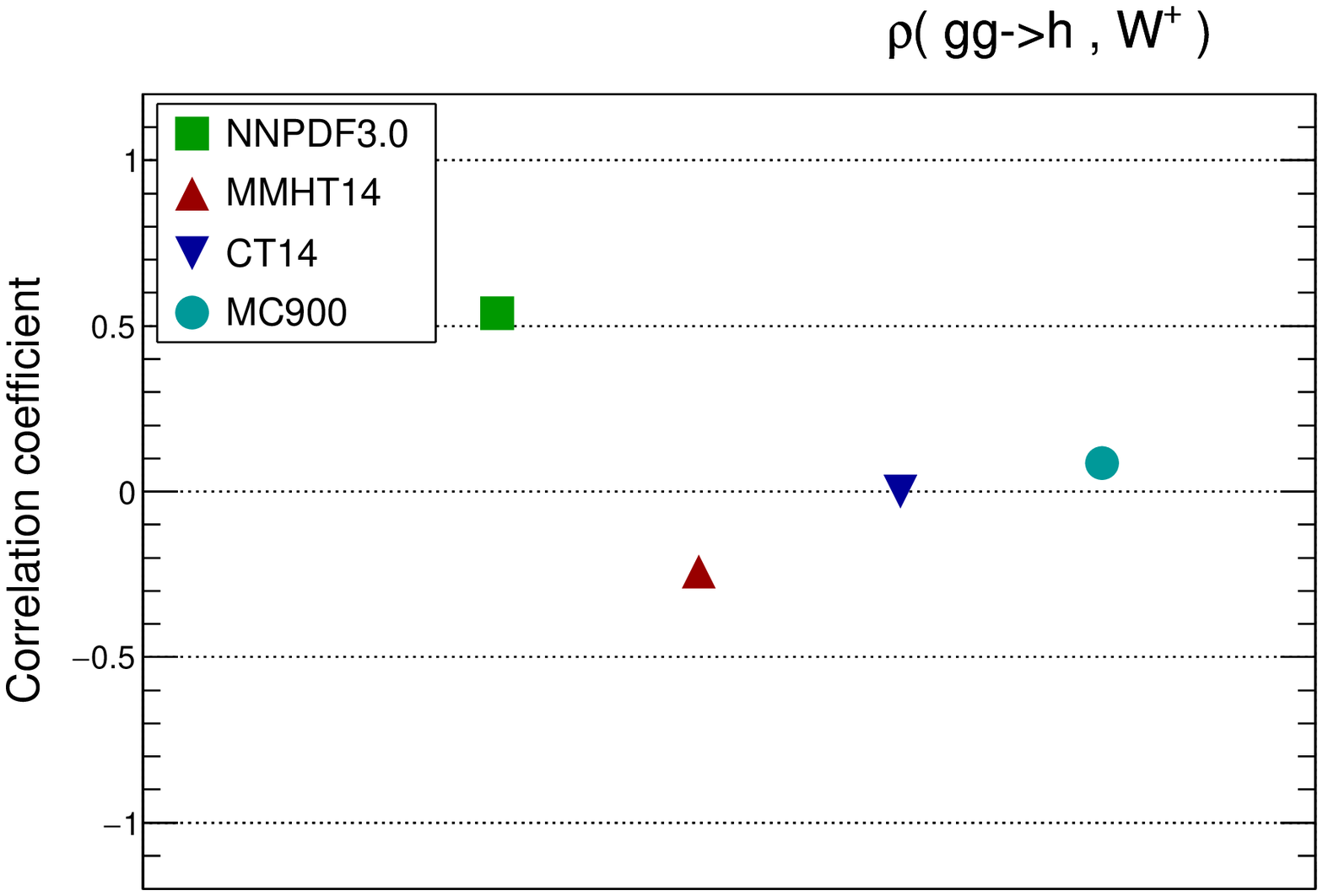}
  \includegraphics[scale=0.36]{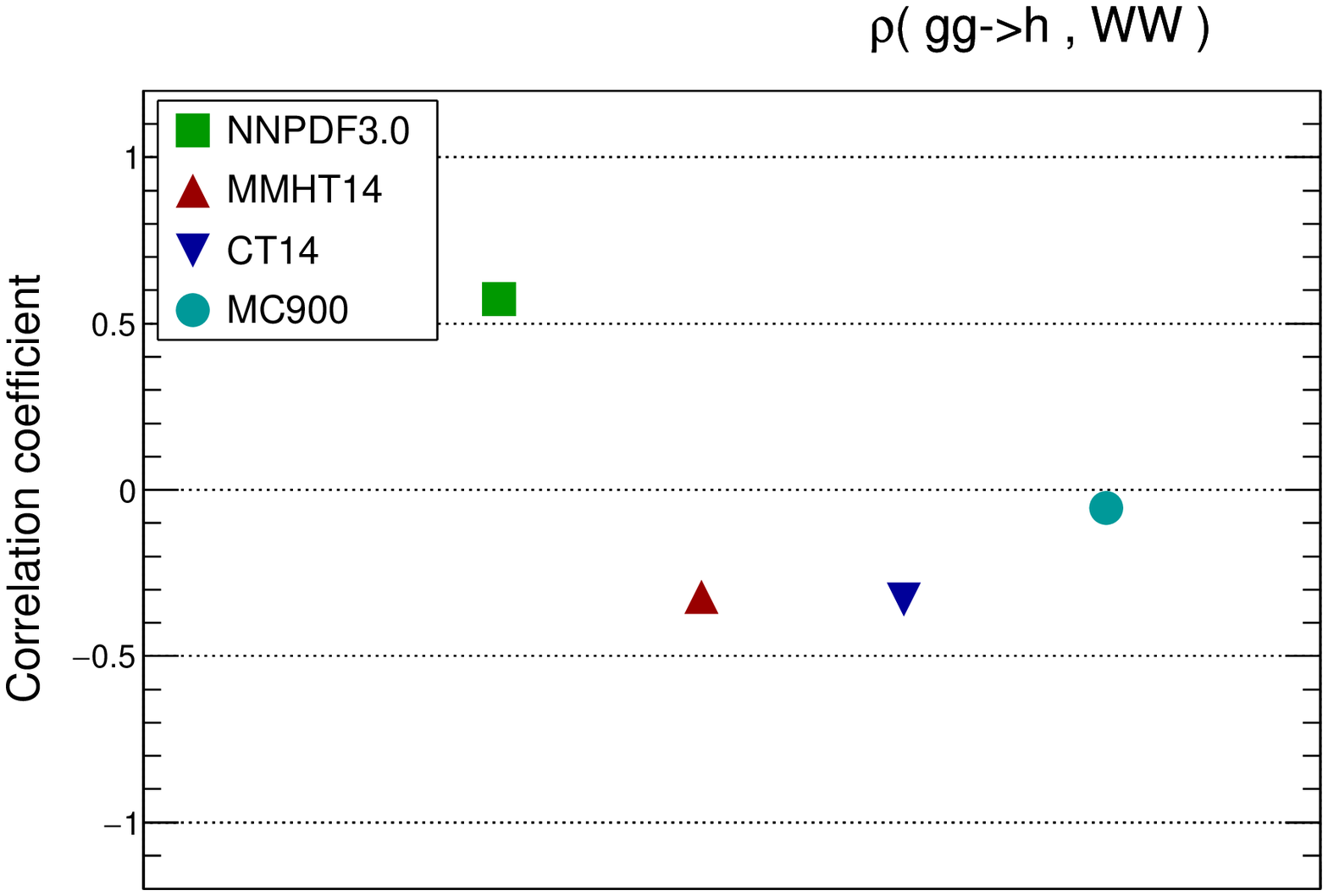}
  \includegraphics[scale=0.36]{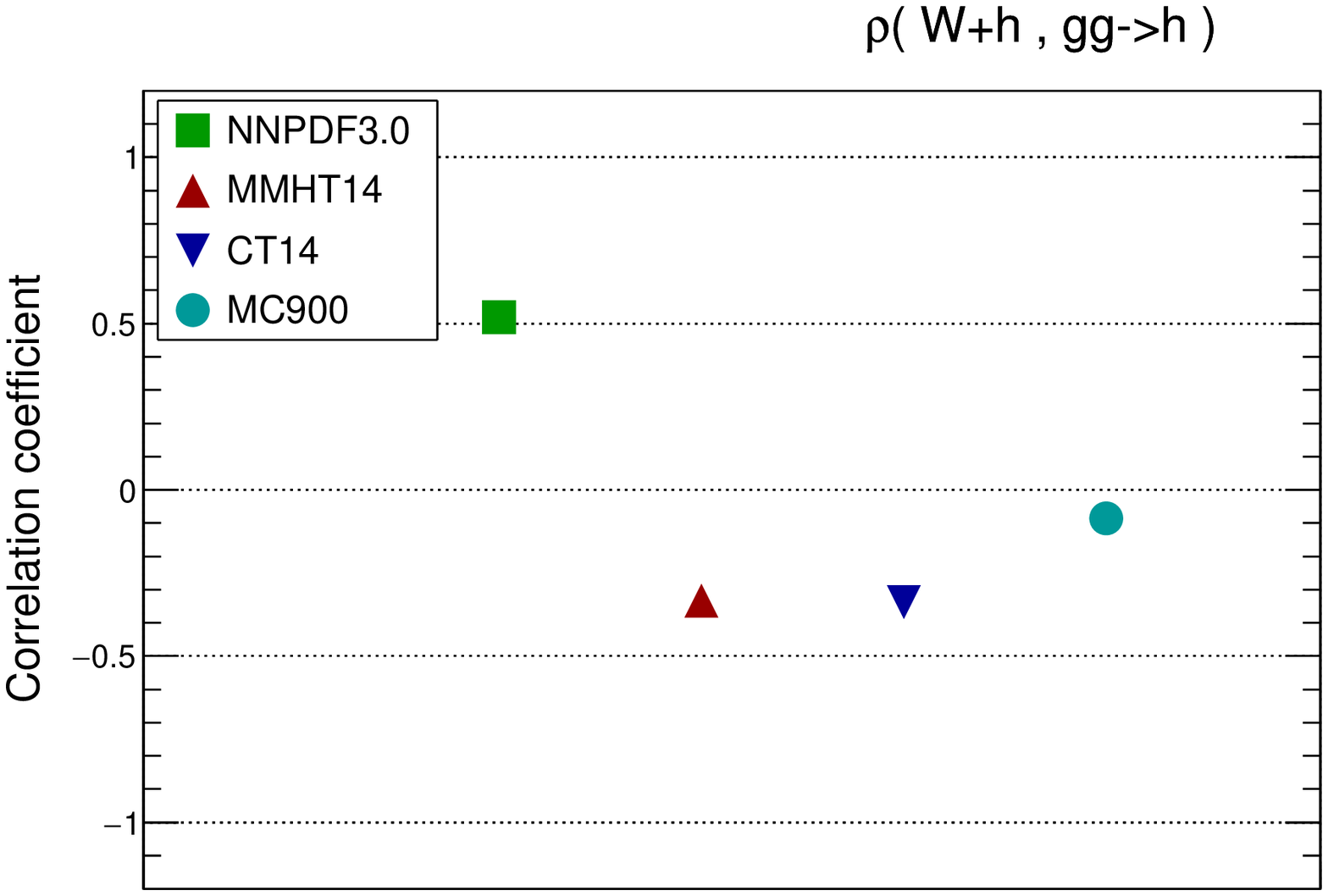}
  \includegraphics[scale=0.36]{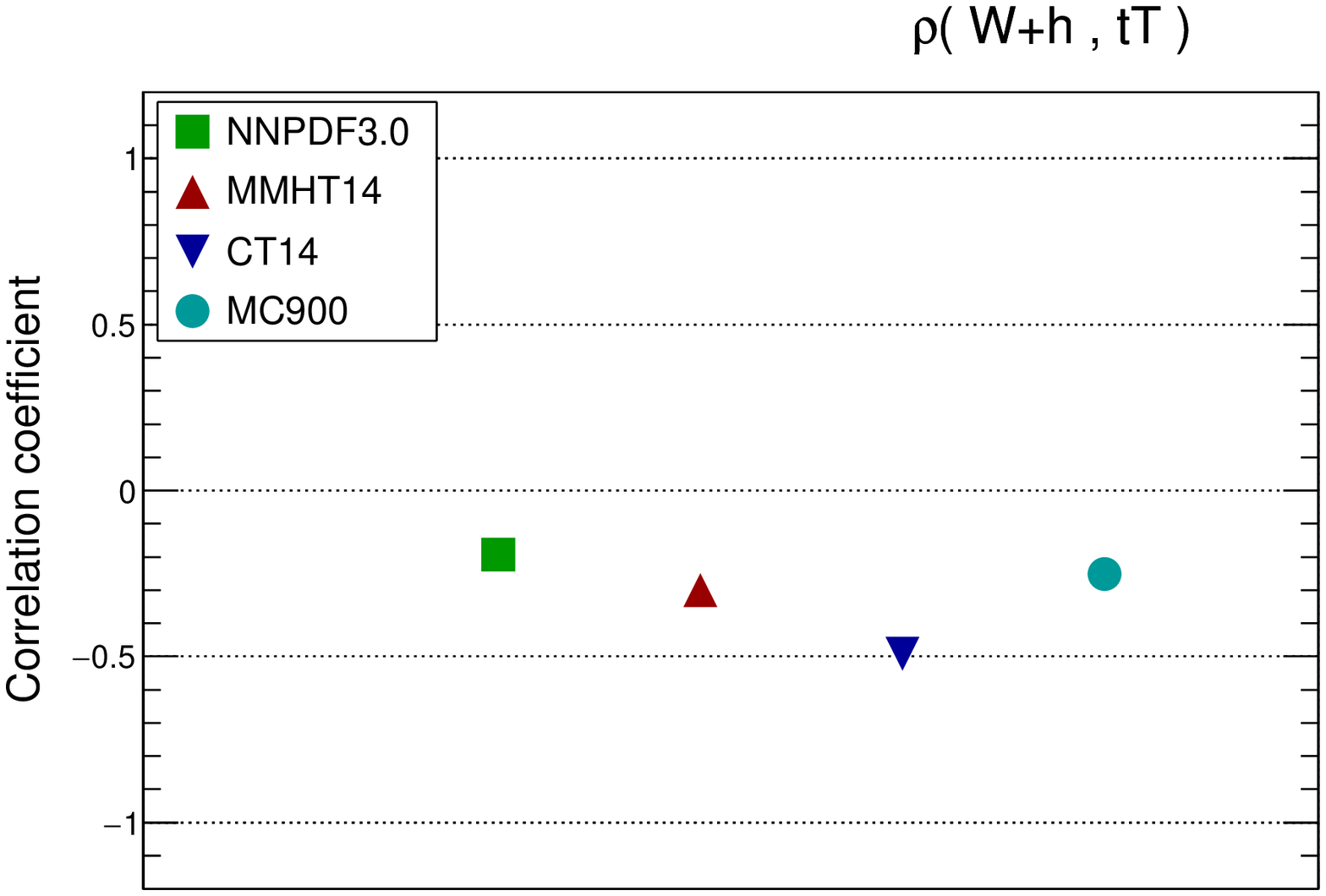}
  \includegraphics[scale=0.36]{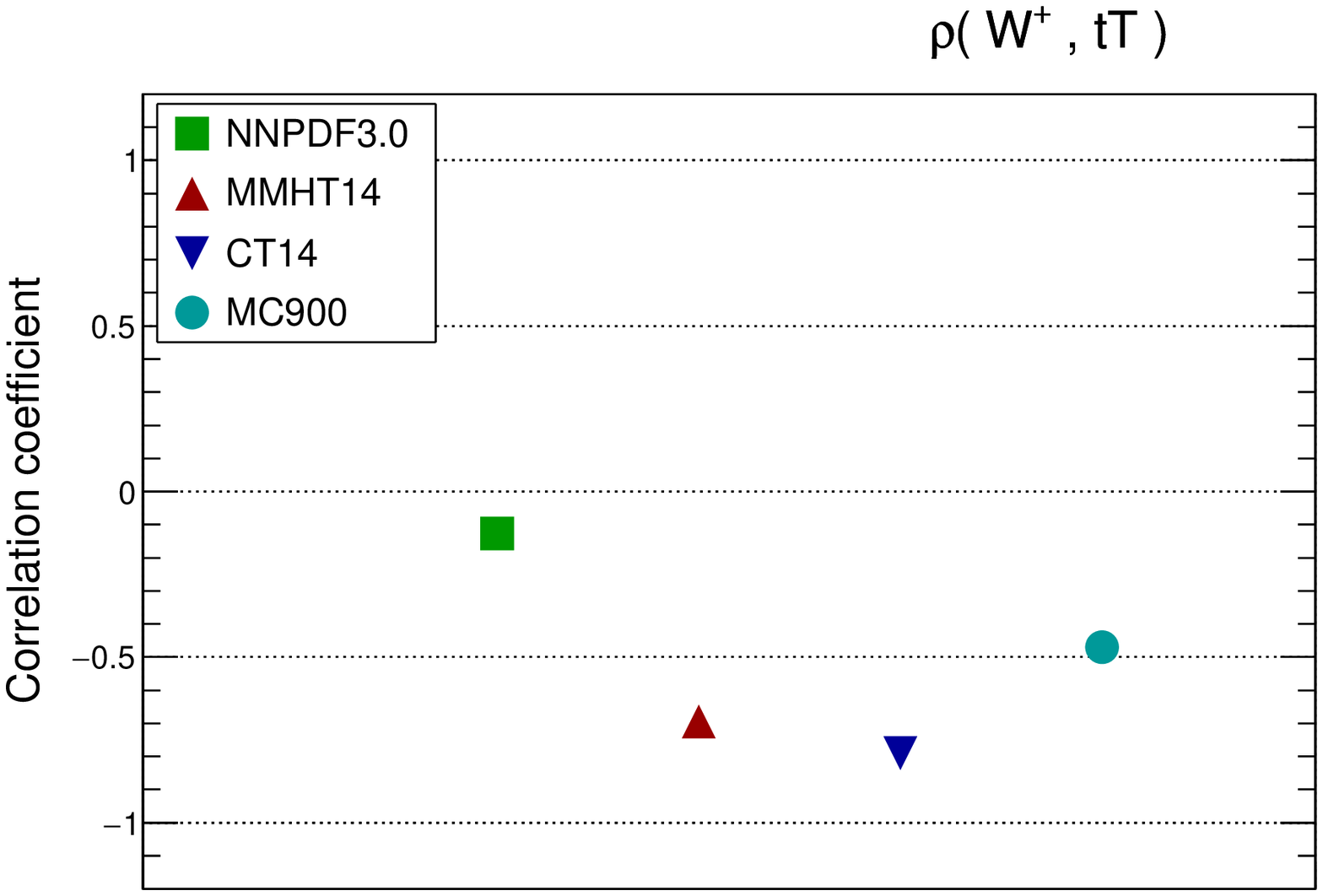}
  \caption{\small Comparison of the correlation
    coefficients between a number
    of representative NLO and NNLO LHC inclusive cross-sections
    computed from  the three individual sets, NNPDF3.0, CT14
    and MMHT14 (using the MC representation
    for the Hessian sets), and with their
    MC combination MC900.
  }
  \label{fig:xsecbenchCorr}
\end{figure}

It is also useful to compare the correlations between LHC
cross-sections computed with
the individual PDF sets and with the MC900 combined set.
A representative set of these correlations is shown in
Fig.~\ref{fig:xsecbenchCorr}, computed using the same
settings as above.
In addition to the processes shown in Fig.~\ref{fig:xsecbenchI}, here
we also show correlations for the $WW$ and $Wh$ production NLO total cross-sections
computed with {\sc\small MFCM}.
For MMHT14 and CT14, correlations are computed from their
Monte Carlo representation.
From the comparison of the
correlation  coefficients shown in Fig.~\ref{fig:xsecbenchCorr}
we note that the correlation coefficients between LHC
cross-sections for the three
global sets, NNPDF3.0, CT14
and MMHT14, can differ substantially more than for
central values and variances.
This effect was also noticed in the Higgs Cross-Section
Working Group study of PDF-induced correlations between
Higgs production channels~\cite{Dittmaier:2012vm}.
By construction, the correlation coefficient for the
combined MC prior produces the correct weighted average of
the correlations from the individual sets.

\clearpage
\section{The compression algorithm}
\label{sec:compression}

In the previous section we have described and validated
the combination of different PDF sets, based on the
Monte Carlo method.
We have shown that the probability distribution
of such a combined PDF set can be represented by $\widetilde{N}_{\rm rep}$
Monte Carlo replicas.
Now in this section we introduce a compression algorithm that aims to
determine,
for a fixed (smaller) number of MC replicas $N_{\rm rep} <
\widetilde{N}_{\rm rep}$,
the optimal subset of the original
representation that most faithfully reproduces the statistical
properties of the combined PDF prior distribution.

First of all, we begin with a presentation of the mathematical
problem, followed by a description of the
technical aspects of the compression strategy, where
we describe the choice of
error function and related parameters that
have been chosen in this work.
Then we apply the compression
method  to the combined Monte Carlo PDFs, producing what
will be dubbed as Compressed Monte Carlo PDFs (CMC-PDFs) in the rest of this
paper.
We also show how the compression strategy can be applied to
native  Monte Carlo PDF sets, using the NNPDF3.0 NLO
set with $\widetilde{N}_{\rm rep}=1000$
as an illustration.
The validation of the compression at the level of parton distributions
and physical observables is then performed in Sects.~\ref{sec:results} and \ref{sec:lhcpheno}.

\subsection{Compression: mathematical framework}

Let us begin by presenting an overview of the mathematical framework for
the problem that we aim to address, namely the compression of
a given probability distribution function.
The starting point is to consider a representation of a
probability distribution $\vec p=(p_1,\ldots,p_n)$, using a finite number $n$ of  instances.
In the case at hand, the number of instances is given
by the number of Monte Carlo replicas $\widetilde N_{\rm rep}$.
Any smaller number set of replicas, $N_{\rm rep} <
\widetilde N_{\rm rep}$, produces a corresponding probability distribution
$\vec q$ that entails a loss of information with respect to
the original distribution $\vec p$.
The problem of optimal compression can be mathematically stated as follows.
We would like to
find the specific subset of the original set of replicas such
that the statistical distance between the original and the compressed
probability distributions is minimal. In other words, we look for
a subset of replicas that delivers a probability distribution as
indistinguishable from the prior set as possible.

A number of different figures of merit to quantify the distinguishability
of probability distribution were proposed many decades ago.
Some of the first efforts  are accounted in the book of Hardy,
Littlewood and Polya \cite{Hardy:1978lp}, where ideas about
strong ordering (majorization) were introduced.
Later on, the problem of
distinguishability was
quantified using the concept of statistical distance among probability
distributions. In particular, the Kolgomorov
distance
\begin{equation}
  K(\vec p, \vec q) = \sum_i |p_i-q_i|\, , \qquad i=1,\ldots,n \, ,
\end{equation}
where the index $i$ runs over the number of instances of $\vec{p}$,
 is a simple and powerful
example of a figure
of merit that quantifies how different a probability distribution is from another one.

With the advent of Information Theory, Shannon introduced the concept of
{\it surprise} of a probability distribution 
 as its {\it distance} to the even prior.  This can be characterized using Shannon
entropy $S(\vec p)=-\sum_i p_i \log p_i$.
It is, then, natural to quantify distinguishability between
two  probability distributions $\vec p$ and $\vec q$
using entropy concepts \cite{Cover:2006}. This leads
to the construction of the Kullback-Leibler divergence
\begin{equation}
 D(\vec p || \vec q)= \sum_{i=1,\ldots,n} q_i \log \frac{q_i}{p_i} \ ,
\end{equation}
which differs from the Kolmogorov distance in the sense
that it weights more the largest probabilities. Later refinements
produced the ideas
of symmetric statistical distances, like the symmetrized Kullback and the Chernhoff distances, used in Quantum
Information nowadays.
As a consequence of these trend of ideas, it is clear
there are very many well-studied
options to define a distance in probability space. Since their
variations are not large, any of them should be suitable
for the problem of Monte Carlo PDF compression, and we present
our specific choice in Sect.~\ref{sec:strategy}.

Let us now be more precise on the way we shall proceed.
If we define $\left\{\vec{p}\right\}$ as the original representation of the
probability distribution (with $\widetilde{N}_{\rm rep}$ replicas)
and $\left\{\vec{q}\right\}$ its compressed version
(with $N_{\rm rep}$ replicas), then given the concept of a distance $d$
between two probability distributions there is an optimal choice
of the subset with $N_{\rm rep}$ replicas defined as
\be
\label{eq:concept}
\left\{\vec{q}\right\}_{\rm opt} \equiv {\rm Min}_{\left\{\vec{q}\right\}}
\lc d \lp \left\{\vec{q}\right\},\left\{\vec{p}\right\} \rp \rc \, .
\ee
Therefore, the mathematical problem at stake is reduced to finding the optimal
subset $\left\{\vec{q}\right\}_{\rm opt}$, by a suitable exploration
of the space of minima of the distance
$d \lp \left\{\vec{q}\right\},\left\{\vec{p}\right\} \rp$.
In this work, this exploration is performed using Genetic Algorithms,
though many other choices would also be suitable.
Fortunately, many choices of subset are equally good minimizations.
From the practical point of view, the specific choice
of the  minimization strategy is not critical.
It is clear that the relevant point is the definition
of a distance between the original and
compressed replica sets.
In this paper we shall take the following approach.

Many
valid definitions of statistical distance differ in the 
way different moments are weighted. Since we are interested
in reproducing real physics, which is dominated by low
moments, we shall explicitly include in our figure of merit
all the distances between means and standard deviations, but also
kurtosis, skewness and correlations, as well as higher moments.
As a consequence, all of them will be minimized, favoring
the role of smaller moments.

\subsection{A compression algorithm for Monte Carlo PDF sets}
\label{sec:strategy}

As we have discussed above,
the most important ingredient for the compression strategy is the
choice of a suitable distance between the prior and the
compressed distributions, Eq.~(\ref{eq:concept}),
or in other words, the definition
of the error function (ERF in the following) for the minimization
problem.
We have explored different possibilities, and the precise
definition of the ERF that will be used in this work can
be written generically as follows
\begin{equation}
\label{eq:erf}
      {\rm ERF}=\sum_{k} \frac{1}{N_{k}}
      \sum_{i}\left(\frac{C^{(k)}_{i}-O^{(k)}_{i}}{O^{(k)}_{i}}\right)^{2},
\end{equation}
where $k$ runs over the number of statistical estimators
used to quantify the distance between the original
and compressed distributions,
$N_{k}$ is a normalization factor,
$O^{(k)}_{i}$ is the value of the estimator $k$
(for example, the mean or the variance) computed
at the generic point $i$ (which could be a given value
of $(x,Q)$ in the PDFs, for instance), and $C^{(k)}_{i}$ is
the corresponding value of the same estimator in the compressed set.
The choice of a normalized ERF is important for the accuracy of the
minimization because some statistical estimators,
in particular higher moments, can
span various orders of magnitude in
different regions of $x$ and $Q^2$.

\begin{figure}[t]
  \centering 
  \includegraphics[scale=0.5]{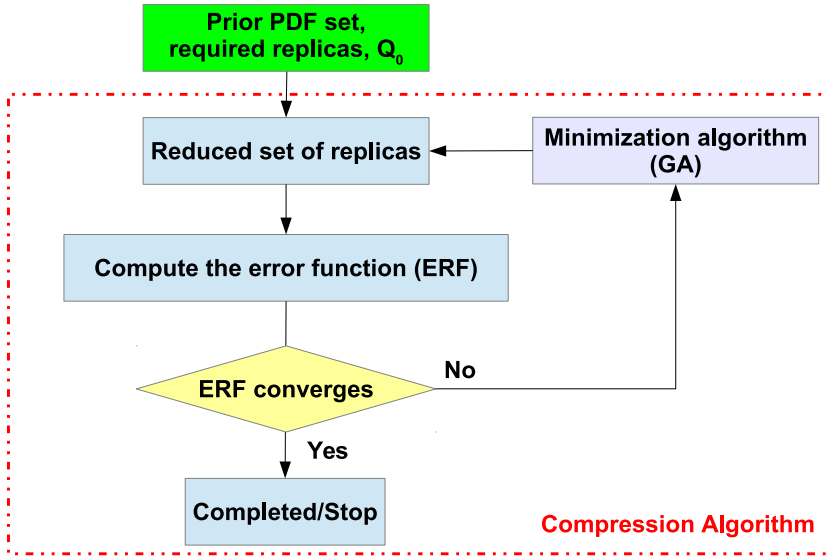}
  \vspace{-1.7cm}
  \caption{\small Schematic
    representation of the
    compression strategy used in this work: a prior PDF set and the number of
    compressed replicas is the input of a GA algorithm which selects
    the best subset of replicas which minimizes the ERF between the
    prior and the compressed set.}
  \label{fig:scheme}
\end{figure}

An schematic diagram for our compression strategy is shown in
Fig.~\ref{fig:scheme}.
The prior set of Monte Carlo PDF replicas, the
desired number of compressed replicas, $N_{\rm rep}$, and the
value of the factorization scale $Q$ at which the PDFs are evaluated, $Q_0$,
are the required parameters for the compression
algorithm.
Note that it is enough to sample the PDFs in a range of
values of Bjorken-$x$ at a fixed value of $Q_0$, since
the DGLAP equation uniquely determines the evolution for
higher scales $Q \ge Q_0$.
The minimization of the error function is
performed using Genetic Algorithms (GAs), similarly
as in the neural network training of the
NNPDF fits.
GAs work as usual by finding candidates for subsets of
$N_{\rm rep}$ leading to smaller values of the error
function Eq.~(\ref{eq:erf}) until some suitable
convergence criterion is satisfied.
The output of this
algorithm is thus the list of the $N_{\rm rep}$ replicas from the prior
set of  $\widetilde{N}_{\rm rep}$ that minimize the error function.
These replicas define the 
 CMC-PDFs for each specific value of $N_{\rm rep}$.
The final step of the process is a series of validation
tests where the CMC-PDFs
are compared to the prior set in terms of parton
distributions at different scales, luminosities and
LHC cross-sections, in a fully automated way.
%

It is important to emphasize that the compression algorithm
only selects replicas from a prior set, and no attempt is
made to use common theoretical settings, {\it i.e.},
the
method for the solution of the DGLAP evolution equations, or
the values of the heavy-quark masses, which are those of
the corresponding original PDF sets.
This important fact automatically ensures that the compressed
set conserves all basic physical requirements of the original
combined set such as the positivity
of physical cross-sections, sum rules and the original
correlations between PDFs.
To avoid problems related to the different treatment of the heavy-quark
thresholds between the different groups, we choose in this work
to compress the combined MC PDF set at a common scale of $Q_0=2$~GeV, while
we use $Q_0=1$~GeV when compressing the native NNPDF3.0 NLO set.

The compression strategy seems conceptually simple:
reducing the size of a Monte Carlo PDF set requiring no
substantial loss of information.
In order to achieve its goal, the compression algorithm
must preserve as much as possible the underlying statistical
properties of the prior PDF set.
However, this conceptual simplicity is followed
by a series of non-trivial issues that have to be addressed
in the practical implementation.
Some of these issues are the sampling of the PDFs
in Bjorken-$x$, the exact definition of the error function,
Eq.~(\ref{eq:erf}), the treatment of PDF correlations
and the settings of the GA minimization.
We now discuss these various issues in turn.

\subsubsection{Definition of the error function for the compression}

In this work we include in the ERF, Eq.~(\ref{eq:erf}), the distances between the prior and
the compressed sets of PDFs for the following estimators:
\begin{itemize}

\item The first four moments of the distribution, which are sampled
  in a grid of $x$ points for $n_{f}$ flavours in terms of central
  value, standard deviation, skewness and kurtosis,
at a fixed value of $Q=Q_0$.
  It is important to
  notice that these estimators are necessary in order to obtain a
  realistic and optimized compressed MC set, but are not sufficient to
  avoid eventual bias of continuity and loss of structure.

\item The output of the Kolmogorov-Smirnov test. This is the simplest
  distance between empirical probability distributions.
  This distance
  complements the  terms in the ERF which contain the first
  four moments, by ensuring that also higher moments
  are automatically adjusted.
  However, if this estimator is used alone
  possible ambiguities arise when defining the regions where the
  distance is computed, leading to large errors when working with few
  replicas.
  
\item The correlation between multiple PDF flavours at different $x$
  points. This information is important for ensuring that
  PDF-induced correlations in physical cross-sections
  are successfully maintained.

\end{itemize}

The final figure of merit used in the compression fit is then the sum
over all these six estimators opportunely weighted by the
corresponding 
normalization factors $N_{k}$ in Eq.~(\ref{eq:erf}).
This normalization
is required due to the fact that the absolute value
of the various estimators can vary among them by several orders
of magnitude.

  \subsubsection{Central values, variances and higher moments}

  Let's denote by $g_{i}^{(k)}(x_{j},Q_0)$ and
  $f_{i}^{(r)}(x_{j},Q_0)$ respectively the
prior and the compressed sets of replicas for a flavor $i$ at the
position $j$ of the $x$-grid containing $N_{x}$ points.
$N_{{\rm rep}}$ is the number of required compressed
replicas.
We then define the contribution to the ERF from
the distances between central values of the prior and compressed distributions
as follows
\begin{equation}
\label{eq:erfcv}
  {\rm ERF}_{\rm CV}=\frac{1}{N_{{\rm CV}}}\sum_{i=-n_{f}}^{n_{f}}\sum_{j=1}^{N_{x}}\left(\frac{f_{i}^{{\rm CV}}(x_{j},Q_0)-g_{i}^{{\rm CV}}(x_{j},Q_0)}{g_{i}^{{\rm CV}}(x_{j},Q_0)}\right)^{2},
\end{equation}
where $N_{\rm CV}$ is the normalization factor for this estimator. We
only include in the sum those points for which the denominator
satisfies $g_{i}^{\rm CV}(x_{j},Q_0)\neq0$. As usual, central values
are computed as the average over the MC replicas, for the compressed
set
\begin{equation}
  f_{i}^{{\rm CV}}(x_{j},Q_0)=\frac{1}{N_{{\rm rep}}}
  \sum_{r=1}^{N_{{\rm rep}}}f_{i}^{(r)}(x_{j},Q_0) \, ,
\end{equation}
while for the prior set we have
\begin{equation}
g_{i}^{{\rm CV}}(x_{j},Q_0)=\frac{1}{\widetilde{N}_{{\rm rep}}}\sum_{k=1}^{\widetilde{N}_{{\rm rep}}}g_{i}^{(k)}(x_{j},Q_0) \, .
\end{equation}

Let us also define
 $r^{t}_{i}(x_{j},Q_0)$ as a random set of replicas
extracted from the prior set, where $t$ identifies an ensemble of
random extractions. The number of random extraction of random sets is
denoted by $N_{{\rm rand}}$.
Now, the normalization factors are extracted for all estimators as the
lower 68\% confidence-level value obtained after $N_{\rm rand}$ realizations of
random sets. In particular for this estimator we have
\begin{equation}
\label{eq:normcv}
  N_{{\rm CV}}=\left.\frac{1}{N_{{\rm rand}}}\sum_{d=1}^{N_{{\rm
        rand}}}\sum_{i=-n_{f}}^{n_{f}}\sum_{j=1}^{N_{x}}\left(\frac{r_{i}^{d,{\rm
          CV}}(x_{j},Q_0)-g_{i}^{{\rm CV}}(x_{j},Q_0)}{g_{i}^{{\rm
          CV}}(x_{j},Q_0)}\right)^{2}\right|_{\text{68\% lower band}}.
\end{equation}

For the contribution to the ERF from
the distance between standard deviation, skewness and kurtosis, we can
built expressions analogous to that of 
 Eq.~(\ref{eq:erfcv}) by replacing the central value
 estimator with the suitable expression for the other statistical
 estimators, which in a Monte Carlo representation can be computed as
\begin{eqnarray}
f_{i}^{{\rm STD}}(x_{j},Q_0)&=&\sqrt{\frac{1}{N_{{\rm rep}}-1}\sum_{r=1}^{N_{{\rm rep}}}\left(f_{i}^{(r)}(x_{j},Q_0)-f_{i}^{{\rm CV}}(x_{j},Q_0)\right)^{2}},\\
f_{i}^{{\rm SKE}}(x_{j},Q_0)&=&\frac{1}{N_{{\rm rep}}}\sum_{r=1}^{N_{{\rm rep}}}\left(f_{i}^{(r)}(x_{j},Q_0)-f_{i}^{{\rm CV}}(x_{j},Q_0)\right)^{3}/\left(f_{i}^{{\rm STD}}(x_{j},Q_0)\right)^{3},\\
f_{i}^{{\rm KUR}}(x_{j},Q_0)&=&\frac{1}{N_{{\rm rep}}}\sum_{r=1}^{N_{{\rm rep}}}\left(f_{i}^{(r)}(x_{j},Q_0)-f_{i}^{{\rm CV}}(x_{j},Q_0)\right)^{4}/\left(f_{i}^{{\rm STD}}(x_{j},Q_0)\right)^{4}\, ,
\end{eqnarray}
for the compressed set, with analogous expressions for the
original prior set.

The normalization factors for these estimators are extracted using the
same strategy presented in Eq.~(\ref{eq:normcv}), by averaging over
random extractions of $N_{\rm rep}$ replicas, exchanging CV by STD, SKE and
KUR respectively.

\subsubsection{The Kolmogorov-Smirnov distance}

As we have mentioned above, the minimization of the Kolmogorov-Smirnov distance
ensures that both lower and higher moments of the prior
distribution are successfully reproduced.
In our case,
we define the contribution to the total ERF from  the Kolmogorov-Smirnov (KS)
distance as follows
\begin{equation} {\rm ERF}_{\rm KS}=\frac{1}{N_{\rm
      KS}}\sum_{i=-n_{f}}^{n_{f}}\sum_{j=1}^{N_{x}}\sum_{k=1}^{(r)}\left(\frac{F_{i}^{k}(x_{j},Q_0)-G_{i}^{k}(x_{j},Q_0)}{G_{i}^{k}(x_{j},Q_0)}\right)^{2}.
\end{equation}
where $F^k_i(x_j,Q_0)$ and $G^k_i(x_j,Q_0)$ are the outputs of the test
for the compressed and the prior set of replicas respectively. The
output of the test consists in counting the number of replicas
contained in the $k$ regions where the test is performed. We count the
number of replicas which fall in each region and then we normalize by
the total number of replicas of the respective set.
Here we have considered six regions defined as multiples of the
standard deviation of the distribution for each flavor $i$ and
$x_{j}$-point.  As an example for the compressed set, the regions are
\begin{equation}
[-\infty,-2f_{i}^{{\rm STD}}(x_{j},Q_0),-f_{i}^{{\rm STD}}(x_{j},Q_0),0,f_{i}^{{\rm STD}}(x_{j},Q_0),2f_{i}^{{\rm STD}}(x_{j},Q_0),+\infty],
\end{equation}
where the values of the PDFs have been subtracted from the corresponding
central value.

In this case, the normalization factor is determined from the output of the KS test for
random sets of replicas extracted from the prior, denoted $R^k_i(x_j,Q_0)$
as follows
\begin{equation}
N_{\rm KS}=\frac{1}{N_{{\rm rand}}}\sum_{d=1}^{N_{{\rm rand}}}\sum_{i=-n_{f}}^{n_{f}}\sum_{j=1}^{N_{x}}\sum_{k=1}^{6}\left(\frac{R_{i}^{k}(x_{j},Q_0)-G_{i}^{k}(x_{j},Q_0)}{G_{i}^{k}(x_{j},Q_0)}\right)^{2},
\end{equation}
and we only include in the sum those points for which
 the denominator satisfies $G_{i}^{k}(x_{j},Q_0)\neq0$.

\subsubsection{PDF correlations}
In addition to all the moments of the prior distribution,
a sensible compression should also maintain the correlations between
values of $x$ and between flavours of the PDFs.
In order to achieve this, correlations are taken into account
in the ERF by means of the trace method.
We
define a correlation matrix $C$ for any PDF set as follows:
\begin{equation}
\label{eq:corr}
C_{ij}=\frac{N_{\rm rep}}{N_{\rm rep}-1}\cdot\frac{\langle ij\rangle-\langle i\rangle\langle
  j\rangle}{\sigma_{i}\cdot\sigma_{j}}\, ,
\end{equation}
where we have defined
\begin{equation}
\langle i\rangle=\frac{1}{N_{{\rm rep}}}\sum_{r=1}^{N_{{\rm
      rep}}}f_{i}^{(r)}(x_{i},Q_0)\, ,\quad\langle ij\rangle=\frac{1}{N_{{\rm rep}}}\sum_{r=1}^{N_{{\rm rep}}}f_{i}^{(r)}(x_{i},Q_0)f_{j}^{(r)}(x_{j},Q_0)\, ,
\end{equation}
and $\sigma$ is the usual expression for the standard deviation
\begin{equation}
\sigma_{i}=\sqrt{\frac{1}{N_{{\rm rep}}-1}\sum_{r=1}^{N_{{\rm rep}}}\left(f_{i}^{(r)}(x_{i},Q_0)-\langle i\rangle\right)^{2}}.
\end{equation}

Now, for each flavor $n_{f}$ we define $N^{\rm corr}_{x}$ points
distributed in $x$ where the correlations are computed. The trace
method consists in computing the correlation matrix $P$ based on
Eq.~(\ref{eq:corr}) for the prior set and then store its inverse
$P^{-1}$. For $n_{f}$ flavours and $N^{\rm corr}_{x}$ points we obtain
\begin{equation}
g={\rm Tr}\left(P\cdot P^{-1}\right)= N^{\rm corr}_{x}\cdot(2 \cdot n_{f} + 1).
\end{equation}

After computing the correlation matrix for prior set, for each
compressed set a matrix $C$ is computed and the trace is determined by
\begin{equation}
f={\rm Tr}\left(C\cdot P^{-1}\right).
\end{equation}
The compression algorithm then includes the correlation ERF by
minimizing the quantity:
\begin{equation}
{\rm ERF}_{\rm Corr}=\frac{1}{N_{\rm Corr}}\left(\frac{f-g}{g}\right)^{2}
\end{equation}
where $N_{\rm Corr}$ is computed as usual from the random sets,
in the same way as Eq.~(\ref{eq:normcv}).

\subsubsection{Choice of GA parameters in {\tt compressor v1.0.0} }
\label{ref:settings}

The general strategy that has been presented in this section has
been implemented in {\tt compressor v1.0.0}, the name of the public code~\cite{compressor} released
together with this paper. A more detailed
description of the code usage is provided in
Appendix~\ref{sec:compressioncode}.
The availability of this code ensures that it will be possible to easily
redo the compression for any further combination of PDF sets
that might be considered in the future.

\begin{table}[H]
  \begin{center}   
    \begin{tabular}{|c||c|}
      \hline
      \multicolumn{2}{|c|}{{\tt compressor v1.0.0}}    \\ \hline
      \hline
      \multicolumn{2}{|c|}{GA Parameters}    \\
      \hline 
      $N_{\rm gen}^{\rm max}$ &  15000  \\
      $N_{\rm mut}$ &  5  \\
      $N_{\rm x}$ &  70  \\    
      $x_{\rm min}$ &  $10^{-5}$  \\
      $x_{\rm max}$ &  $0.9$  \\
      $n_{f}$ &  7  \\
            $Q_0$ &  user-defined  \\
      $N^{\rm corr}_{\rm x}$ & 5 \\
      $N_{\rm rand}$ & 1000 \\
      \hline
    \end{tabular}   \hskip10pt \begin{tabular}{|c|c|}
      \hline
      \multicolumn{2}{|c|}{{\tt compressor v1.0.0}}    \\ \hline
      \hline 
      $N_{\rm rep}^{\rm mut}$ &   $P_{\rm mut}$ \\
      \hline
      1 & 30\% \\
      2 & 30\% \\
      3 & 10\% \\
      4 & 30\% \\
      \hline
    \end{tabular}
  \end{center}
  \caption{\small Left table: setting of the compression algorithm
    used in this work.
    Right table: mutation rates used in the Genetic Algorithm
    minimization.}
  \label{tab:gapars}
\end{table}

This said, there is a certain flexibility in the choice of
settings for the compression, for example in the choice
of parameters for the Genetic Algorithms, the sampling of
the PDFs in $x$ or the choice of common scale for the compression
$Q_0$.
The compression setup used in this paper is presented in
Table~\ref{tab:gapars} together with the optimal set of GA
parameters and mutation probability rates, determined
by trial and error.

As mentioned before, in the present work the compression of CMC-PDFs
is performed at a scale of $Q_0=2$ GeV while in the next section we
use $Q_0=1$ GeV for the native NNPDF3.0 NLO set.
The ERF includes only the contribution of the $n_f=7$ light partons:
$u$, $\bar{u}$, $d$, $\bar{d}$, $s$, $\bar{s}$, and $g$.
Concerning the sampling of the PDFs in $x$,
we have limited the range of $x$ points to the
region where data is available, i.e. $x\sim[10^{-5},0.9]$, by
selecting 35 points logarithmically spaced between $[10^{-5}, 0.1]$ and 35
points linearly spaced from $[0.1, 0.9]$.
Note that this is different to the Meta-PDF approach, where for each
PDF a different range $\lc x_{\rm min}, x_{\rm max}\rc$ is used for the
fit with the meta-parametrisation, restricted to the regions
where experimental constraints are available for each flavor.

The correlation matrix
is then computed for the $n_f$ input
PDFs in $N^{\rm corr}_{\rm x}=5$ points in $x$, generating a
correlation matrix of 35 entries.
Increasing the number of points for the calculation of the correlation
matrix
would be troublesome since numerical instabilities due to the presence
of large correlations between neighboring points in $x$ would be introduced.

The Genetic Algorithm minimization is performed for a fixed length
of 15k generations.
Note that as opposed to the neural network learning in the NNPDF fits,
in the compression problem there is no risk of over-learning, since
the absolute minimum of the error function always exists.
On the other hand, we find that after a few thousand generations the ERF
saturates and no further improvements are achieving by running the code
longer, hence the maximum number of GA generations
$N_{\rm gen}^{\rm max}=15$k
used in this work.

\subsection{Results of the compression for native MC PDF sets}

\begin{figure}[t]
\centering 
\includegraphics[scale=0.52]{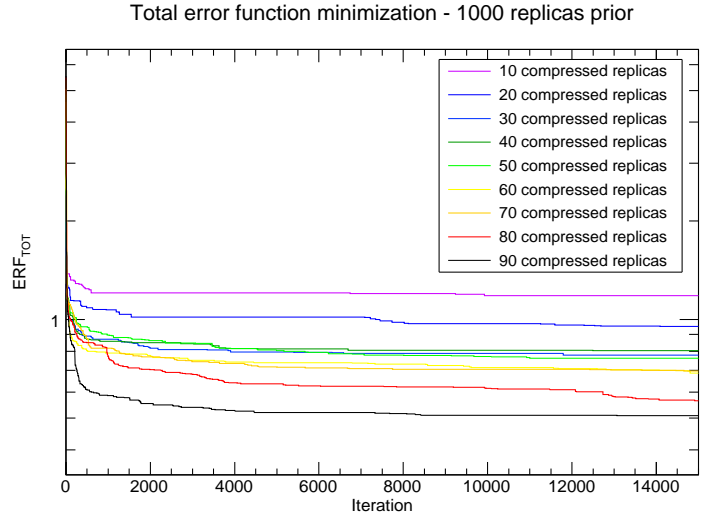}
\caption{\small The value of the total
  error function, Eq.~(\ref{eq:erf}), for the compression
  of the 1000 replica set of NNPDF3.0 NLO, as a function of the number
  of GA generations, for different values of the
  number of replicas in the compressed set $N_{\rm rep}$.
  After 15k iterations, the error function saturates and no further
  improvement of the error function would be achieved for longer training.}
\label{fig:erfmc}
\end{figure}

In order to illustrate the performance of the compression algorithm, we
consider here the compression of a native Monte Carlo set of PDFs at
$Q_0=1$ GeV, based on the prior set with $\widetilde{N}_{\rm rep}=$1000
replicas of NNPDF3.0 NLO.
In Fig.~\ref{fig:erfmc} we show the dependence of the total ERF as a
function of the number of iterations of the GA for $N_{\rm
  rep}=10,20,30,40,50,60,70,80$ and $90$.
We observe that the first 1k
iterations are extremely important during the minimization, while after
15k iterations the total error function is essentially flat for any required
number of compressed replicas.
For each compression, the final value
of the error function is different, with deeper minima being
achieved as we increase the number of compressed replicas,
as expected.
The flatness of the ERF as a function of the number of iterations
confirms that the
current parameters provide a suitably efficient minimization
strategy.

\begin{figure}
\centering 
\includegraphics[scale=0.4]{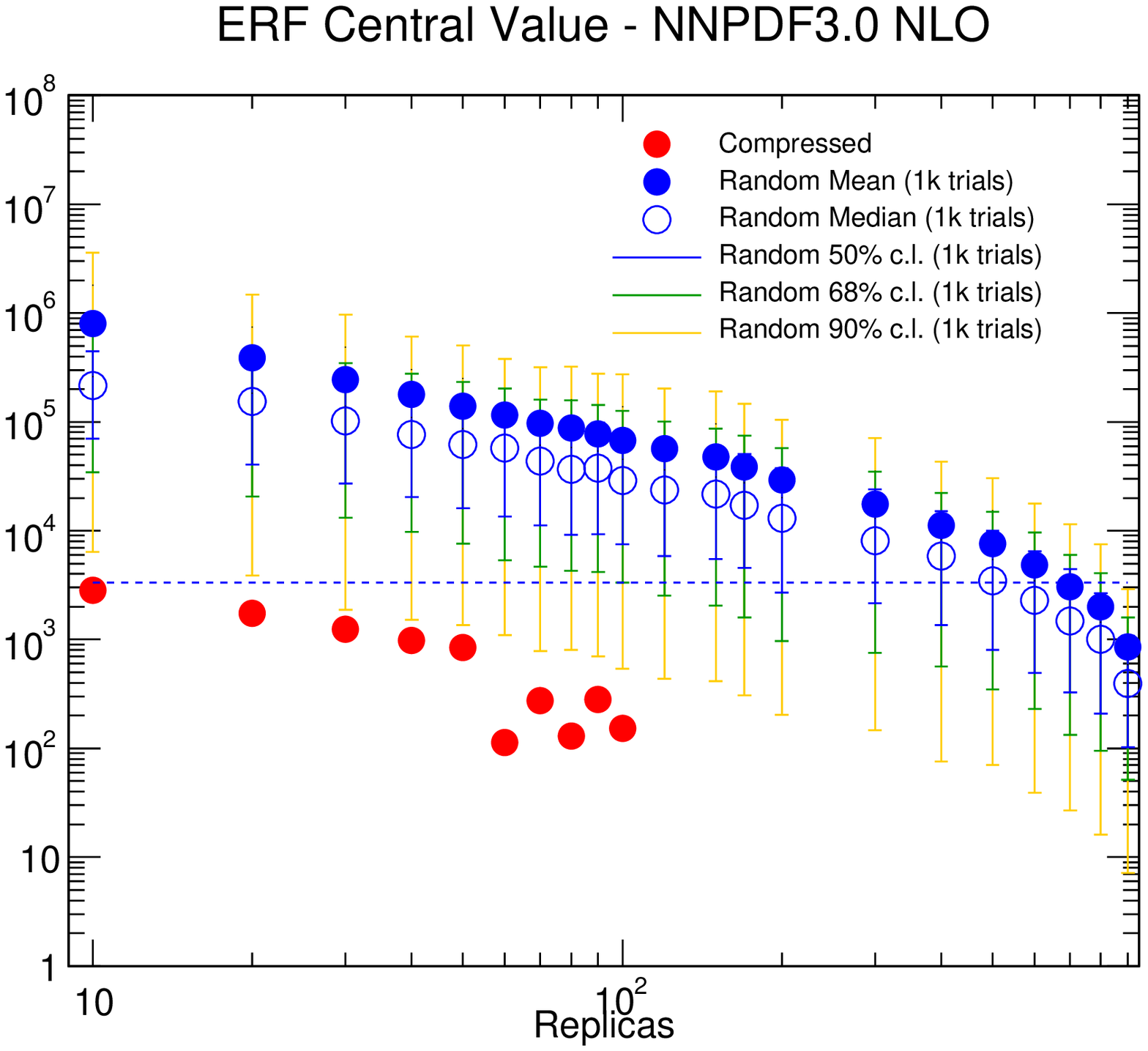}\includegraphics[scale=0.4]{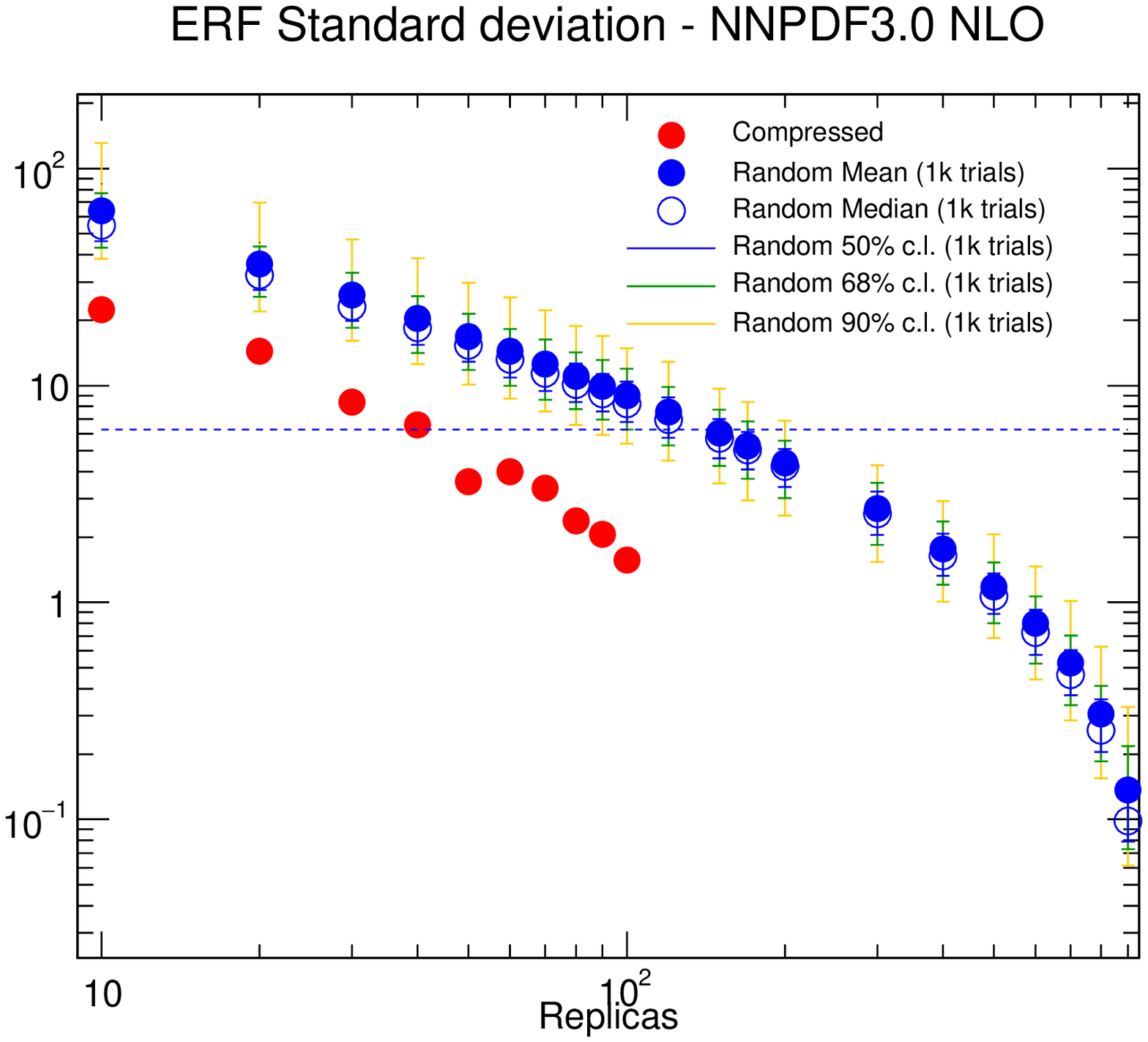}
\includegraphics[scale=0.4]{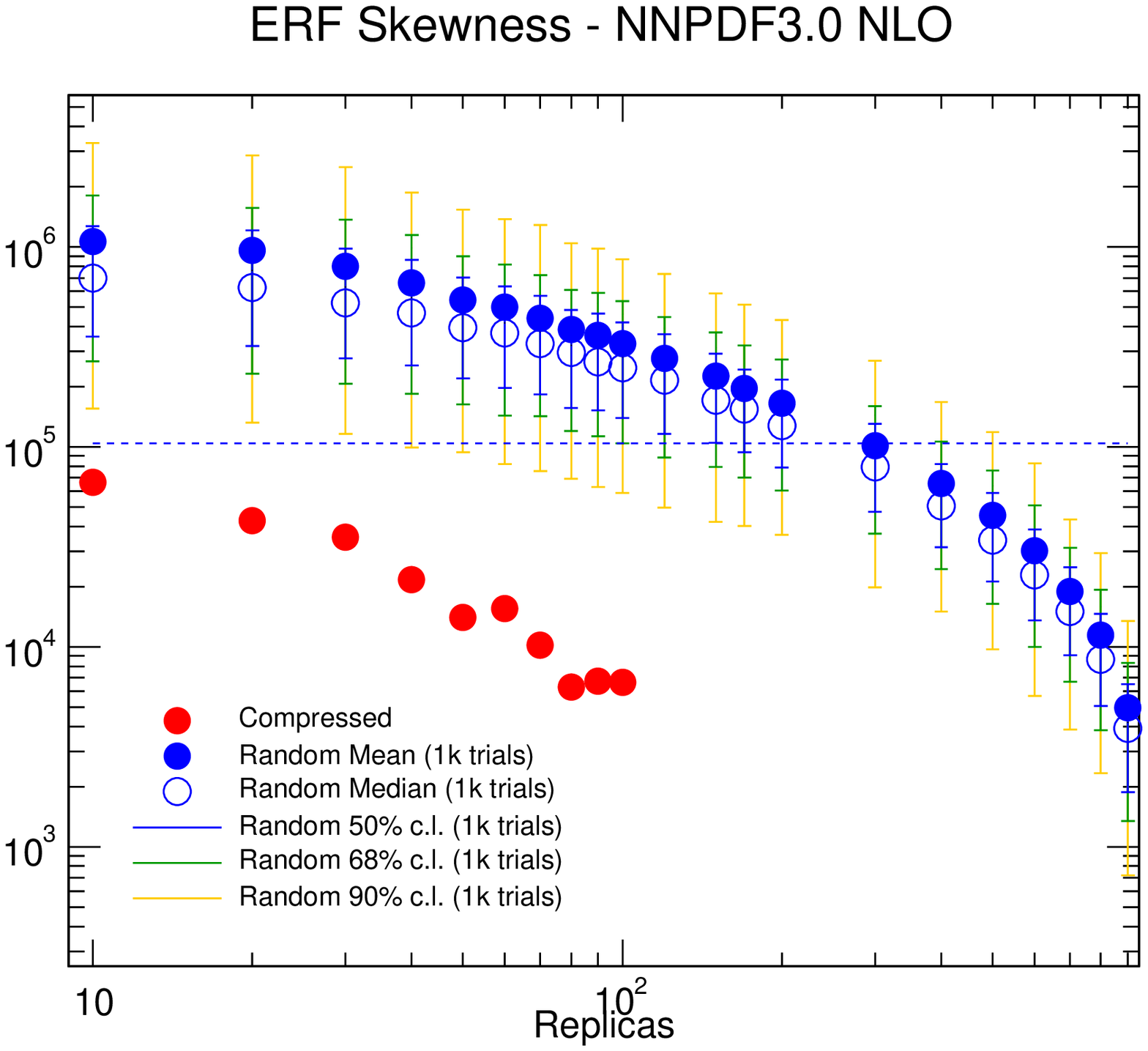}\includegraphics[scale=0.4]{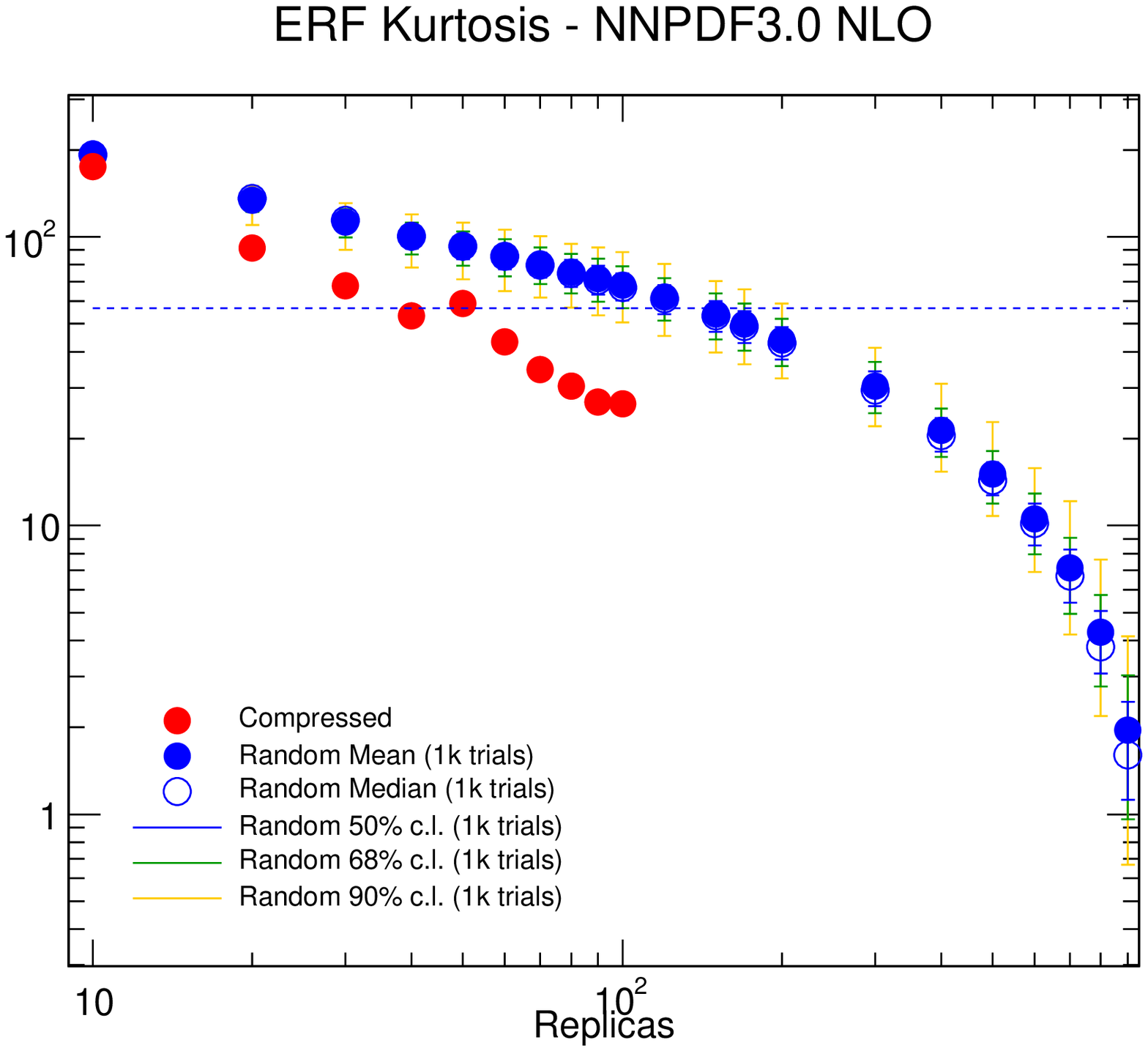}
\includegraphics[scale=0.4]{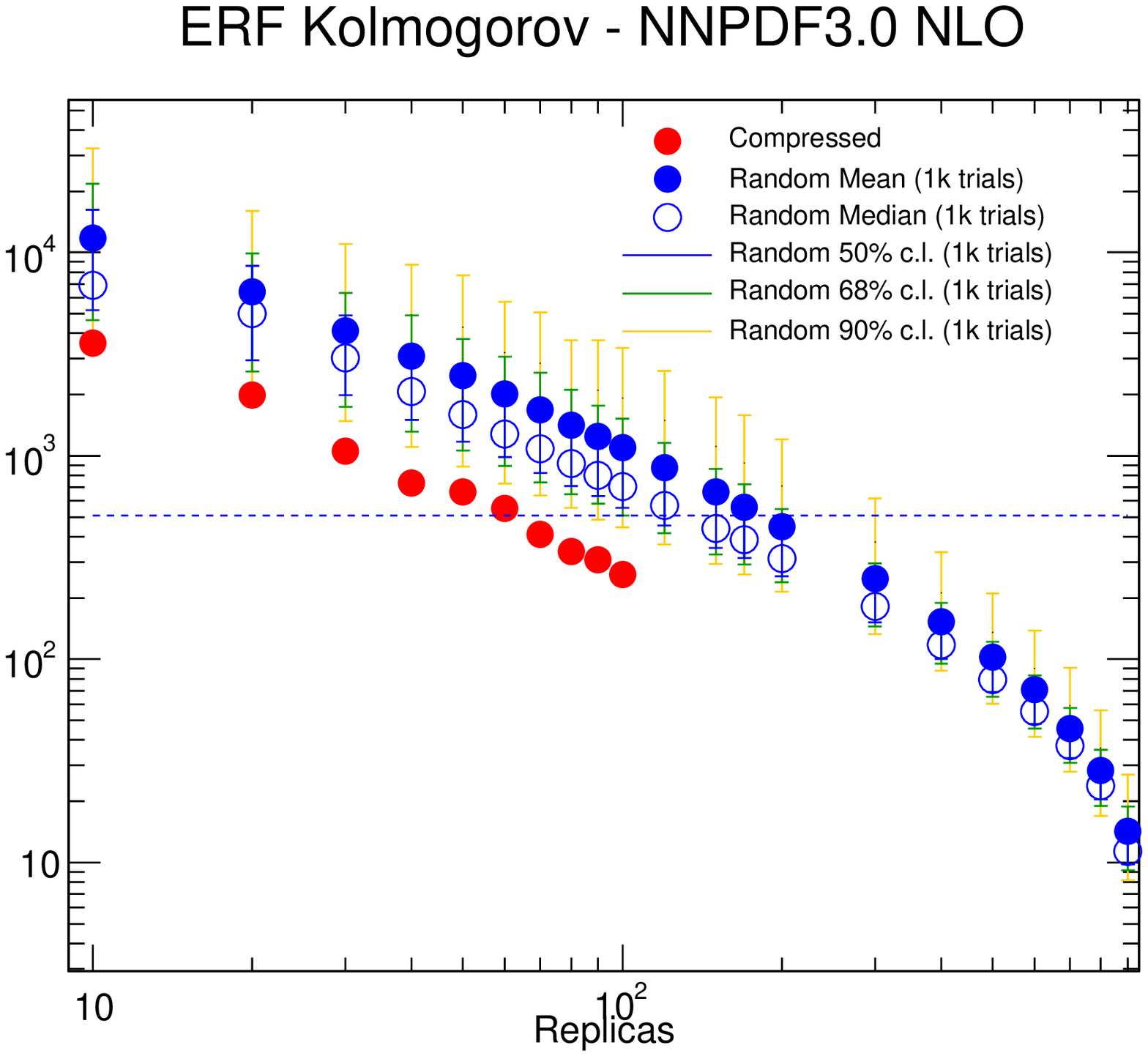}\includegraphics[scale=0.4]{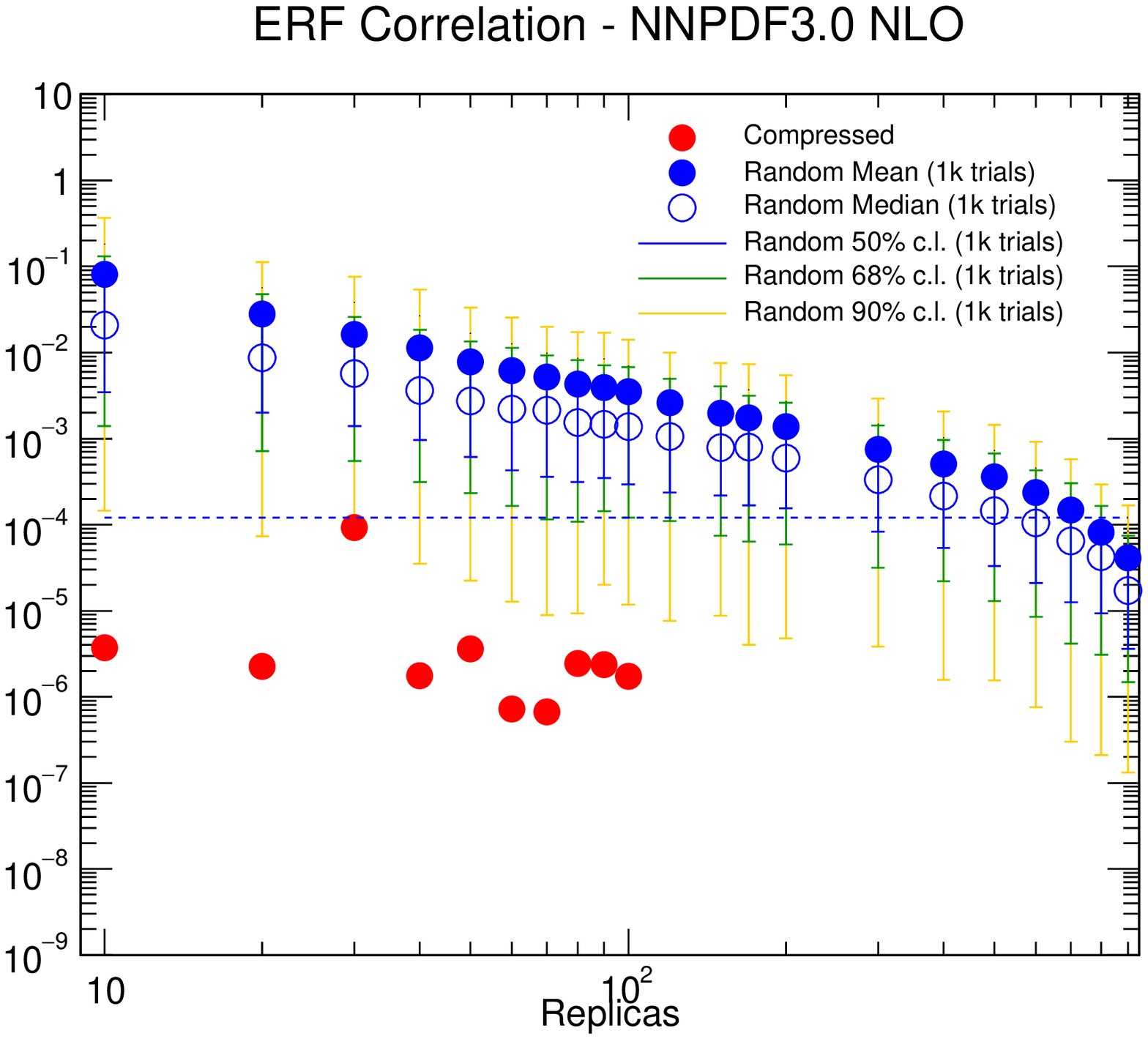}
\caption{\small The various contributions to the ERF,
  Eq.~(\ref{eq:erf}), for the compression of the
  NNPDF3.0 NLO set with $\widetilde{N}_{\rm rep}=1000$ replicas.
  For each value of $N_{\rm rep}$, we show the value of each
  contribution to the ERF for the best-fit result
  of the compression algorithm (red points).
  We compare the results of the compression with the values of the ERF
  averaged over
  $N_{\rm rand}=1000$ random partitions of  $N_{\rm rep}$
  replicas (blue points),
  as well as the 50\%, 68\% and 90\% confidence-level intervals
  computed over these random partitions.
  The dashed horizontal line is the 68\% lower band of the
  ERF for the average of the random partitions with $N_{\rm rep}=100$, and is
  inserted for illustration purposes only.}
\label{fig:erfnnpdf30}
\end{figure}

In order to quantify the performance of the compression algorithm, and
to compare it with that of a random selection of the reduced set of
replicas, Fig.~\ref{fig:erfnnpdf30} shows the various contributions to
the ERF, Eq.~(\ref{eq:erf}), for the compression of the NNPDF3.0 NLO set
with $\widetilde{N}_{\rm rep}=1000$ replicas.
For each value of $N_{\rm rep}$, we show the value of each
contribution to the ERF for the best-fit result of the compression
algorithm (red points).
We compare the results of the compression with the values of the ERF
averaged over $N_{\rm rand}=1000$ random partitions of $N_{\rm rep}$
replicas (blue points), as well as the 50\%, 68\% and 90\%
confidence-level intervals computed over these random partitions.
Various observations can be made from the inspection of
Fig.~\ref{fig:erfnnpdf30}.
First of all, the various contributions to the ERF tend to zero when
the number of compressed or random replicas tends to the size of the
prior set, as expected for consistency.
For the random partitions of $N_{\rm rep}$ replicas the mean value and
the median values averaged over $N_{\rm rand}$ trials are not
identical, emphasizing the importance of taking confidence levels.
From Fig.~\ref{fig:erfnnpdf30} we also confirm that the compression
algorithm is able to provide sets of PDFs with smaller ERF values for
all estimators that outperform random selections with a much larger
number of replicas.
To emphasize this point, the dashed horizontal line in
Fig.~\ref{fig:erfnnpdf30} corresponds to the lower limit of the 68\%
confidence level of the ERF computed over $N_{\rm rand}=1000$ random partitions with
$N_{\rm rep}=100$, and is inserted for illustration purposes only.
It
indicates that the NNPDF3.0 NLO PDF set with $\widetilde{N}_{\rm rep}=1000$ can now be
compressed down to $N_{\rm rep}=50$ replicas in a way that reproduces
better the original distribution that most of the random partitions of
$N_{\rm rep}=100$ replicas.

The results of Fig.~\ref{fig:erfnnpdf30} 
confirm that the compression algorithm outperforms essentially any
random selection of replicas for the construction of a reduced set,
and provides an adequate representation of the prior probability
distribution with a largely reduced number of replicas.
Similar results are obtained when compressing the CMC-PDFs, as we will discuss
in Sect.~\ref{sec:resultsCMCPDFs}.
  %

\section{The Compressed Monte Carlo PDF sets}
\label{sec:results}

In this section we present the results
for the CMC-PDFs, first discussing
the compression of a native Monte Carlo PDF
set, in this case NNPDF3.0 with $\widetilde{N}_{\rm rep}=1000$, and then the
compression of the MC combination for NNPDF3.0, CT14 and MMHT14
with $\widetilde{N}_{\rm rep}=900$.
In both cases, we compare the PDFs from the prior and compressed
sets, for different values of the number of replicas $N_{\rm rep}$ of the latter.
We also verify that correlations between PDFs
are successfully reproduced
by the compression.
The phenomenological validation
of the CMC-PDF sets at the level of LHC observables is addressed
in Sect.~\ref{sec:lhcpheno}.

\subsection{Compression of native MC PDF sets}

\begin{figure}[t]
  \centering 
  \includegraphics[scale=0.35]{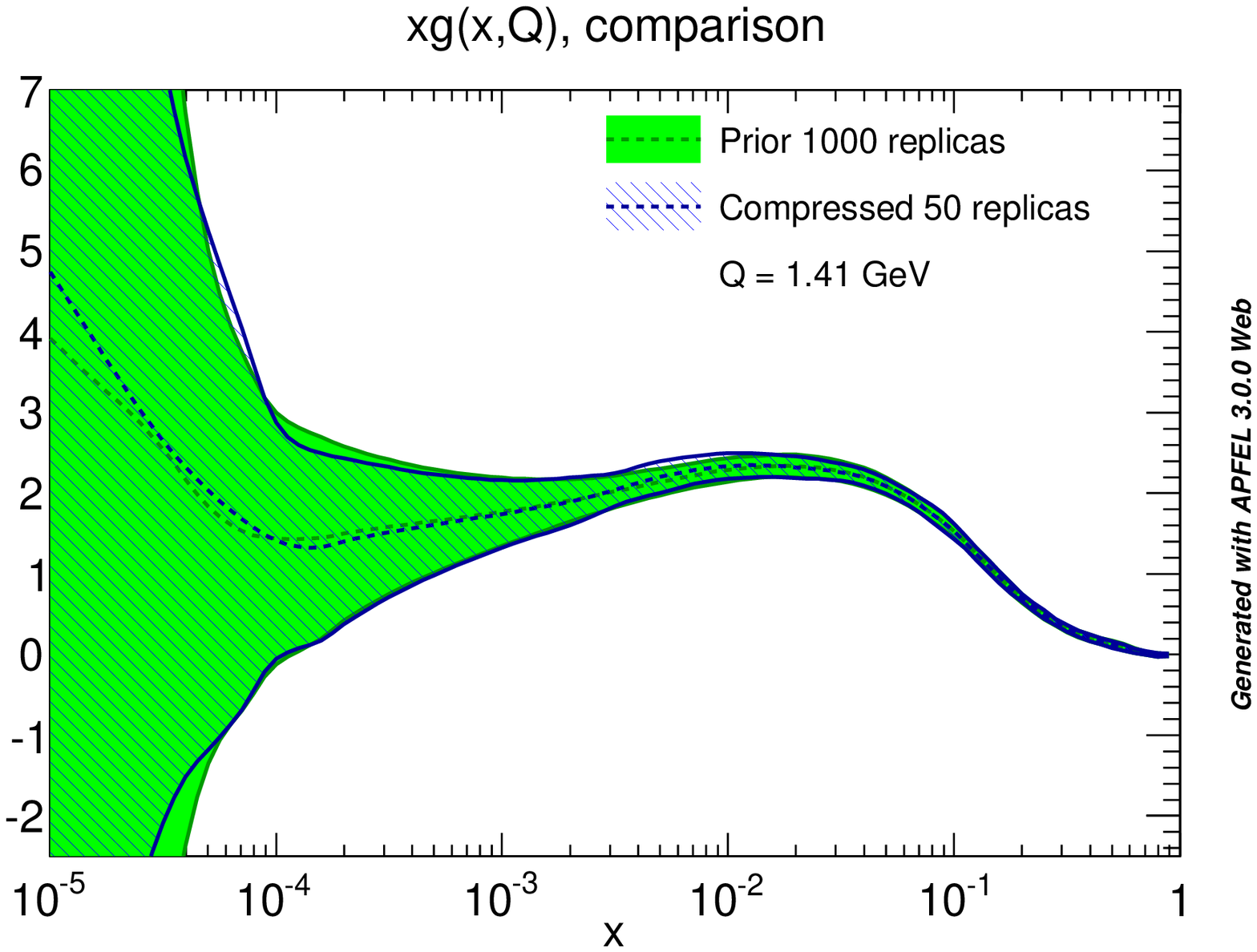}\includegraphics[scale=0.35]{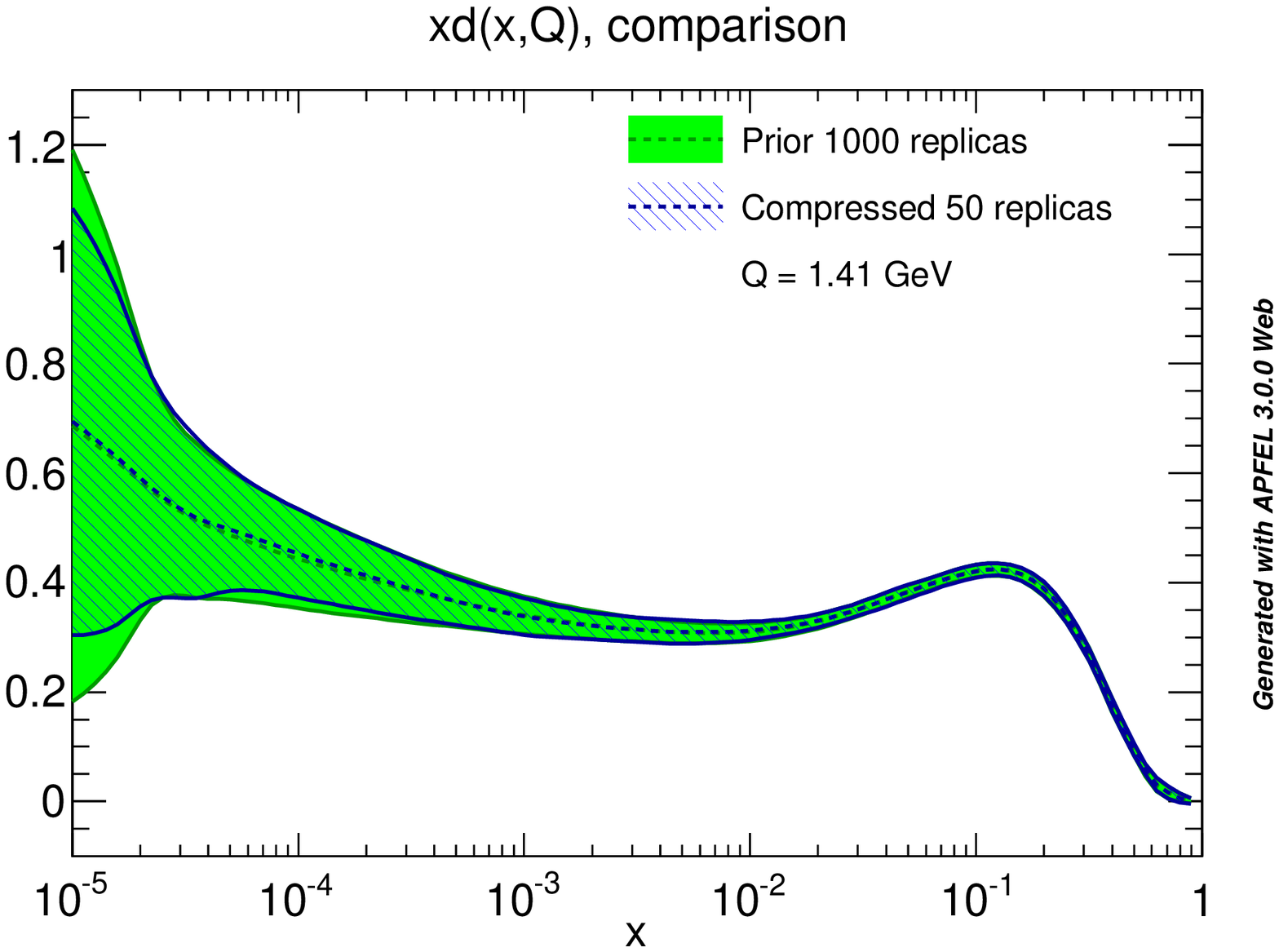}
  \includegraphics[scale=0.35]{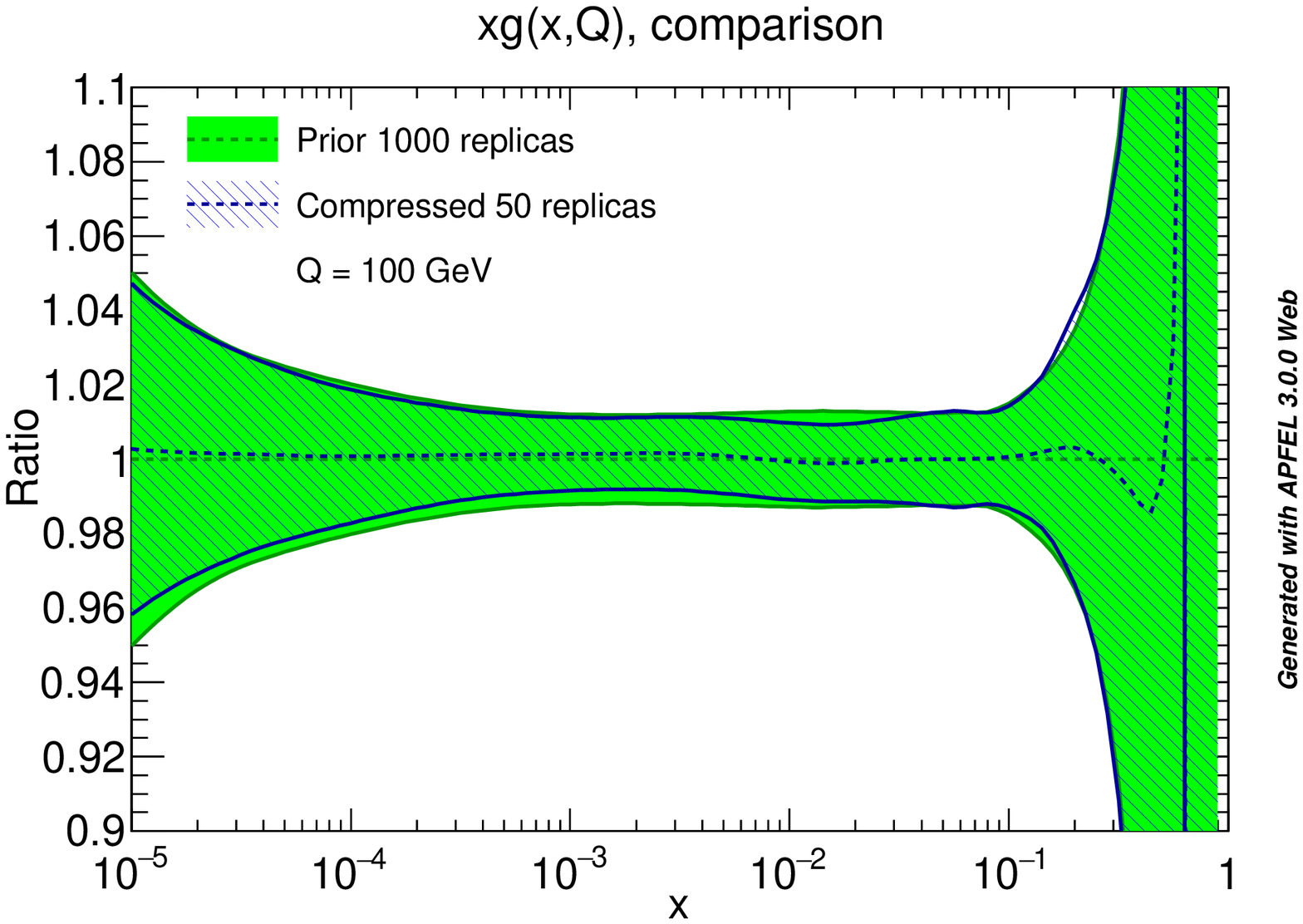}\includegraphics[scale=0.35]{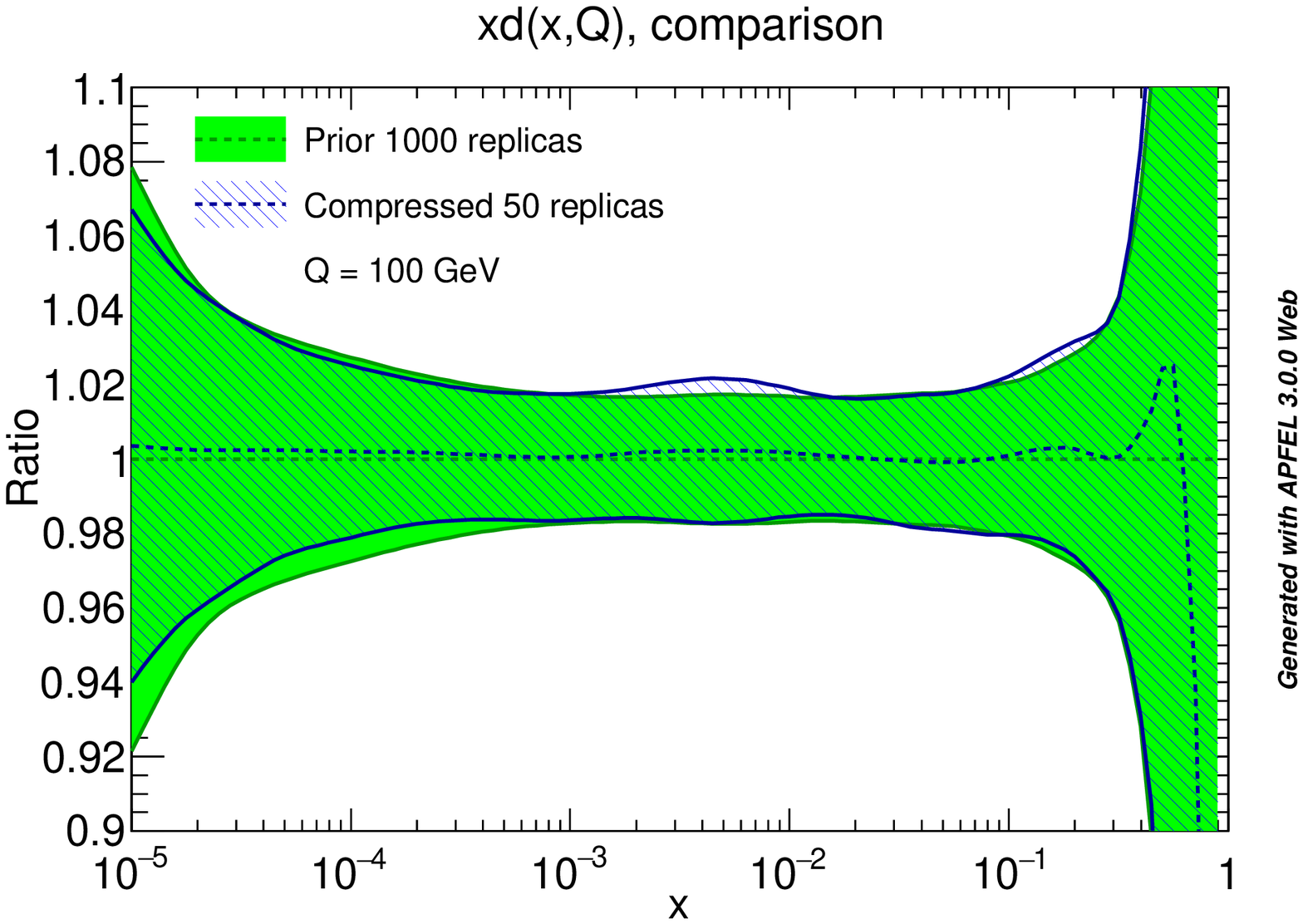}
  \caption{\small Upper plots: comparison of the prior NNPDF3.0 NLO set
    with $\widetilde{N}_{\rm rep}=1000$ and the compressed set with $N_{\rm rep}=50$ replicas,
    for the gluon and the down quark at the scale $Q^2=2$ GeV$^2$.
    Lower plots: the same comparison this time at a typical
    LHC scale of $Q=100$ GeV, normalized to the central value of the prior set.
  }
  \label{valnnpdf}
\end{figure}

\begin{figure}[t]
  \centering 
  \includegraphics[scale=0.5]{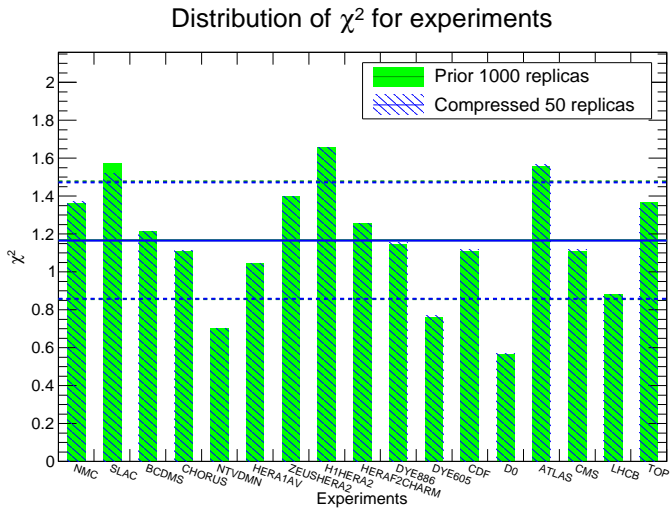}
  \caption{\small Distribution of $\chi^2$ for all the experiments
    included in the NNPDF3.0 fit, comparing the original and the compressed
    PDF sets. }
  \label{valnnpdf3}
\end{figure}

First of all, we show the results for the compression
of a native MC PDF set, for the case of the NNPDF3.0 NLO set with
$\widetilde{N}_{\rm rep}=1000$ replicas.
In Fig.~\ref{valnnpdf} we compare the original and
the compressed gluon and down quark at $Q^2=2$ GeV$^2$, using
$N_{\rm rep}=50$ in the compressed set.
Excellent agreement can be seen at the level of
central values and variances.
The comparison is also shown
at a typical LHC scale of $Q=100$ GeV, finding similar agreement.
The plots in this section have been obtained using the
{\sc\small APFEL-Web} online PDF plotter~\cite{Bertone:2013vaa,Carrazza:2014gfa}.
The result that the central values of the original set are
perfectly reproduced by the compressed set can also be seen
from Fig.~\ref{valnnpdf3}, where we show the distribution of $\chi^2$ for all the experiments
included in the NNPDF3.0 fit, comparing the original and the compressed
PDF set, and find that they are indistinguishable.

\begin{figure}[t]
  \centering 
  \includegraphics[scale=0.35]{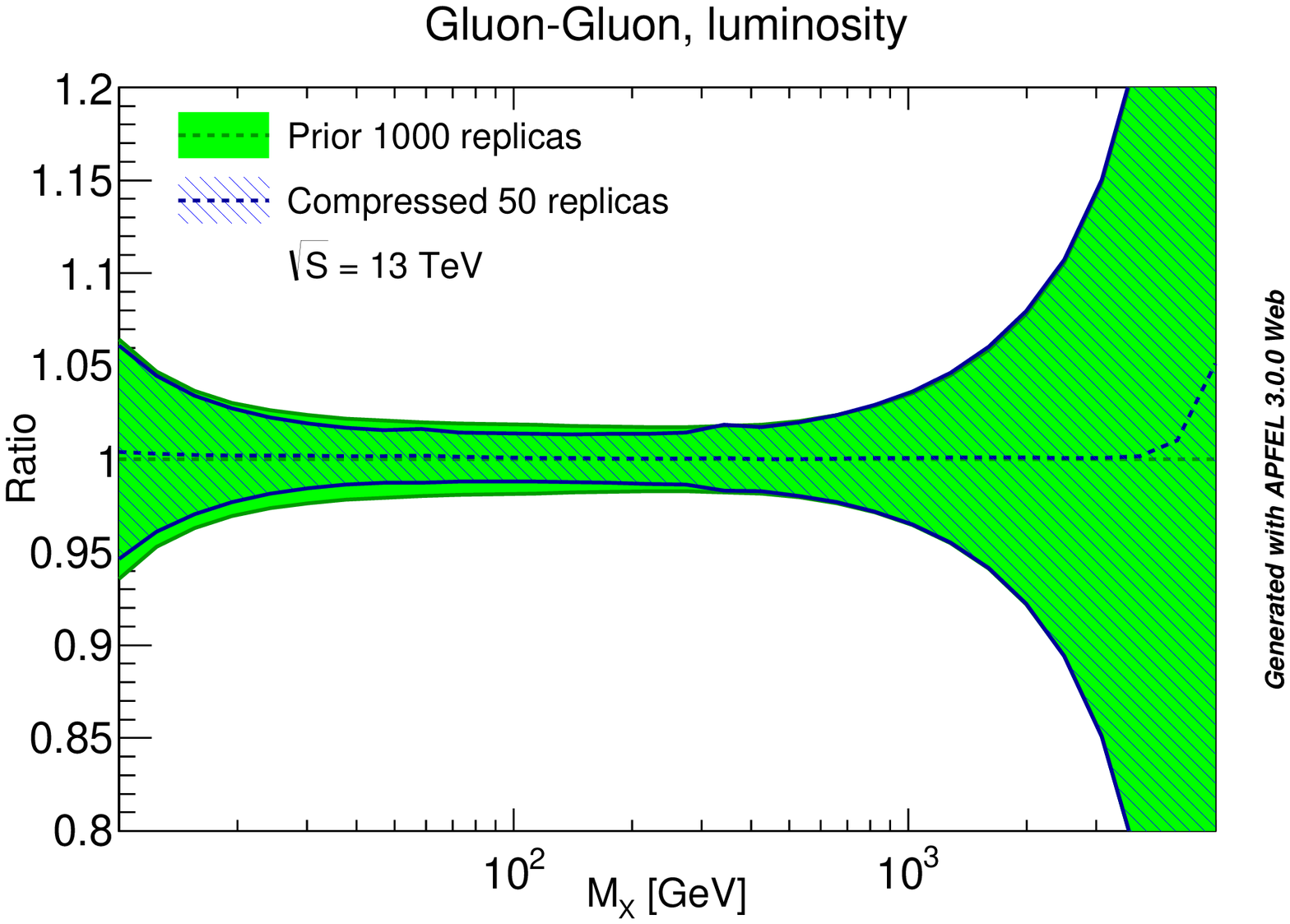}
  \includegraphics[scale=0.35]{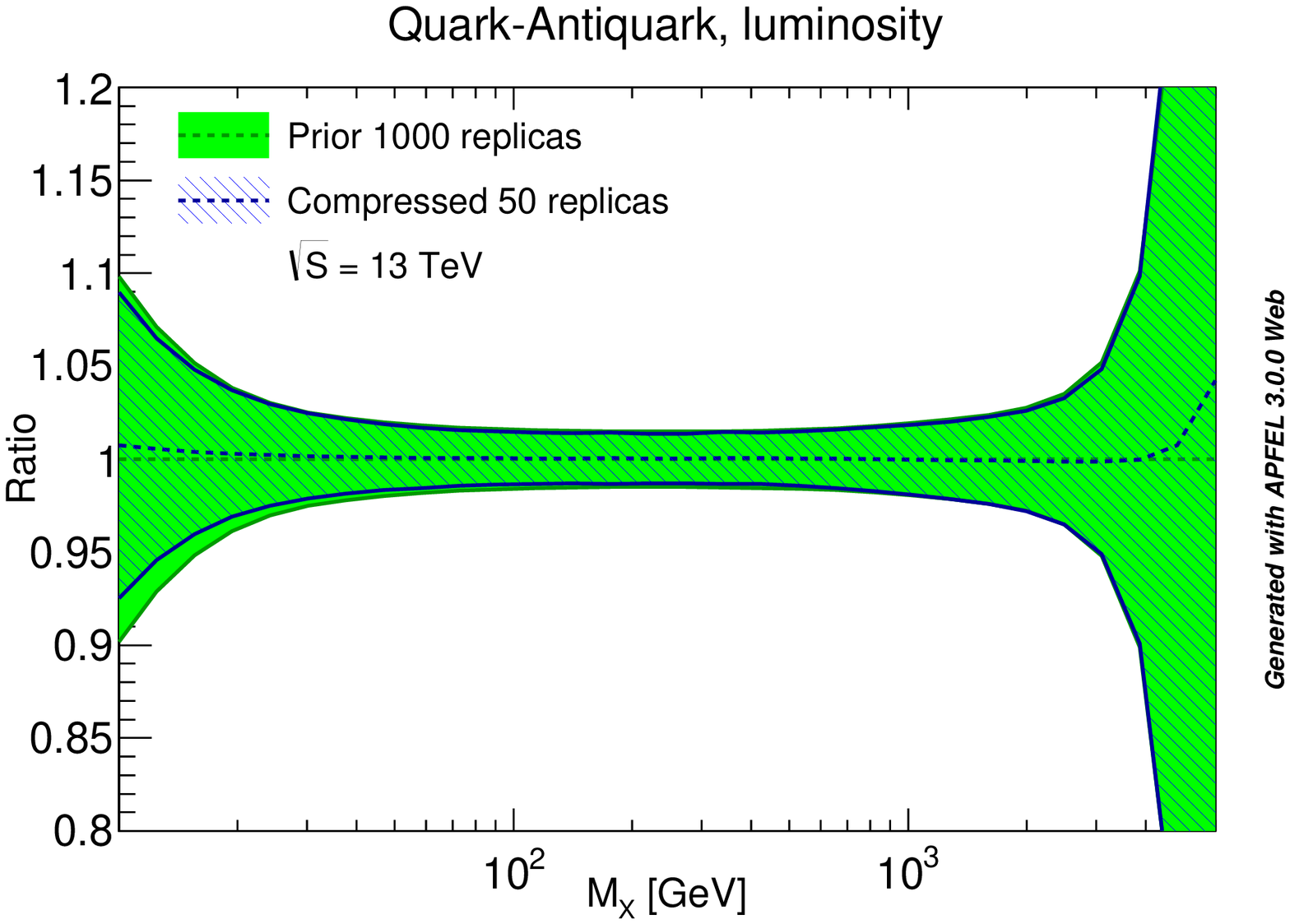}
  \includegraphics[scale=0.35]{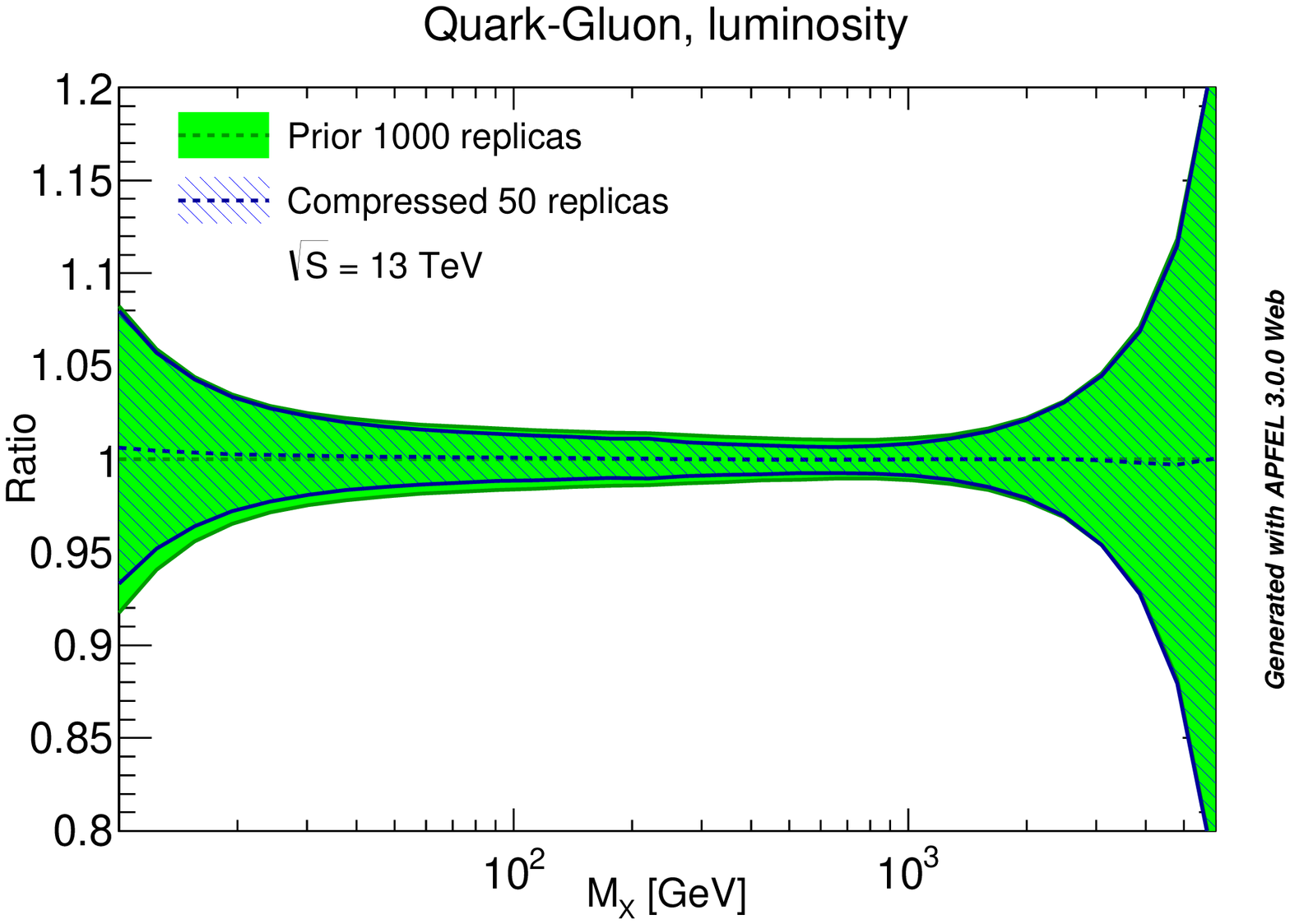}
  \includegraphics[scale=0.35]{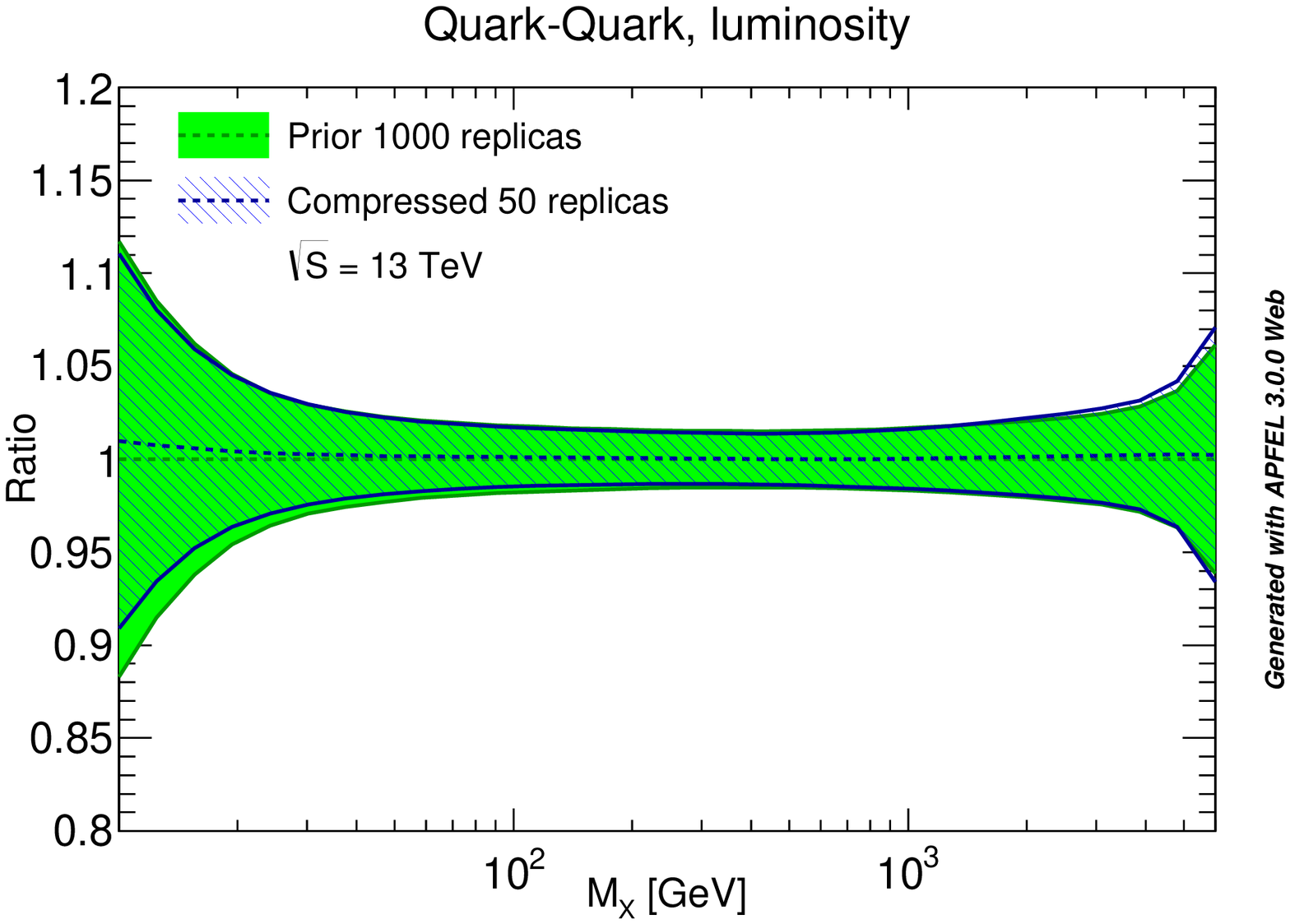}
  \caption{\small Comparison of PDF luminosities between the original and compressed
    NNPDF3.0 set, for the LHC 13 TeV as a function of the invariant mass of the final
    state $M_X$.
    From top to bottom and left to right, we show the gluon-gluon, quark-antiquark, quark-gluon and
    quark-quark luminosities.
  }
  \label{valnnpdf2}
\end{figure}

Next, we compare in Fig.~\ref{valnnpdf2} the various PDF
luminosities between the original and the compressed set
at the LHC with centre-of-mass energy of $\sqrt{s}=13$ TeV.
We show the gluon-gluon, quark-antiquark, quark-gluon and
quark-quark luminosities.
As in the case of the individual PDF flavours, good agreement is found
in all the range of possible final state invariant masses $M_X$.
Note that the agreement is also good in regions, like small $M_X$
and large $M_X$, where the underlying PDF distribution
is known to be non-Gaussian.

\begin{figure}[th]
  \centering 
  \includegraphics[scale=0.52]{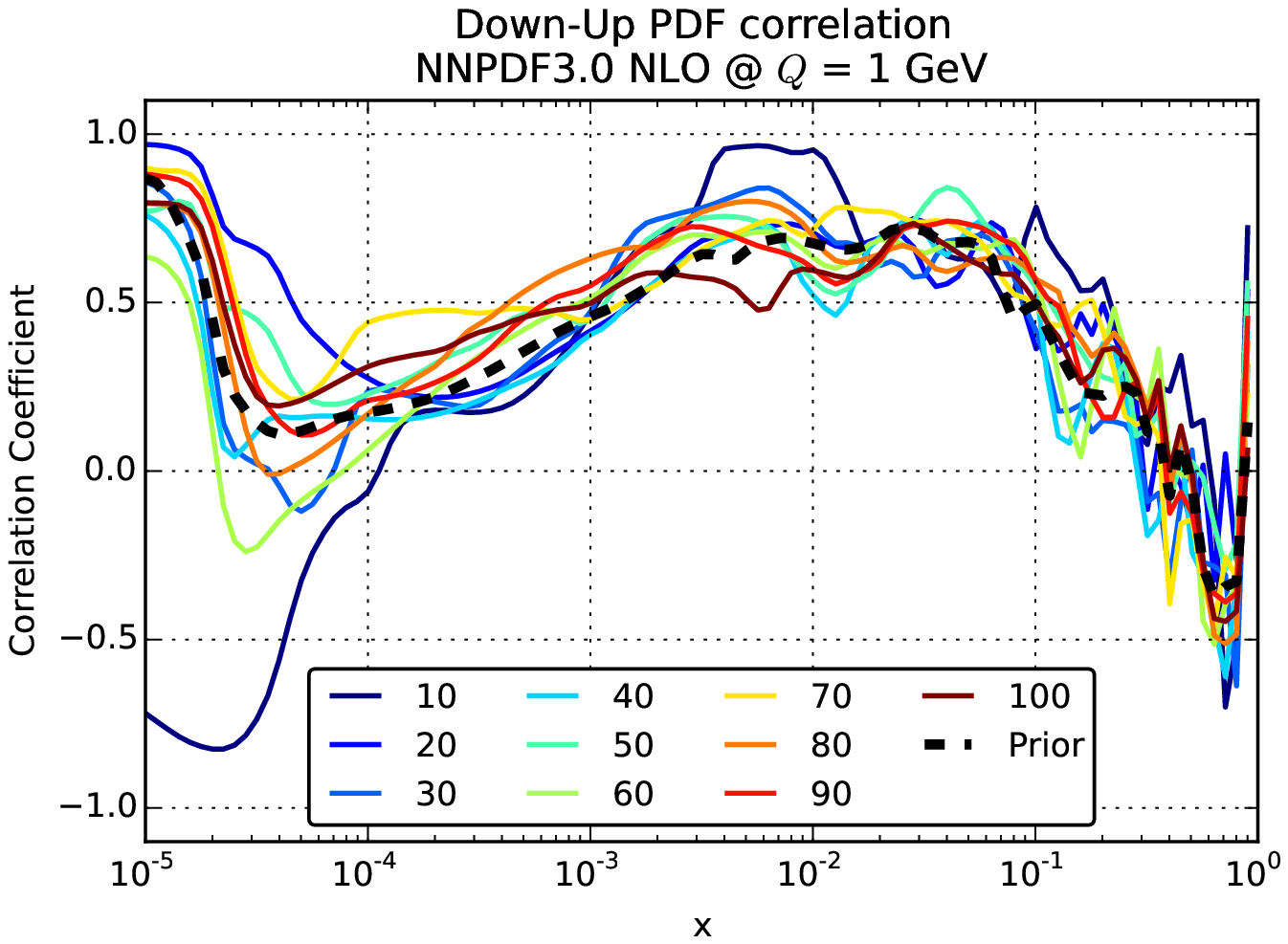}
  \includegraphics[scale=0.52]{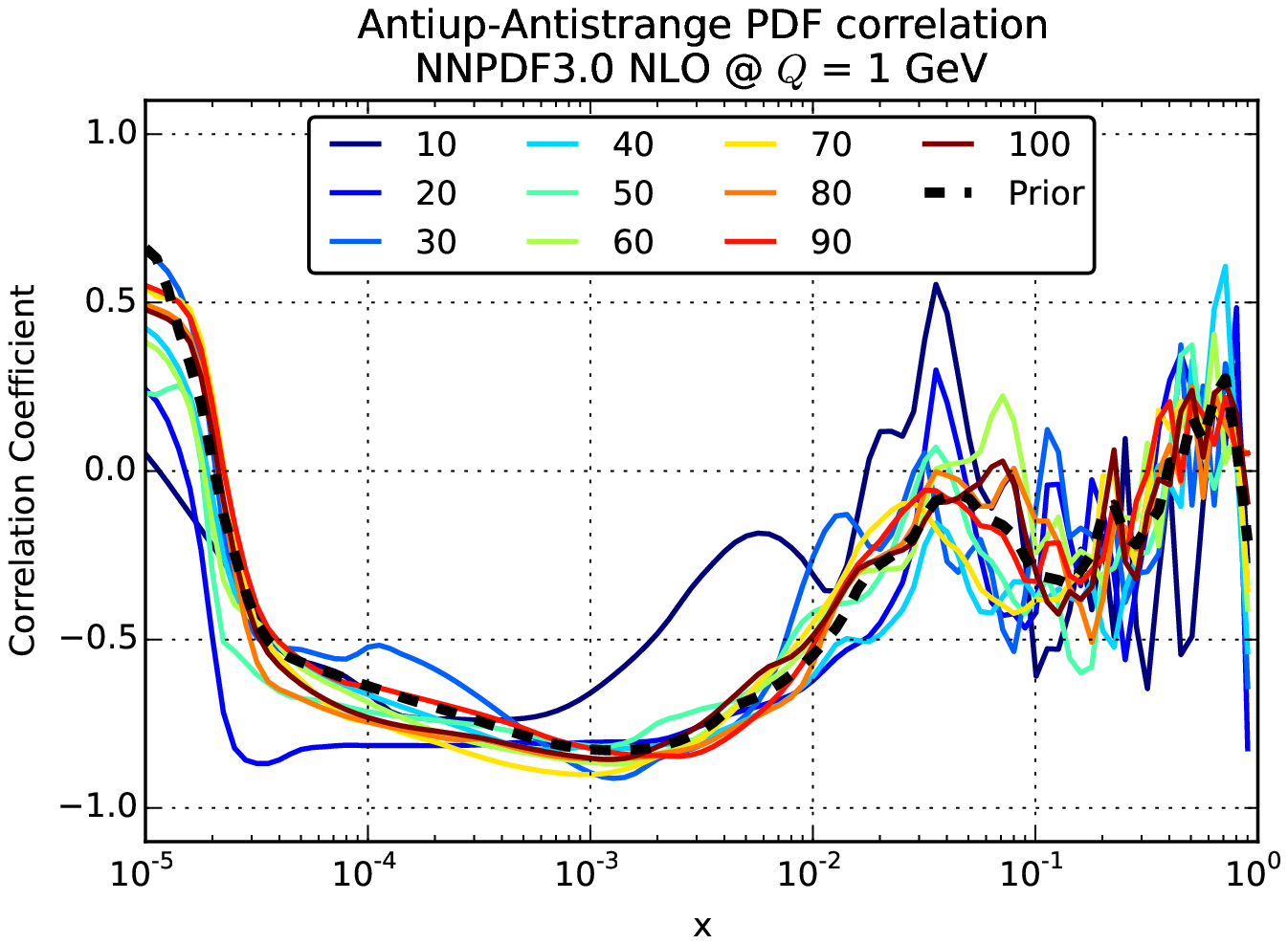}
  \includegraphics[scale=0.52]{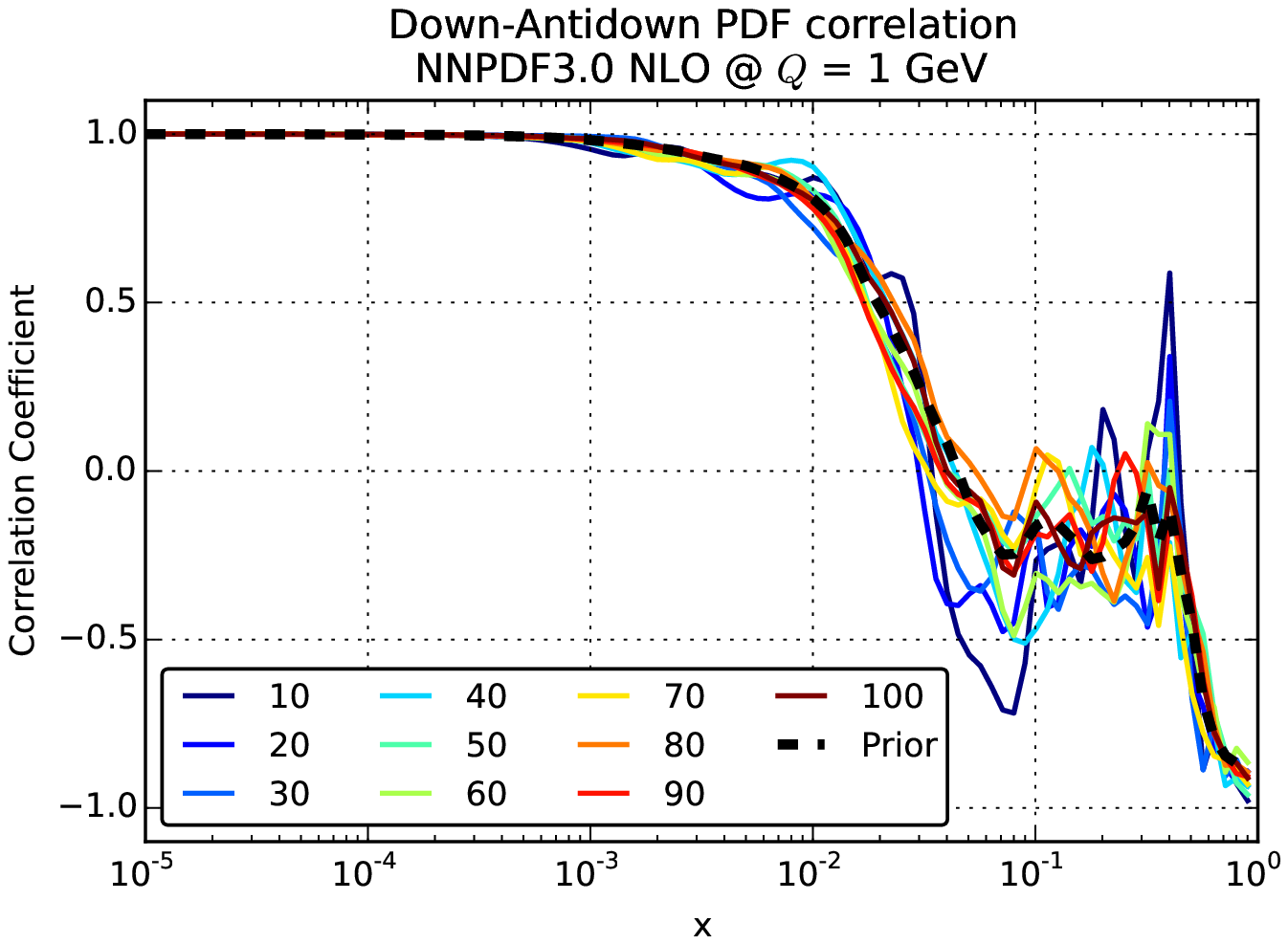}
  \includegraphics[scale=0.52]{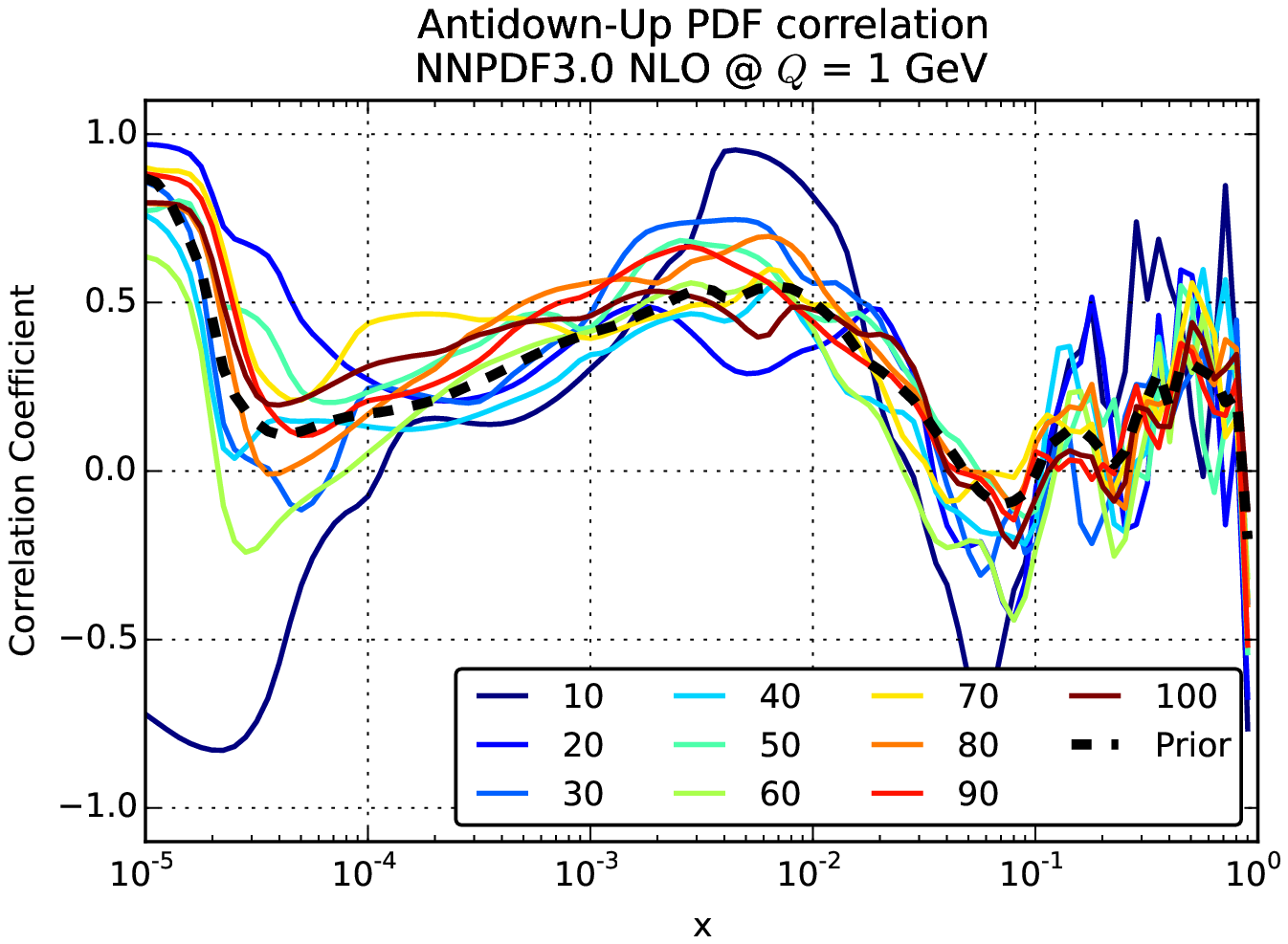}
  \caption{\small Comparison between the PDF correlations among different
    PDF flavours, as a function of Bjorken-$x$,
    for $Q=$1 GeV, for the original NNPDF3.0 set
    with $\widetilde{N}_{\rm rep}=1000$ replicas and the compressed sets
    for various values of $N_{\rm rep}$.
    From top to bottom and from left to right, we show the correlations
    between up and down quarks, between up and strange anti-quarks,
    between down quarks and down anti-quarks, and between up quarks
    and down anti-quarks.
  }
  \label{valnnpdf4}
\end{figure}

%
It is also important to verify that not only central values and
variances are reproduced, but also that higher moments and
correlations are well reproduced by the compression.
Indeed, one of the main advantages of the
$N_{\rm rep}=1000$ replica sets of NNPDF as compared
to the  $N_{\rm rep}=100$ sets is that correlations
should be reproduced more accurately in the former case.
In Fig.~\ref{valnnpdf4} we show the results for the correlation
coefficient between different PDFs, as a function of Bjorken-$x$,
for $Q=1$ GeV.
We compare the results of the original $\widetilde{N}_{\rm rep}=1000$ replica set,
together  with the results of the compressed sets for
a number of $N_{\rm rep}$ values.
From top to bottom and from left to right we show the correlations
between up and down quarks, between up and strange anti-quarks,
between down quarks and down anti-quarks, and between up quarks
and down anti-quarks.
The correlations between PDF flavours have been computed
using the suitable expression for Monte Carlo sets~\cite{Demartin:2010er}.
As we can see correlations are reasonably well reproduced, already
with $N_{\rm rep}=50$ the results of the compressed set and of the prior
are very close to each other.

\begin{figure}[th]
  \centering 
  \includegraphics[scale=0.7]{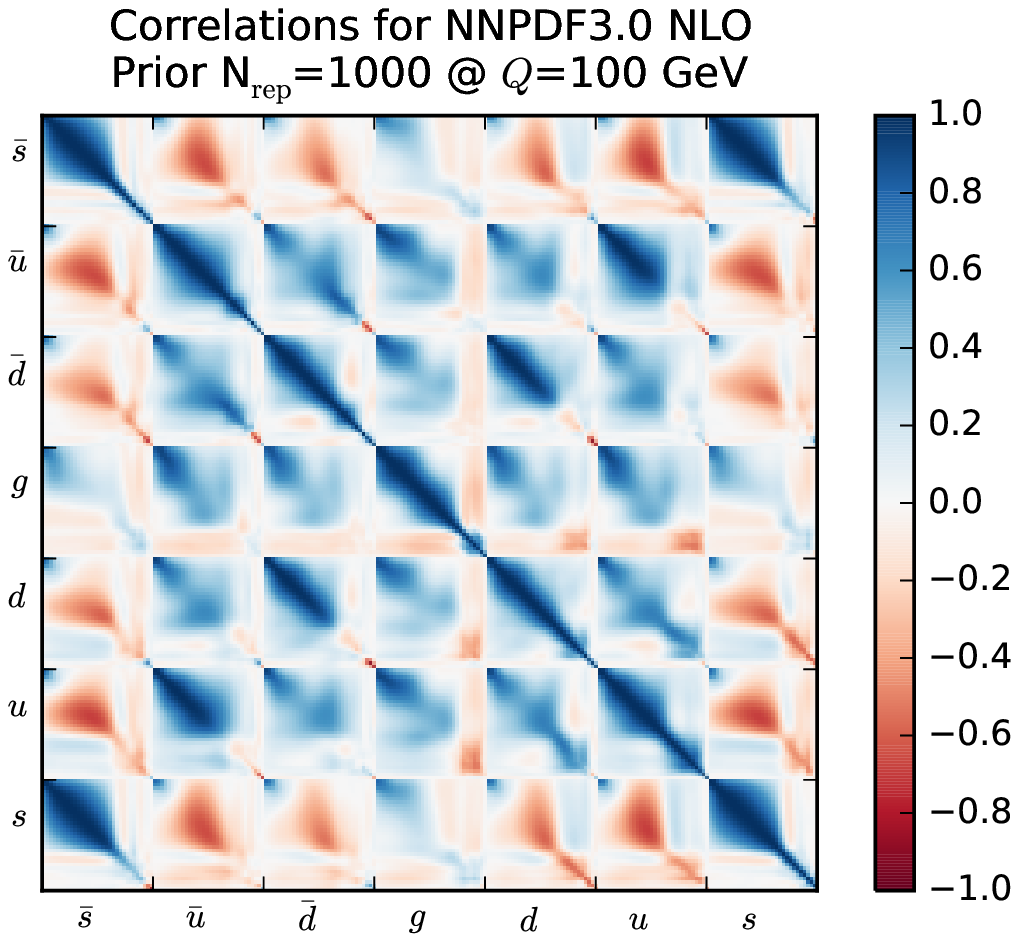}
  \includegraphics[scale=0.7]{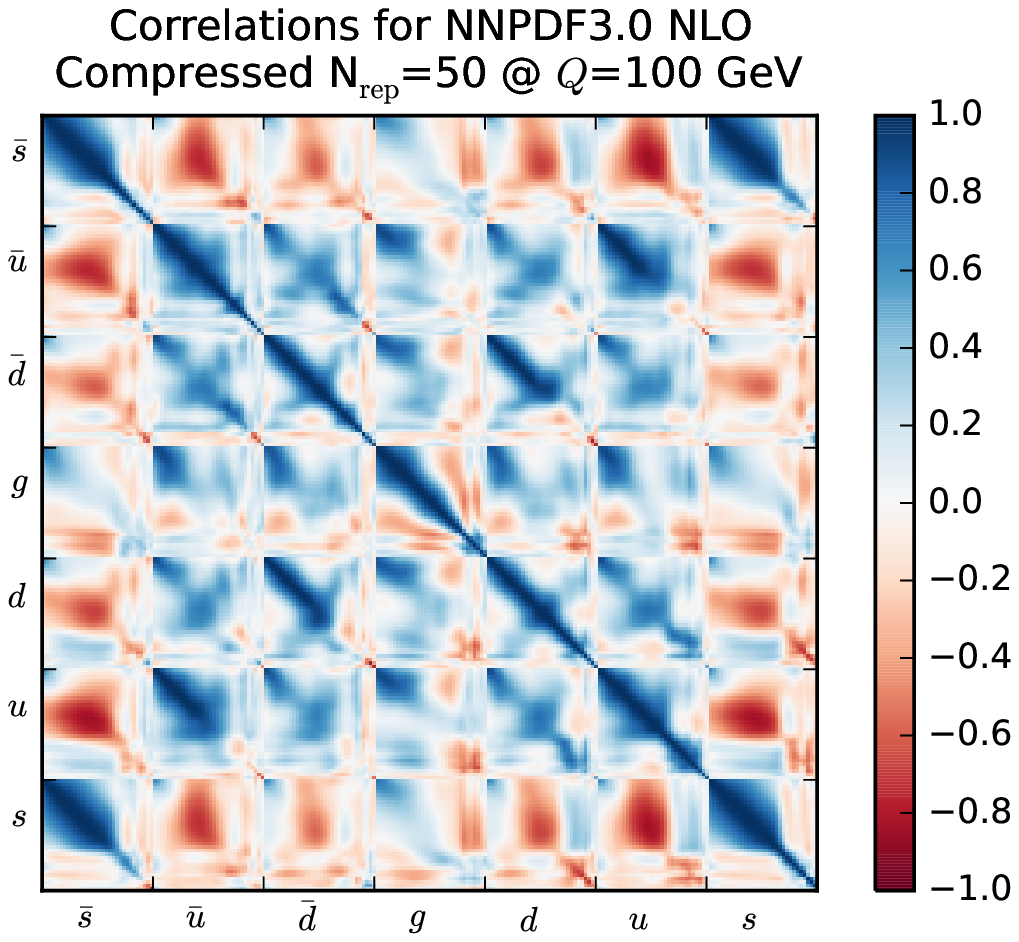}
  \includegraphics[scale=0.7]{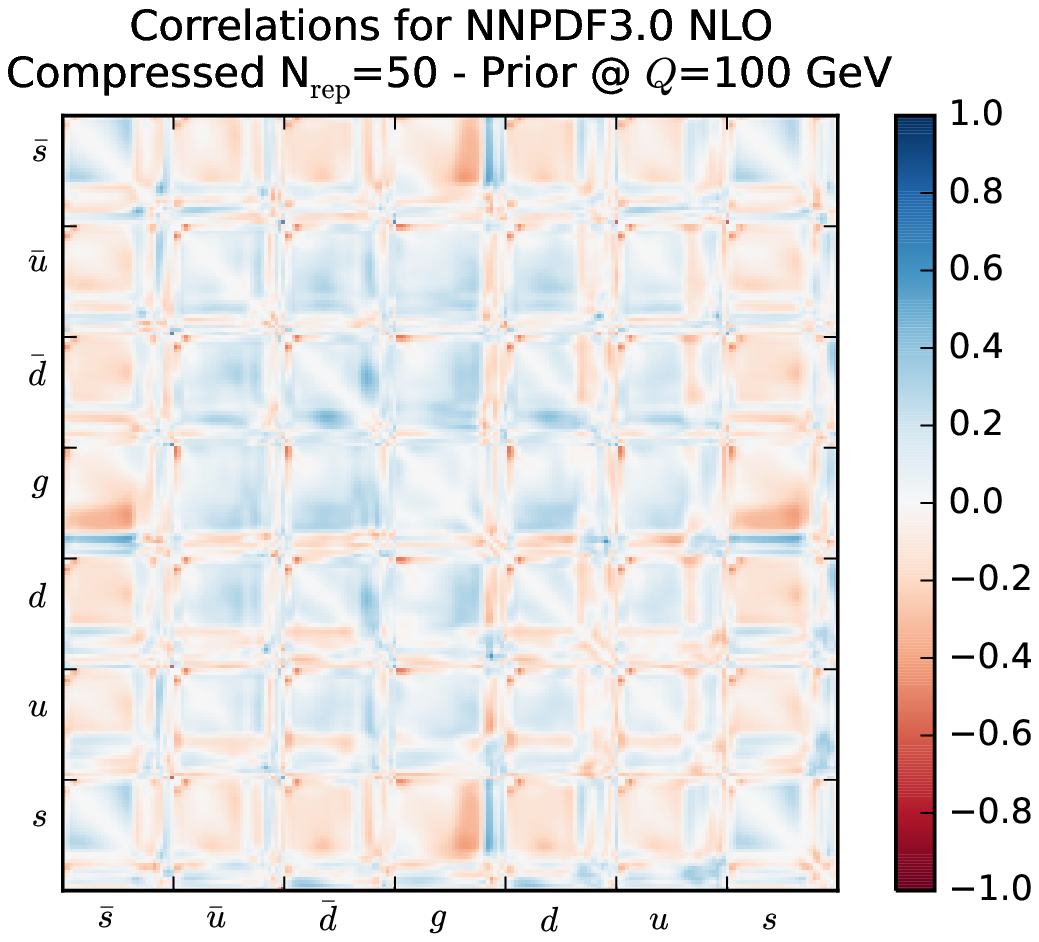}
  \caption{\small The correlation matrix of the NNPDF3.0 set with $\widetilde{N}_{\rm
      rep}=1000$ at $Q=100$ GeV. On the right, the same matrix for the
    NNPDF3.0 compressed set with $N_{\rm rep} = 50$ replicas.
  The bottom plot represents the difference between the two matrices.
    See text for more details.
  }
  \label{fig:pycorrelation}
\end{figure}

Another illustration of the fact that PDF correlations are maintained
in the compression is provided by Fig.~\ref{fig:pycorrelation}, where
we show the correlation matrix of the NNPDF3.0 set at a scale of
$Q=100$ GeV, comparing the prior with $\widetilde{N}_{\rm rep}=1000$
with the compressed set with $N_{\rm rep} = 50$ replicas.
The correlation matrices presented here are defined in a grid of
$N_x=50$ points in $x$, logarithmic distributed between $[10^{-5},1]$
for each flavor ($\bar{s},\bar{u},\bar{d},g,d,u,s$).
To facilitate the comparison, in the bottom plot we show the
differences between the correlation coefficients in the two cases.
Is clear from this comparison that the agreement of the PDF correlations
reported in Fig.~\ref{valnnpdf4} holds for the complete set of
possible PDF combinations, in all the relevant range of Bjorken-$x$.

Having validated the compression results for a native
MC set, we now turn to discuss the results of the compression
for a combined MC PDF set.

\clearpage

\subsection{Compression of the CMC-PDFs}
\label{sec:resultsCMCPDFs}

Now we turn to a similar validation study but this time for
the CMC-PDFs.
As we have discussed in Sect.~\ref{sec:combination}, the combined
MC set has been constructed by adding together $N_{\rm rep}=300$ replicas
of NNPDF3.0, MMHT14 and CT14 each, for a total of $\widetilde{N}_{\rm rep}=900$ replicas.
Starting from this prior set, the compression algorithm
has been applied as discussed in Sect.~\ref{sec:compression},
and we have produced CMC-PDF sets
for a number of values of $N_{\rm rep}$ from 5 to 250 replicas,
using the settings from  Sect.~\ref{ref:settings}.

\begin{figure}[h]
  \centering 
  \includegraphics[scale=0.35]{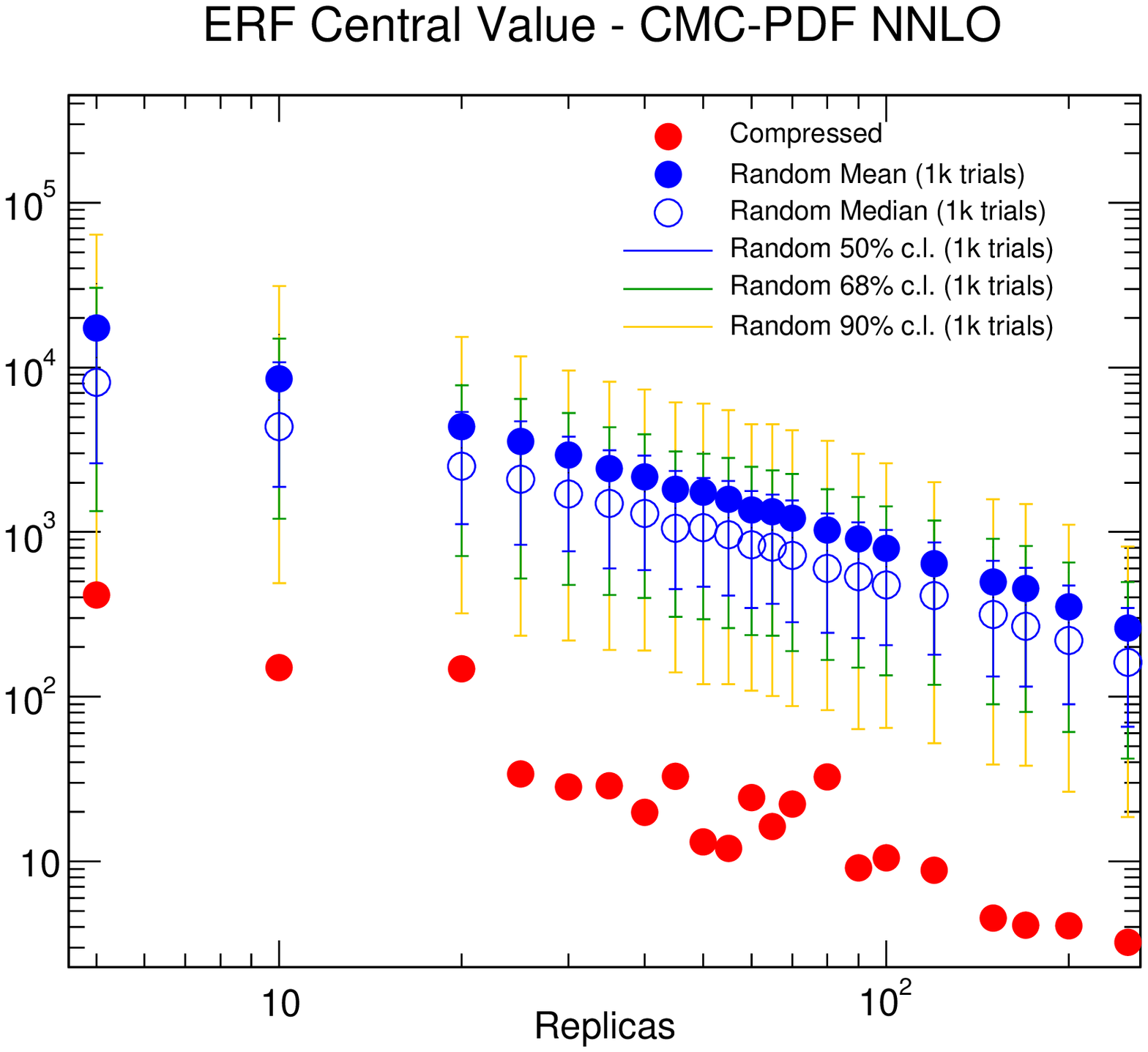}\includegraphics[scale=0.35]{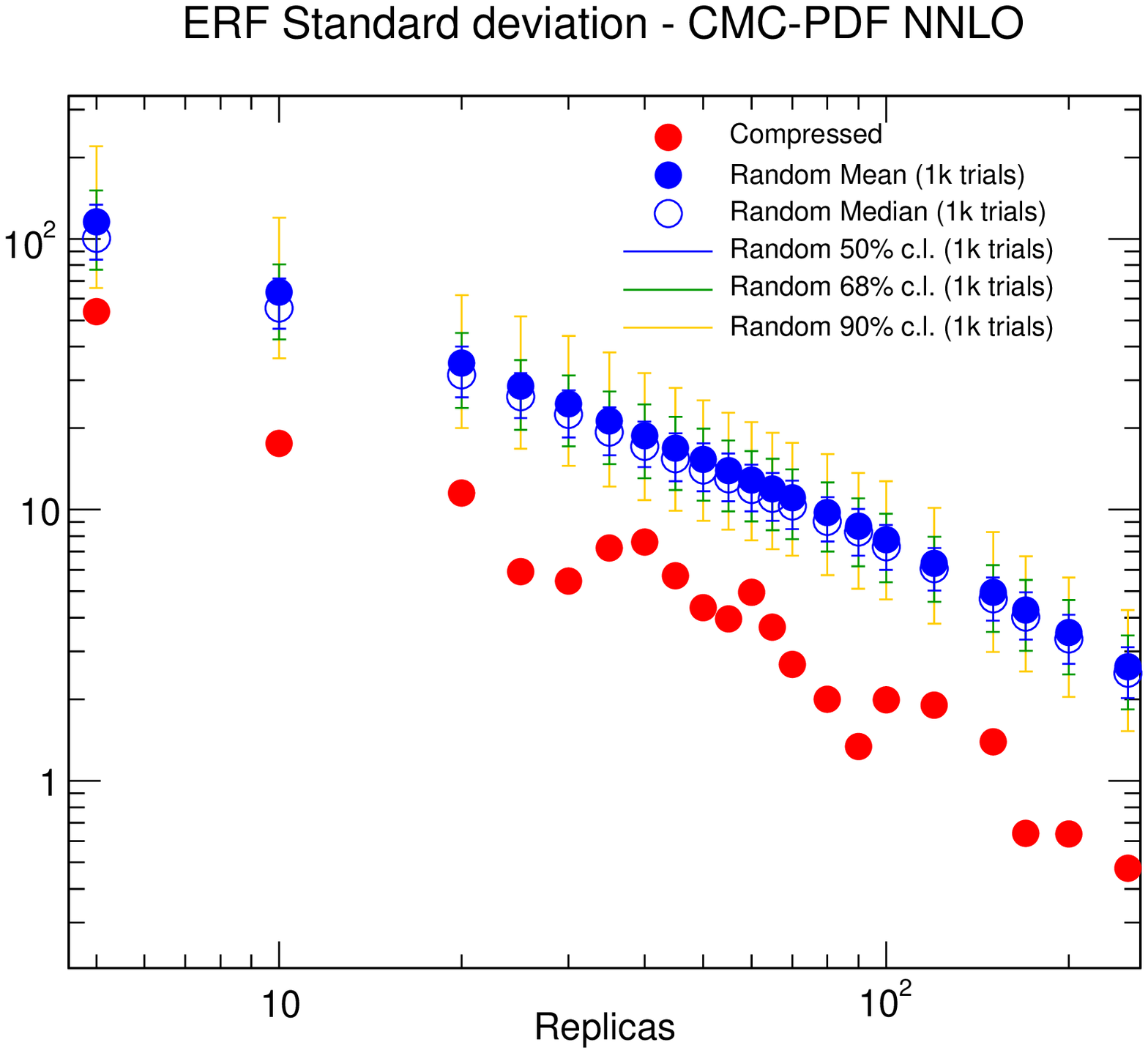}
  \includegraphics[scale=0.35]{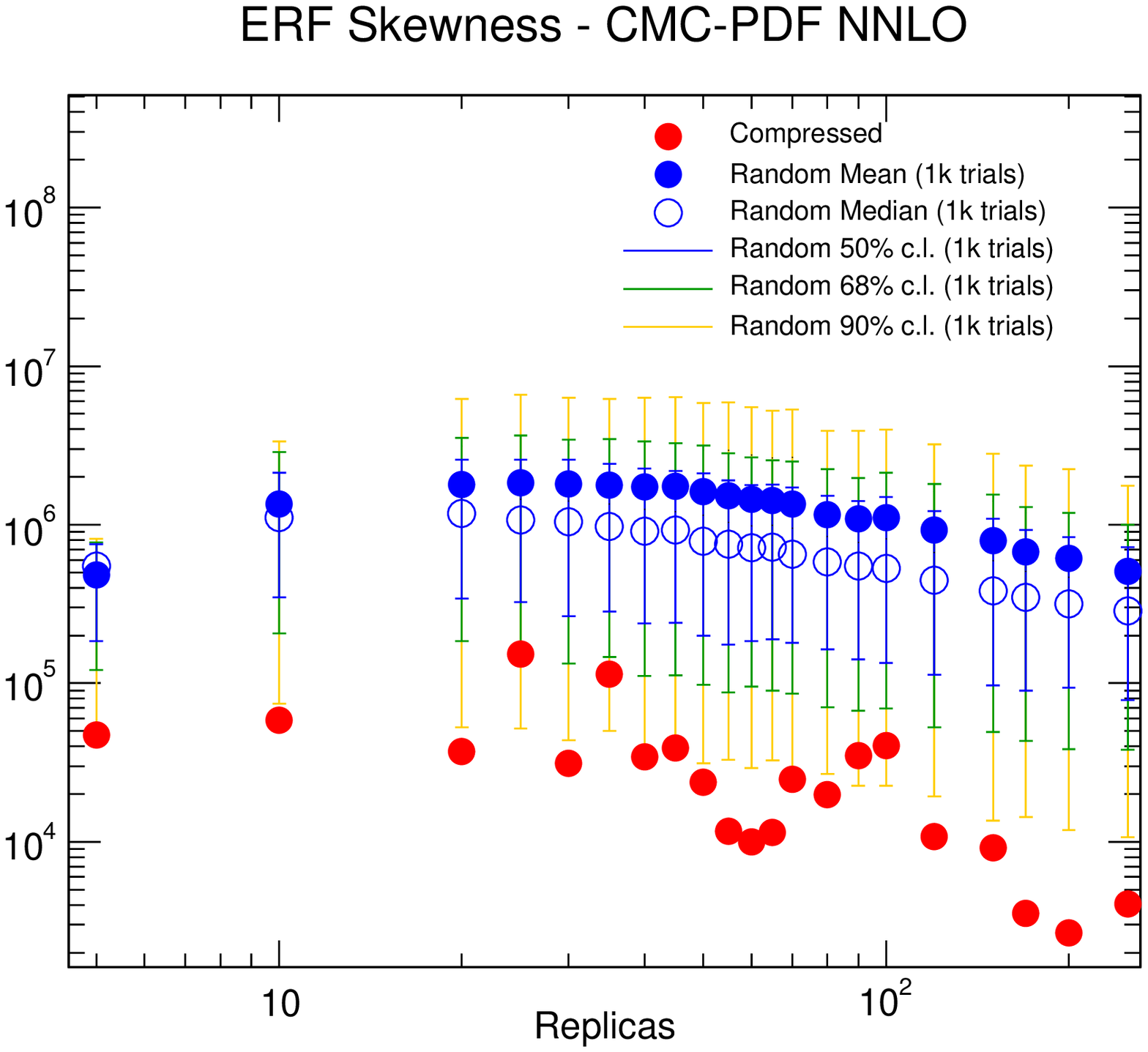}\includegraphics[scale=0.35]{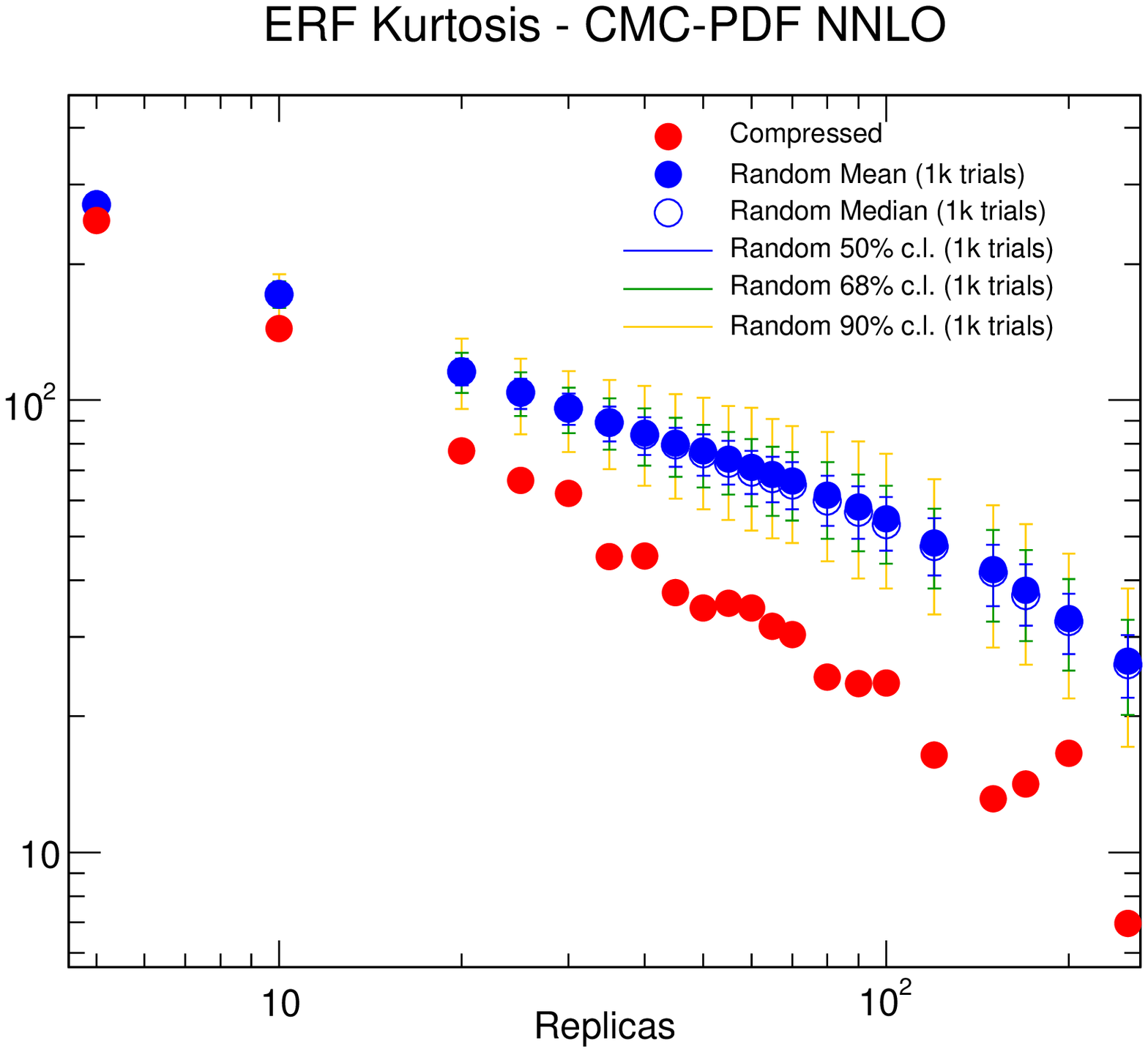}
  \includegraphics[scale=0.35]{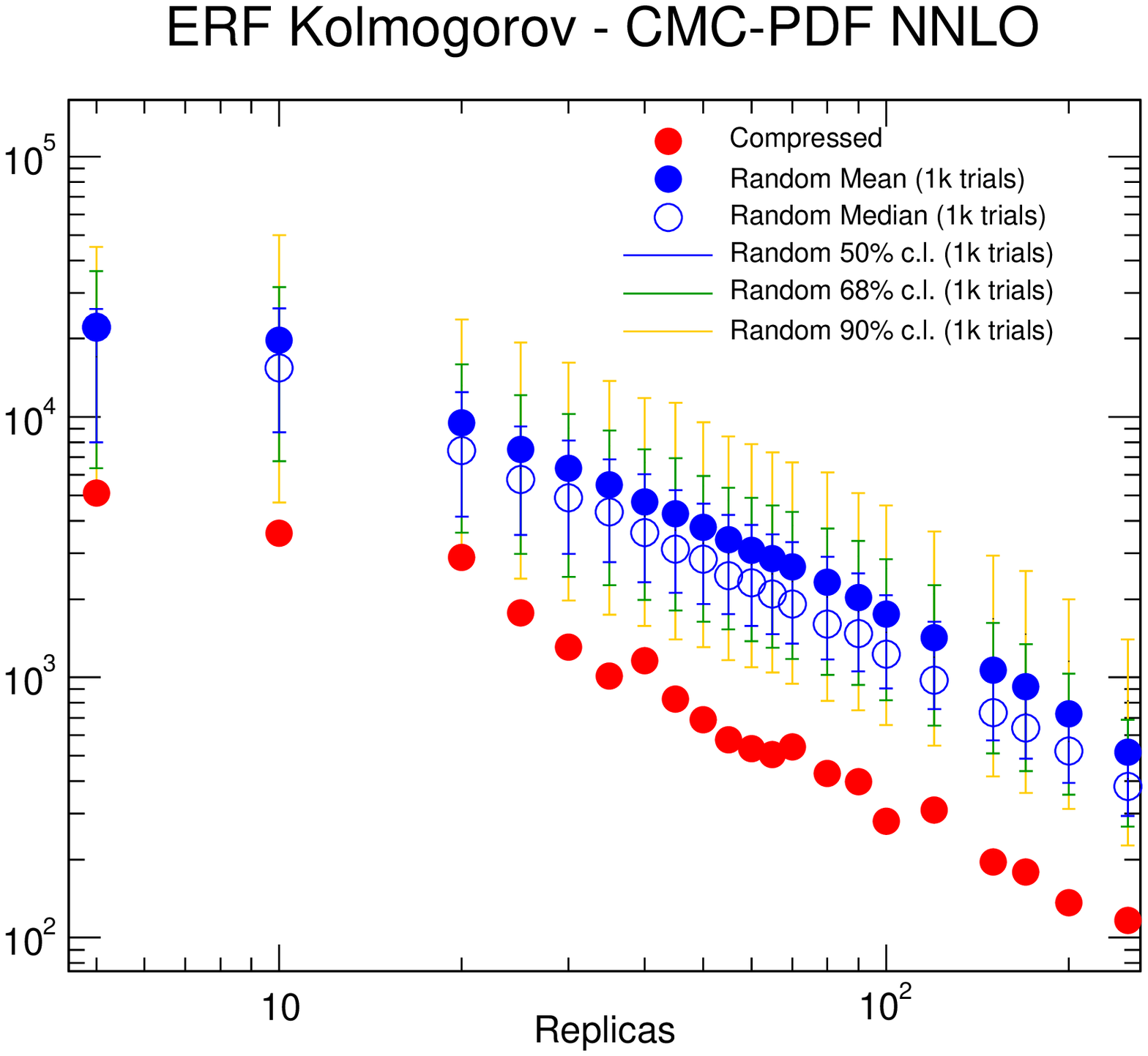}\includegraphics[scale=0.35]{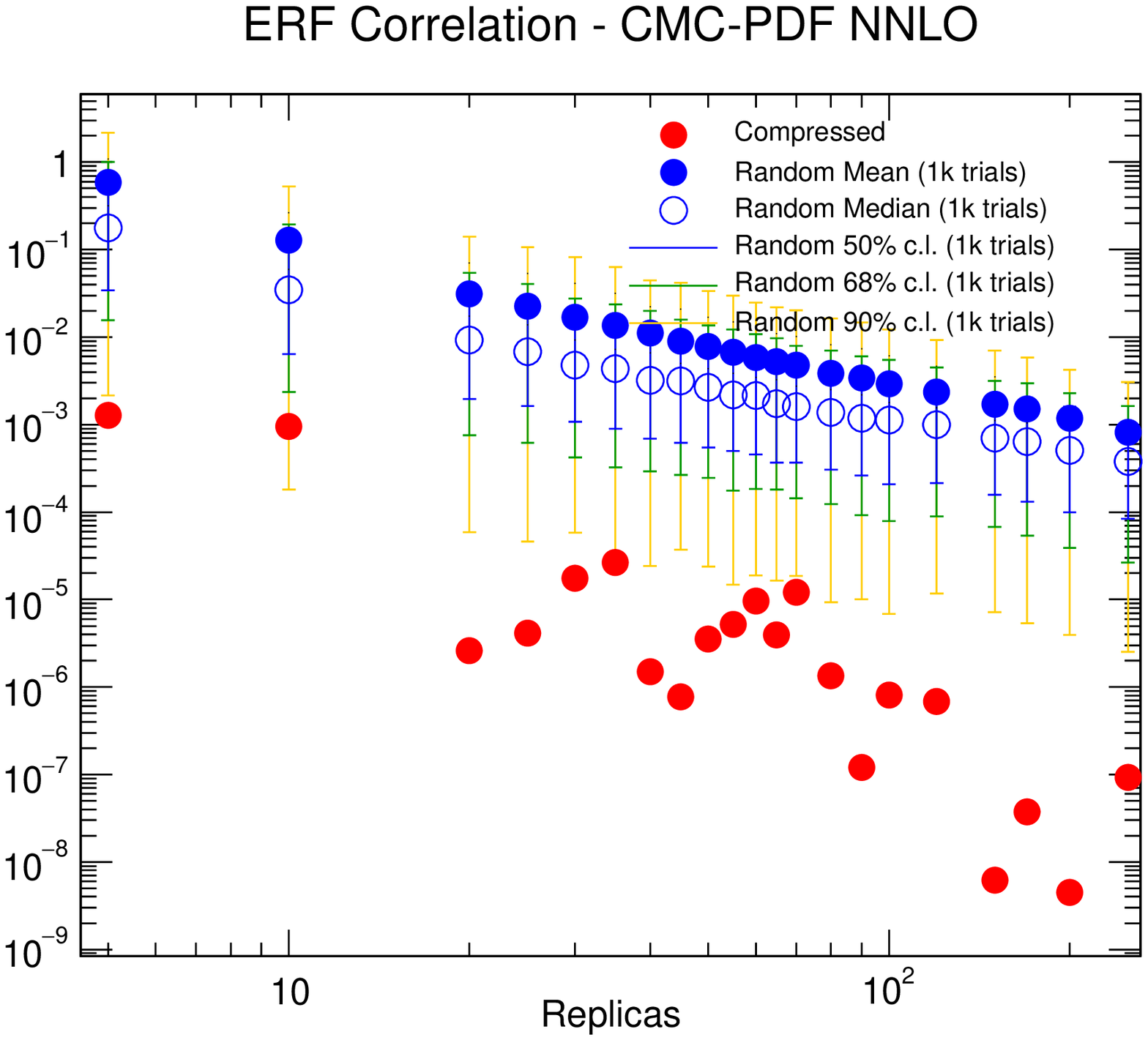}
  \caption{\small Same as Fig.~\ref{fig:erfnnpdf30} for the CMC-PDFs,
    starting from the prior with $\widetilde{N}_{\rm rep}$=900 replicas.}
  \label{fig:erfcmcpdf}
\end{figure}

We have verified that
the performance of the compression algorithm is similar regardless of the
prior.
To illustrate this point, in Fig.~\ref{fig:erfcmcpdf} we show the
corresponding version of Fig.~\ref{fig:erfnnpdf30}, namely
the various contributions to
the error function, for the case of
compression of the CMC-PDF sets.
We see that also in the case of the CMC-PDF sets the compression
improves the ERF as compared to random selections by an order of
magnitude or even more.

\begin{figure}[t]
  \centering 
  \includegraphics[scale=0.45]{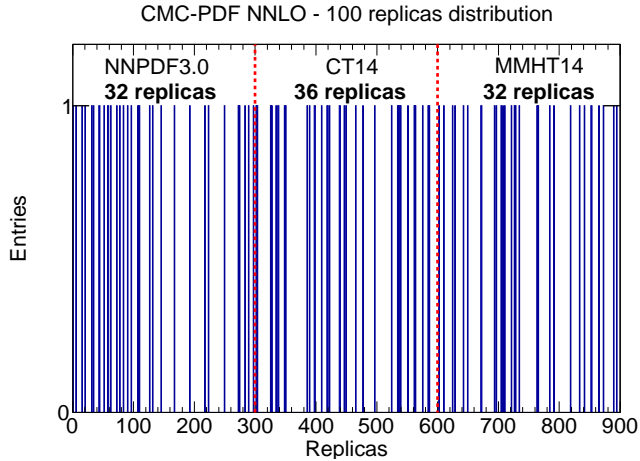}
  \caption{\small Replicas of the original
    combined set of $\widetilde{N}_{\rm rep}=900$ replicas
    selected for the compression with
    ${N}_{\rm rep}=100$ replicas, classified for each of the three
    input PDF sets.
  }
  \label{fig:replicascompression}
\end{figure}

It is interesting to determine, for a given compression, how many replicas
are selected from each of the three PDF sets that enter the combination.
Given that originally we assign equal weight to the three sets,
that is, the same number of replicas, we expect that if the compression
algorithm is unbiased the number of replicas from each set after the
compression should also be approximately the same.
We have verified that this is indeed the case, for instance,
in Fig.~\ref{fig:replicascompression} we show, for a compression
with $N_{\rm rep}=100$ replicas, how the replicas of the original distribution
are selected: we see that a similar number
has been selected from NNPDF3.0, CT14 and MMHT14: 32, 36 and 32
replicas respectively, in agreement
with our expectations.

\begin{figure}[h]
  \centering 
  \includegraphics[scale=0.35]{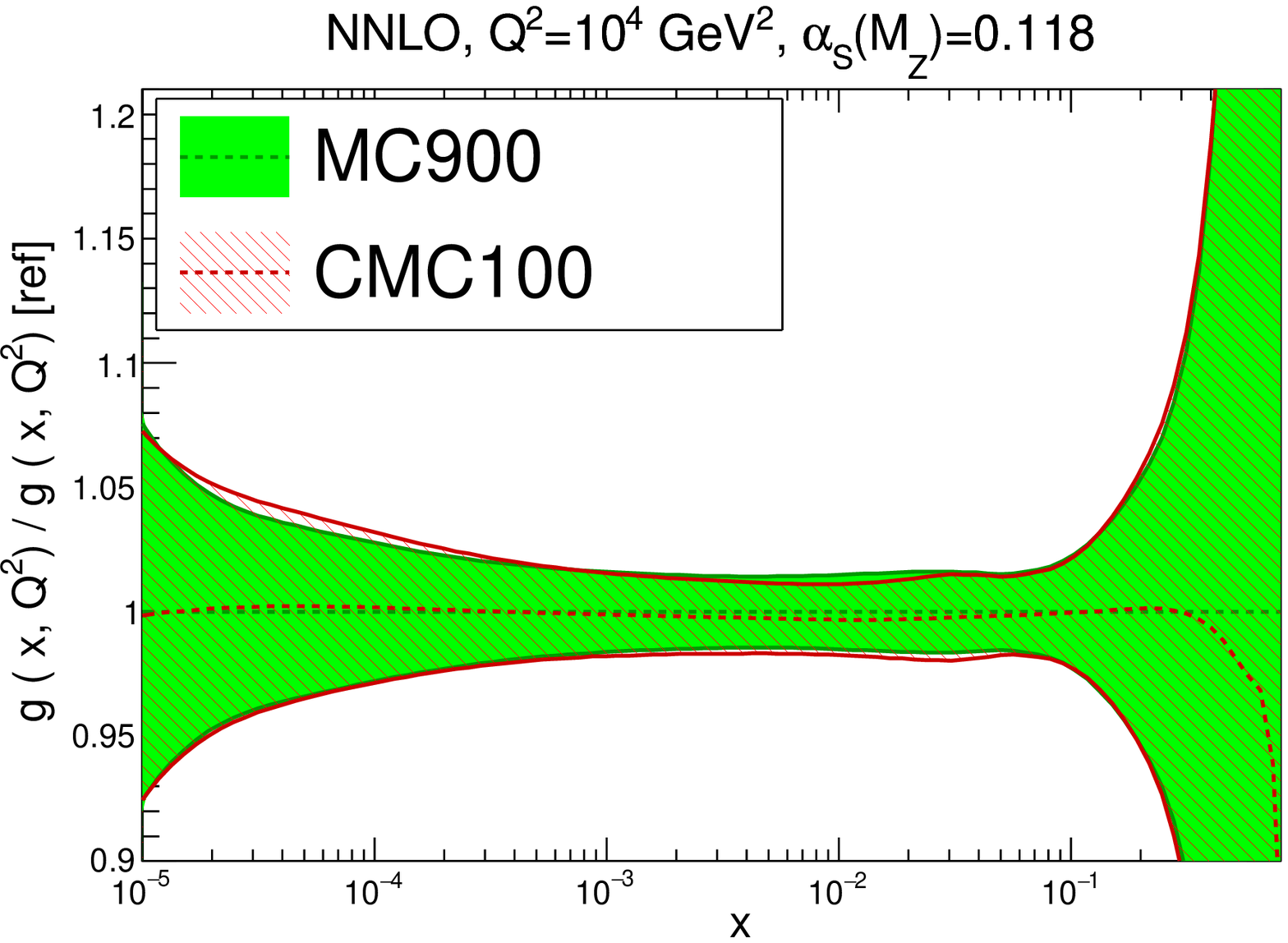}
  \includegraphics[scale=0.35]{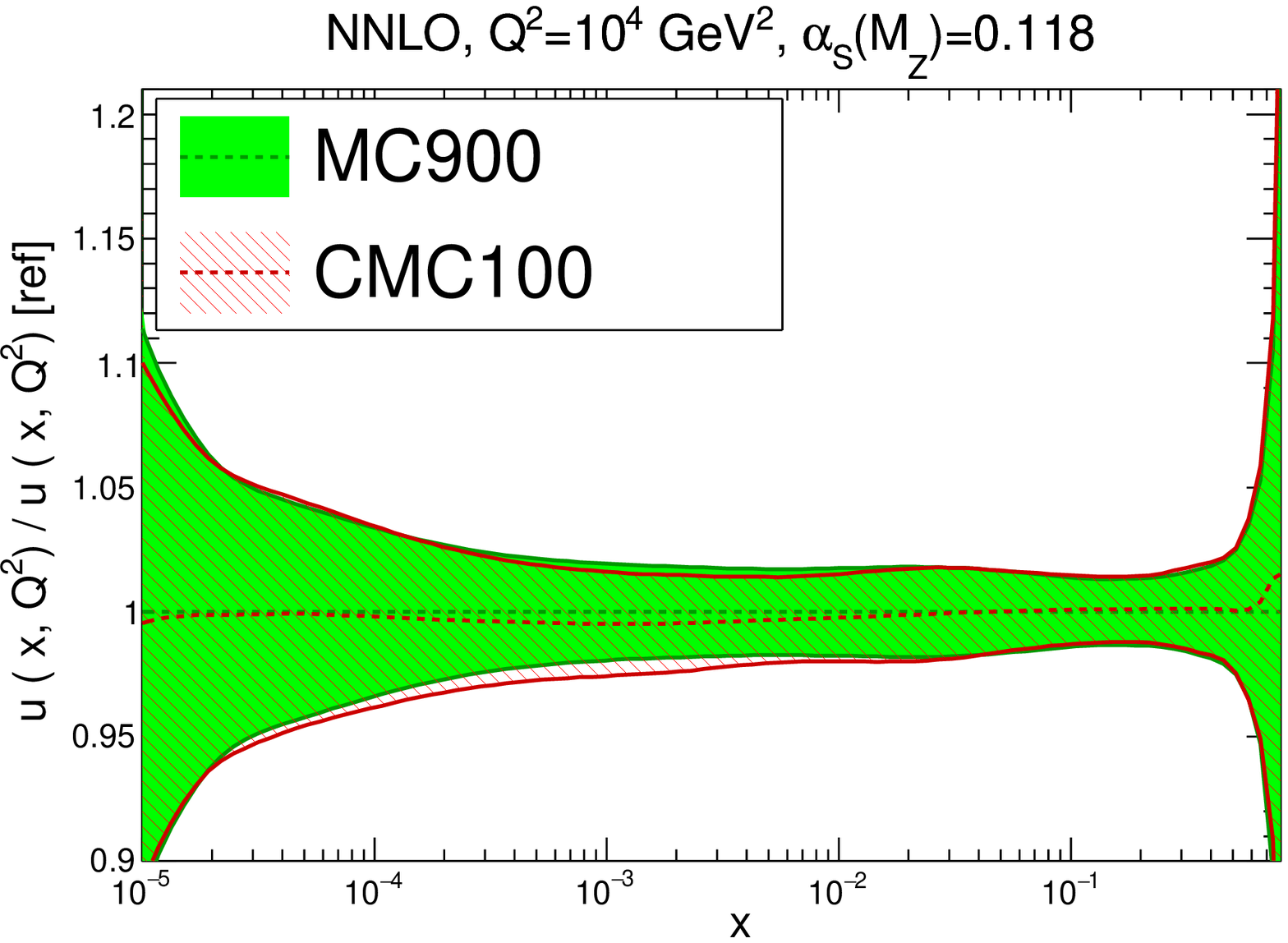}
  \includegraphics[scale=0.35]{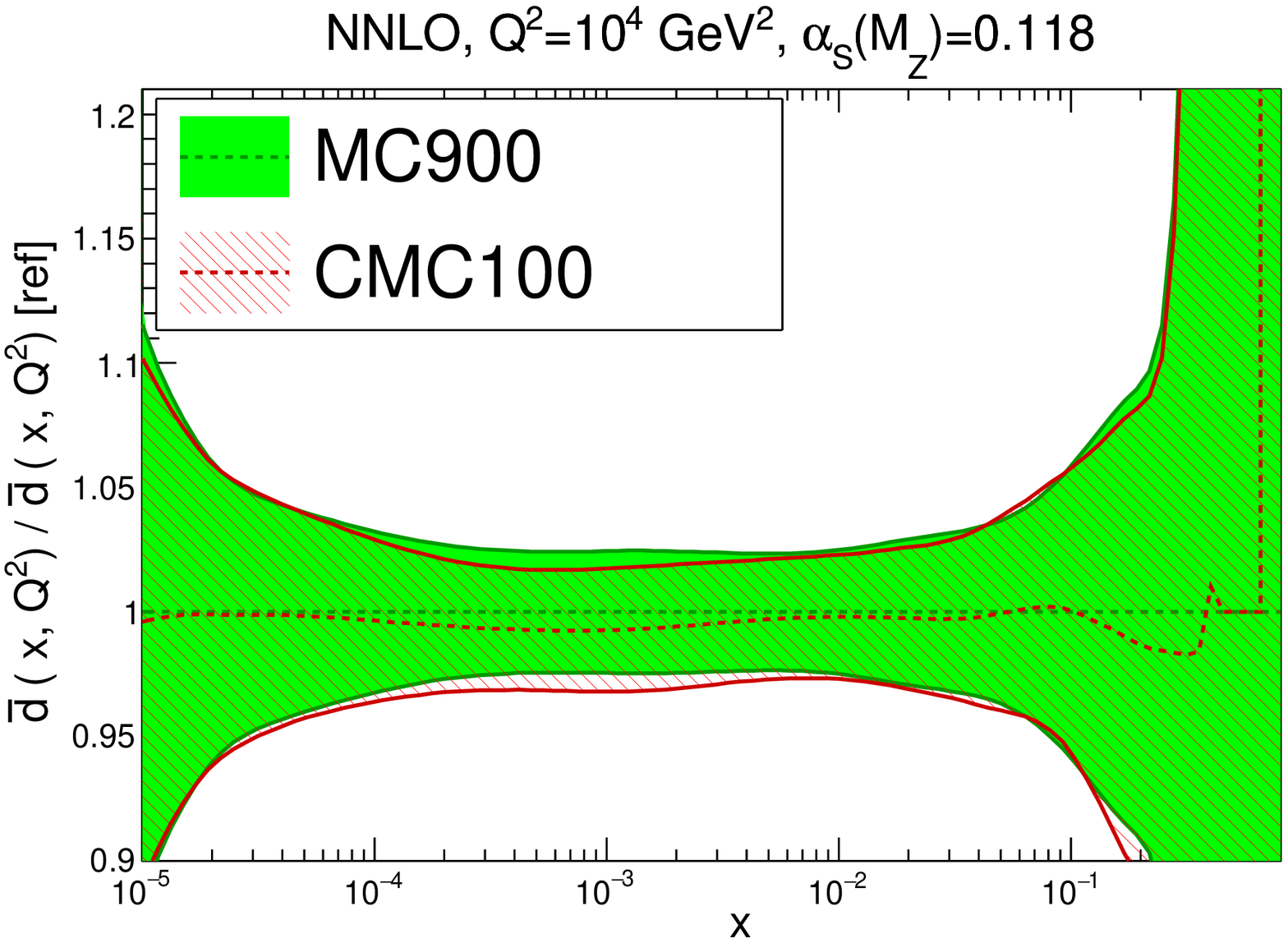}
  \includegraphics[scale=0.35]{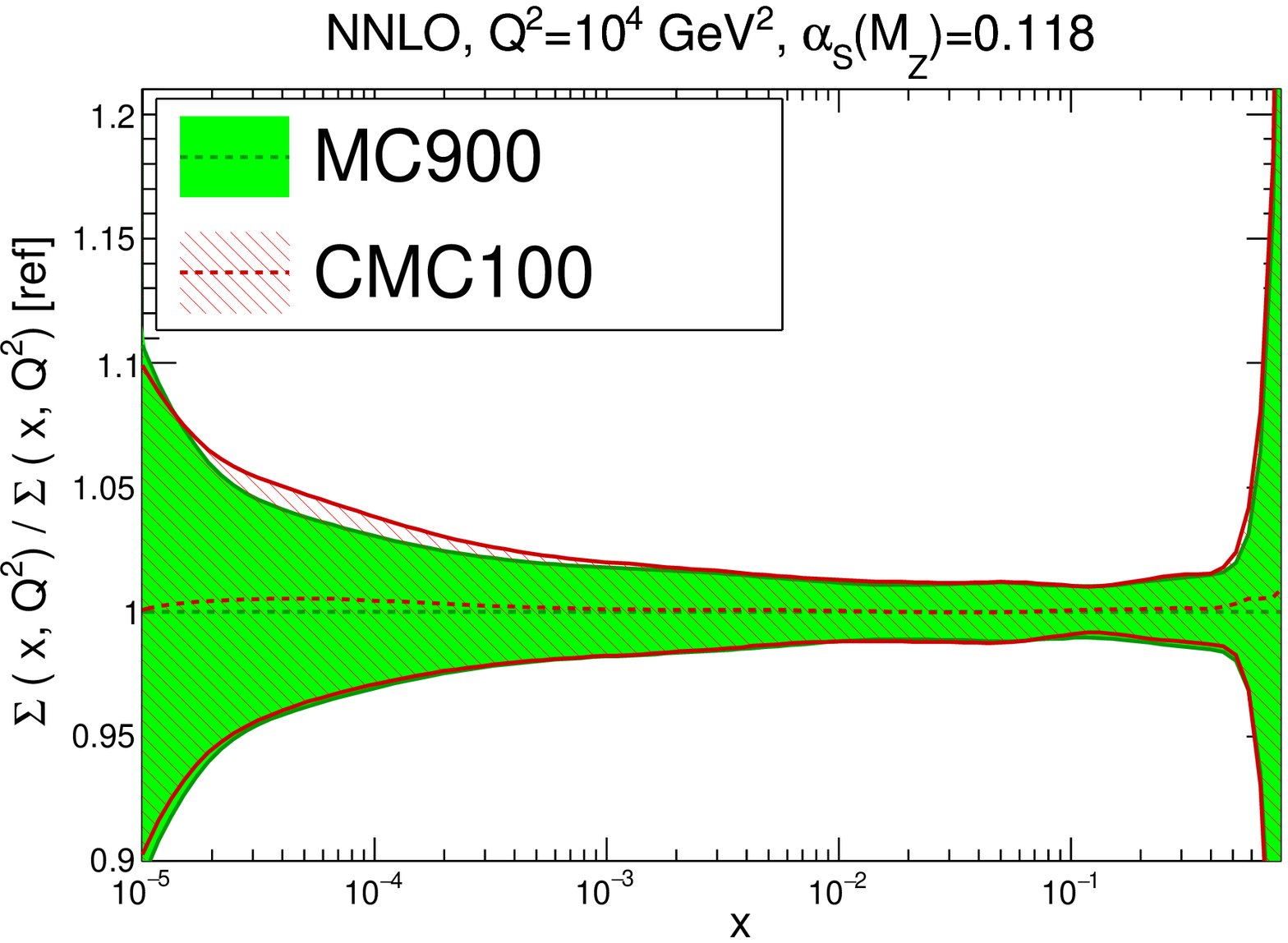}
  \caption{\small
    Comparison of the PDFs between the original  Monte Carlo
    combination of NNPDF3.0, CT14 and MMHT14, MC900,
    with the compressed CMC100 PDFs.
    We show the gluon, up quark, down antiquark, and total
    quark singlet,
    as ratios to the prior for $Q^2=10^4$ GeV$^2$.
  }
  \label{fig:cmcpdf-validation1}
\end{figure}

We now address the comparison between the MC900 prior
and the
new CMC-PDFs.
For illustration, we will show results for
$N_{\rm rep}=100$, with the understanding that using a larger
number of replicas would improve even further the agreement
with the prior.
In Fig.~\ref{fig:cmcpdf-validation1} we show the 
comparison of the PDFs between the original  Monte Carlo
combination of NNPDF3.0, CT14 and MMHT14,
with $\widetilde{N}_{\rm rep}=900$ replicas, with the corresponding
compressed set with $N_{\rm rep}=100$ replicas.
We show the gluon, up quark, down antiquark, and strange quark,
as ratios to the prior set at a typical
LHC scale of $Q=100$ GeV.
We see that in all cases the agreement is sufficiently good.

\begin{figure}[h]
  \centering 
  \includegraphics[scale=0.35]{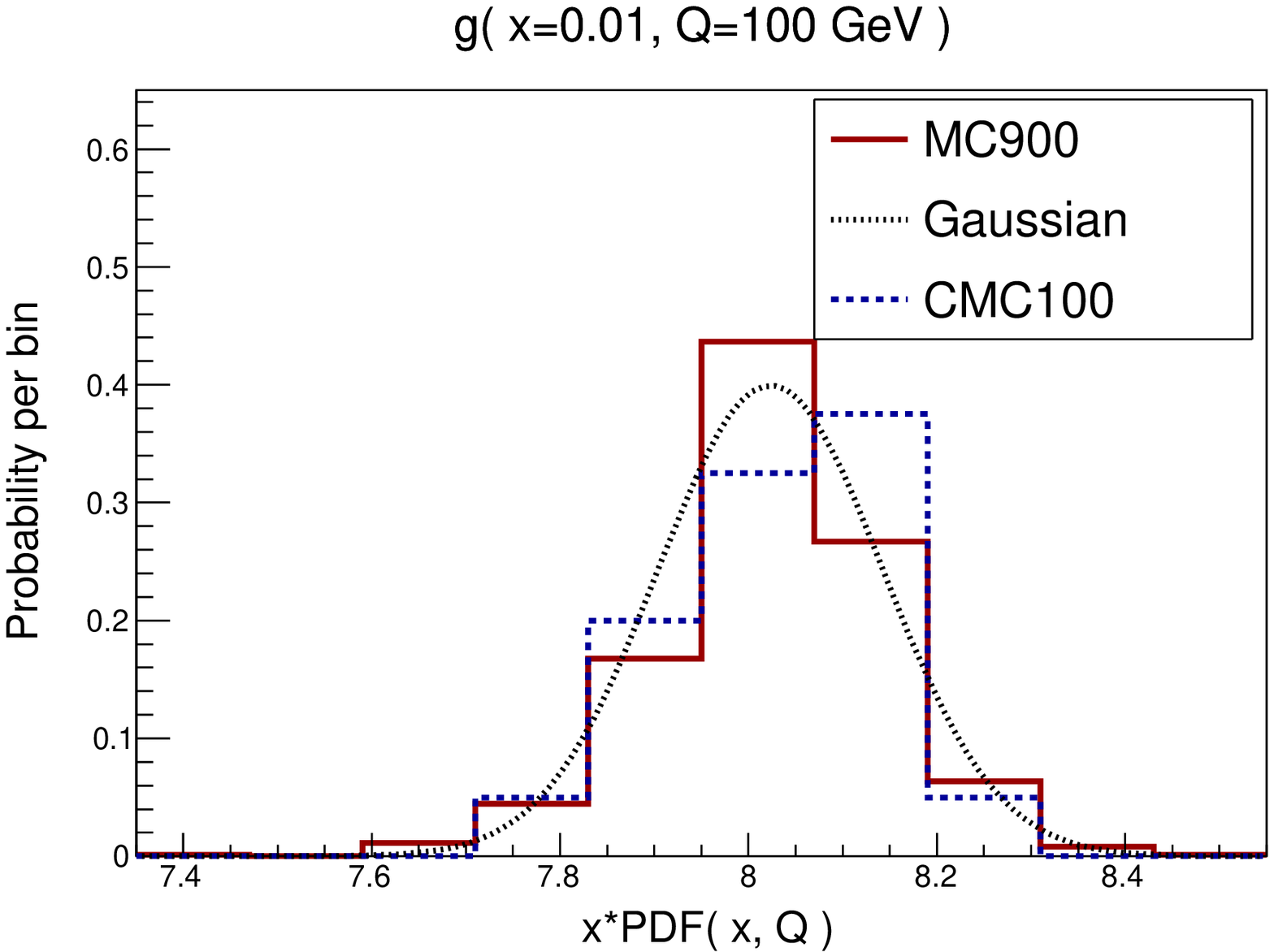}
  \includegraphics[scale=0.35]{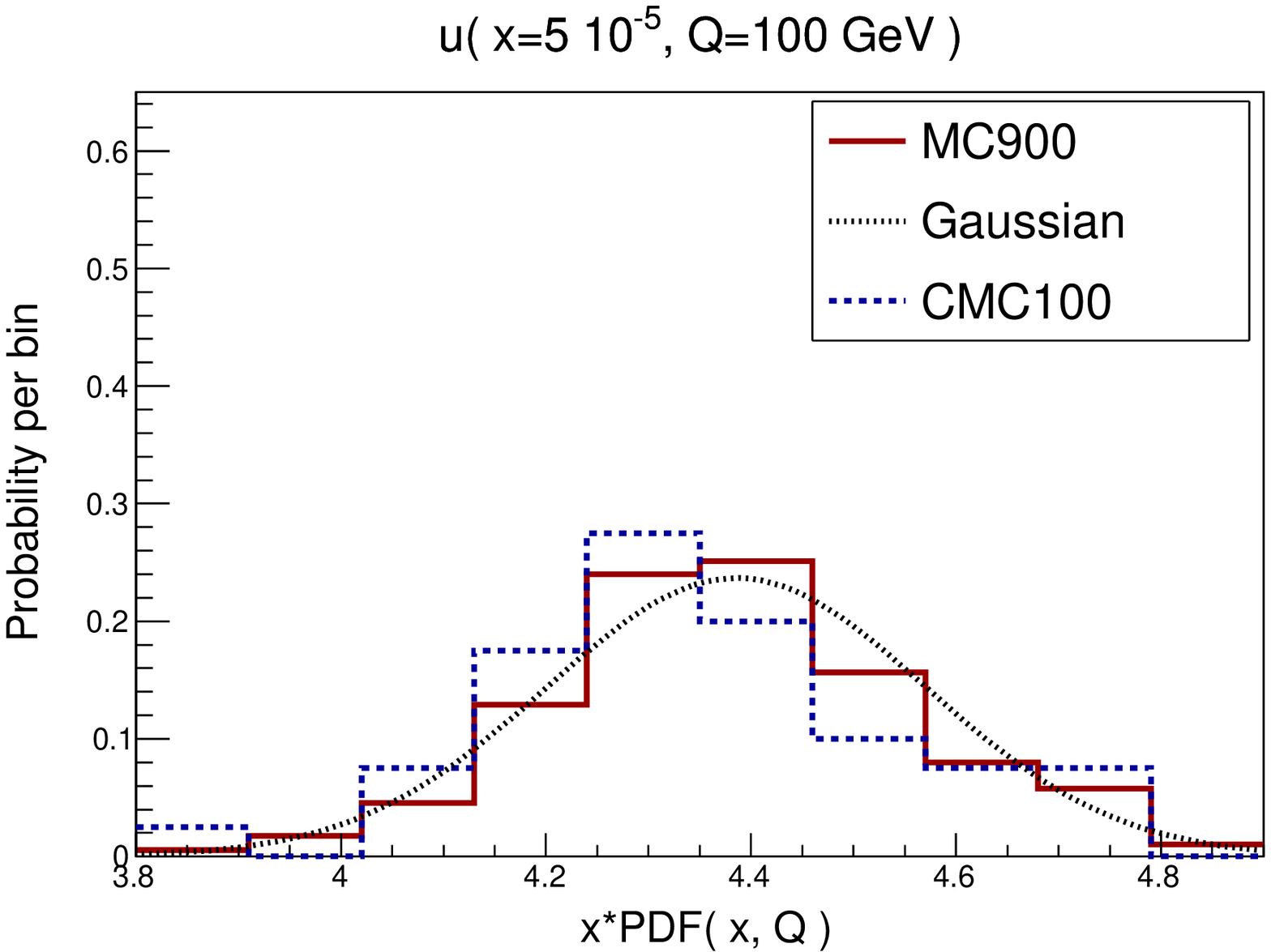}
  \includegraphics[scale=0.35]{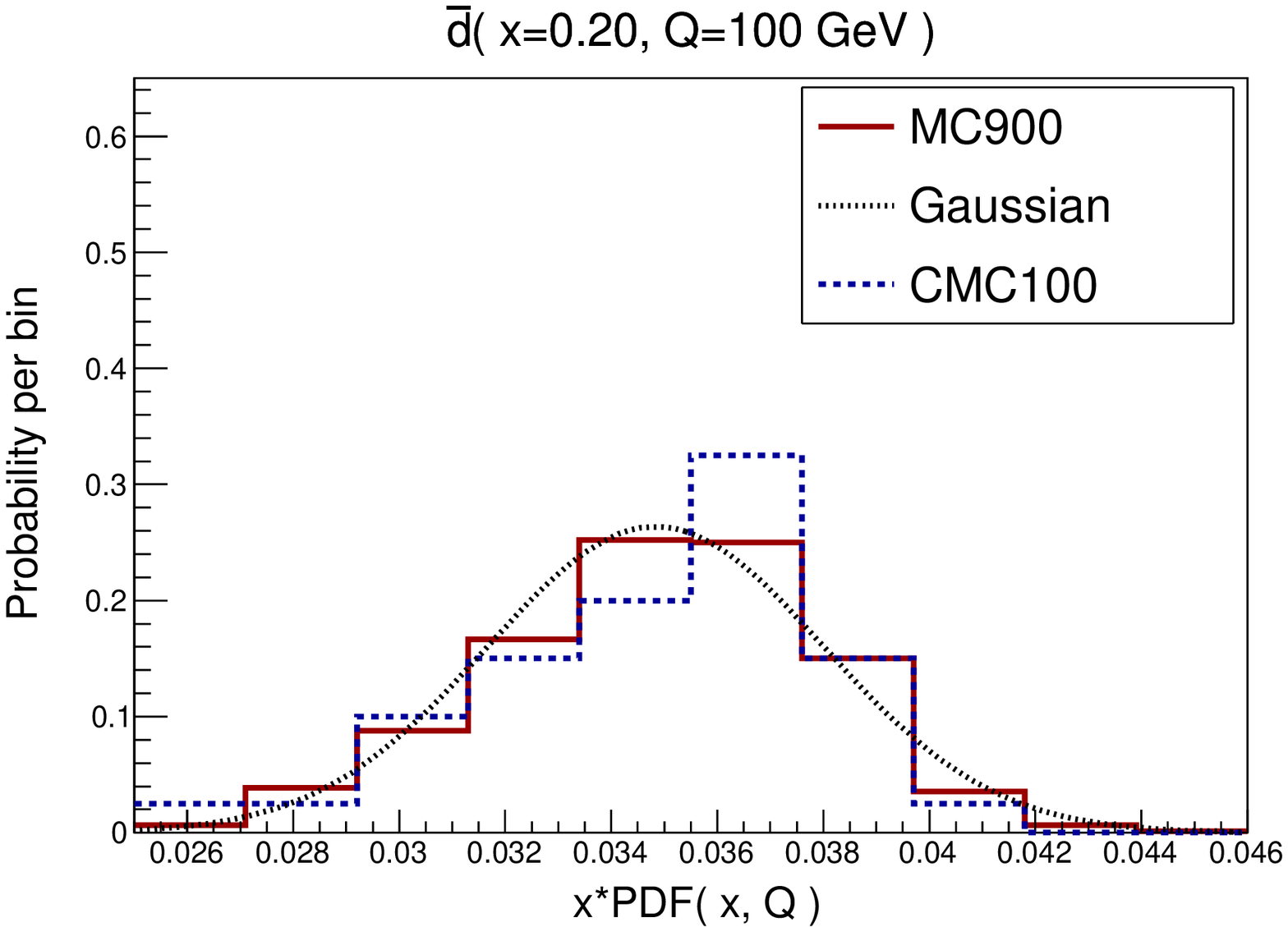}
  \includegraphics[scale=0.35]{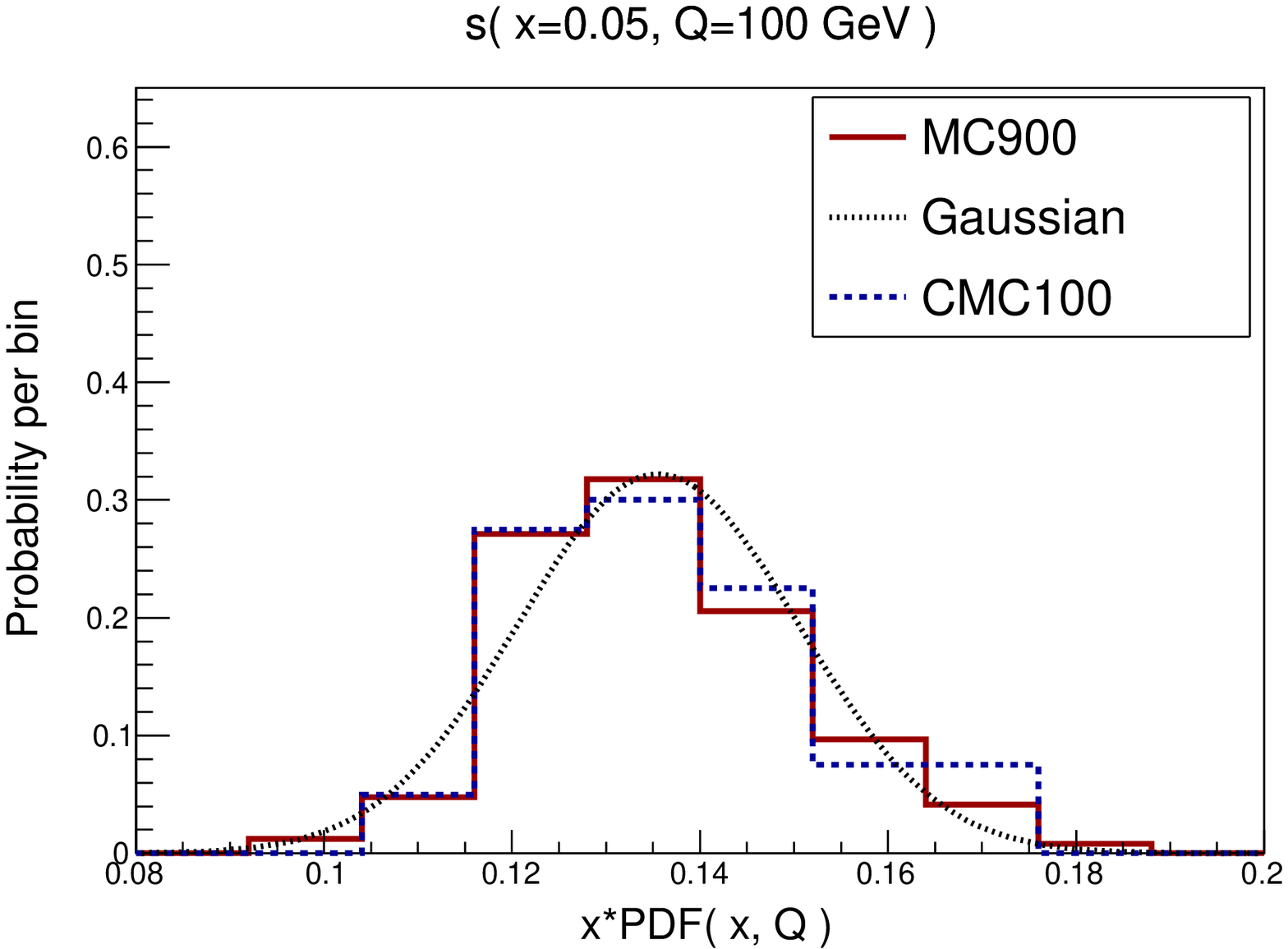}
  \caption{\small
    Same as Fig.~\ref{fig:PDFcomp_histo} now with the comparison
    between MC900 and CMC100.
    The Gaussian curve 
has the same mean and variance as the MC900 prior.
  }
  \label{fig:PDFcomp_histo2}
\end{figure}

In Fig.~\ref{fig:PDFcomp_histo2}
we show the same as in Fig.~\ref{fig:PDFcomp_histo}, namely
the histograms representing the distribution of the values
of the PDFs over the
Monte Carlo replicas for different flavours and values of $(x,Q)$,
now comparing the original and compressed CMC-PDFs with
$\widetilde{N}_{\rm rep}=900$ and $N_{\rm rep}=100$ respectively.
As was done in Fig.~\ref{fig:PDFcomp_histo}, in Fig.~\ref{fig:PDFcomp_histo2} we also show a Gaussian
with mean and variance determined from the prior
$\widetilde{N}_{\rm rep}=900$ CMC-PDF.

\begin{figure}[h]
  \centering 
  \includegraphics[scale=0.33]{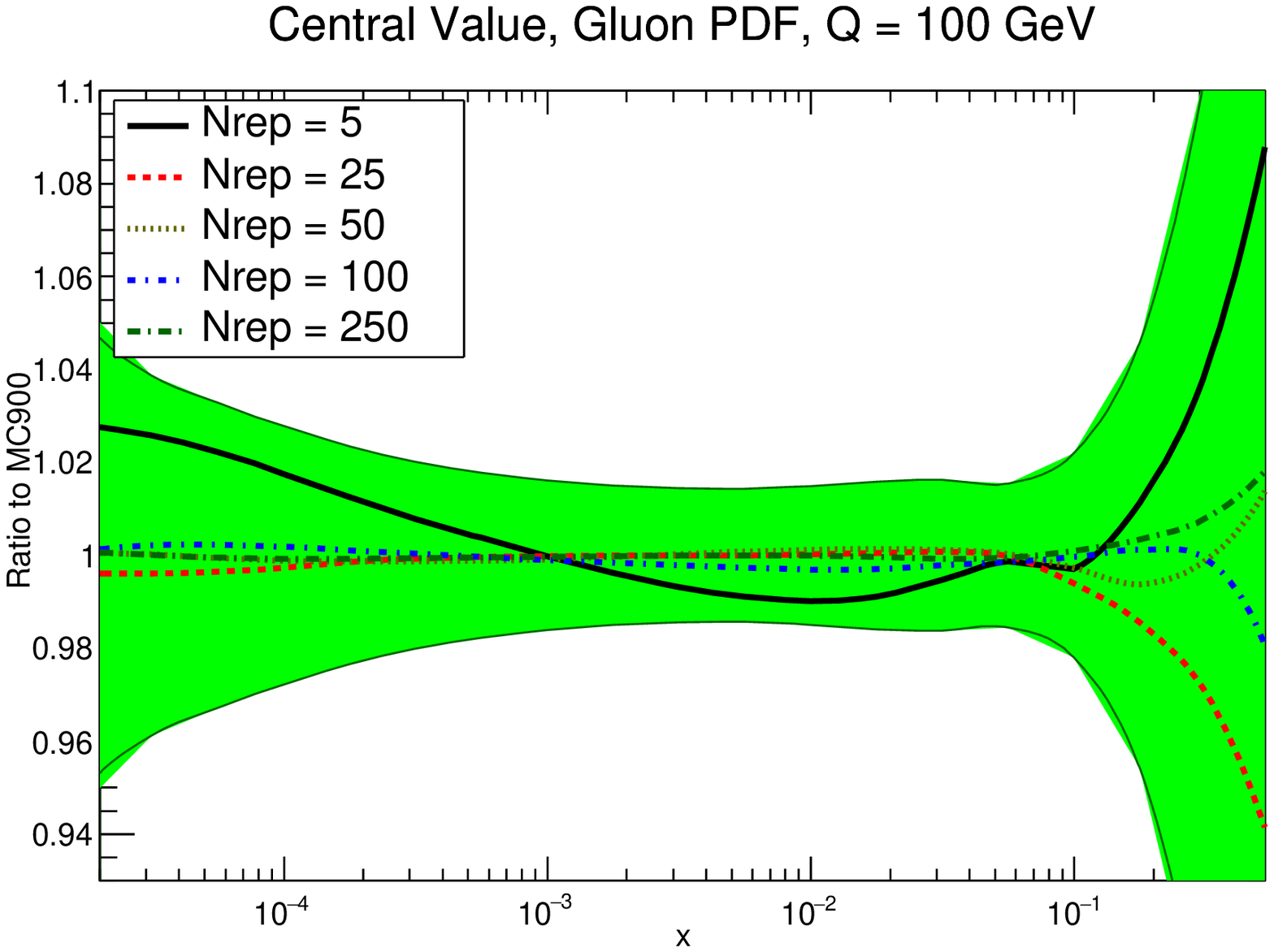}
  \includegraphics[scale=0.33]{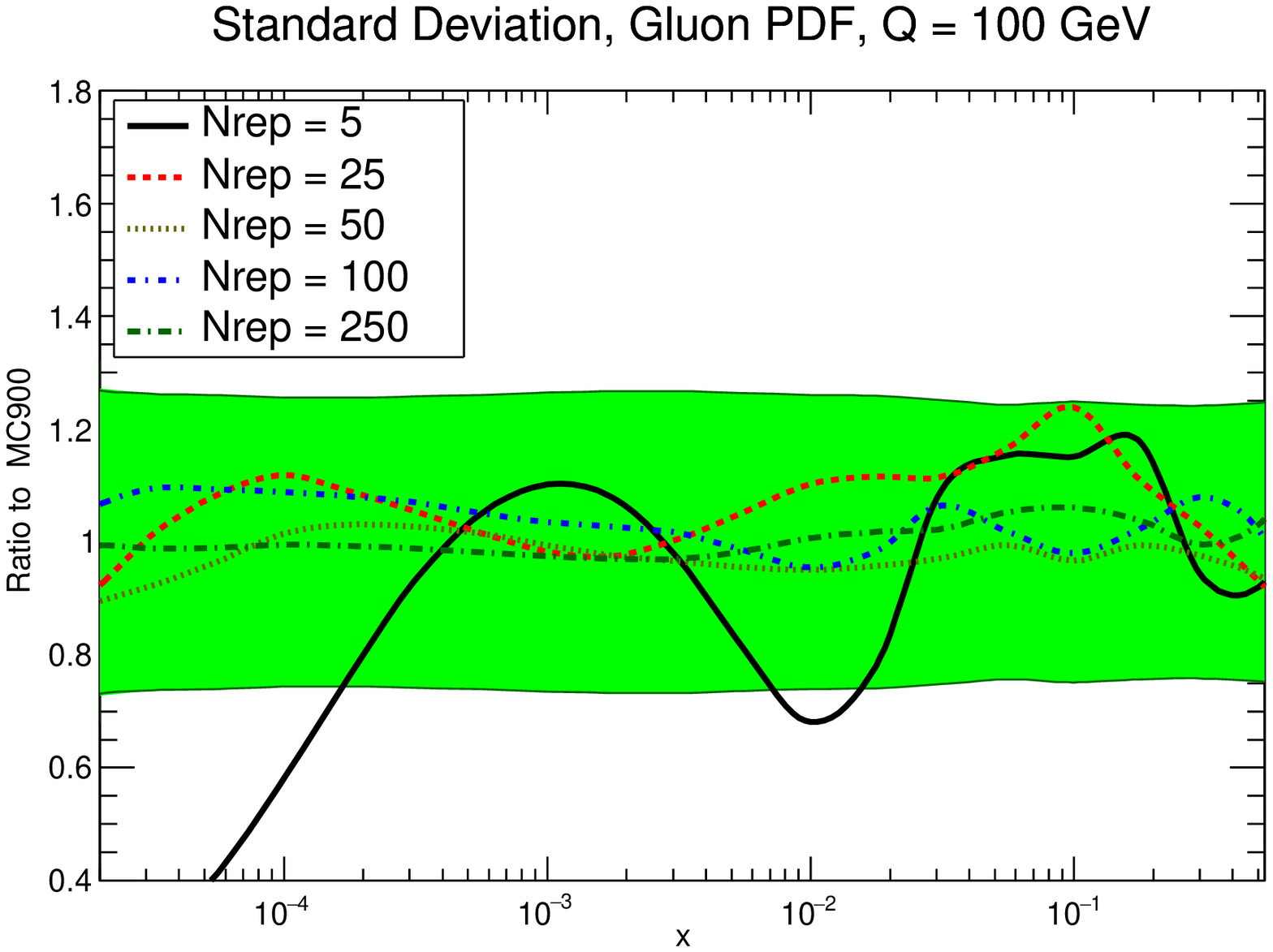}
  \includegraphics[scale=0.33]{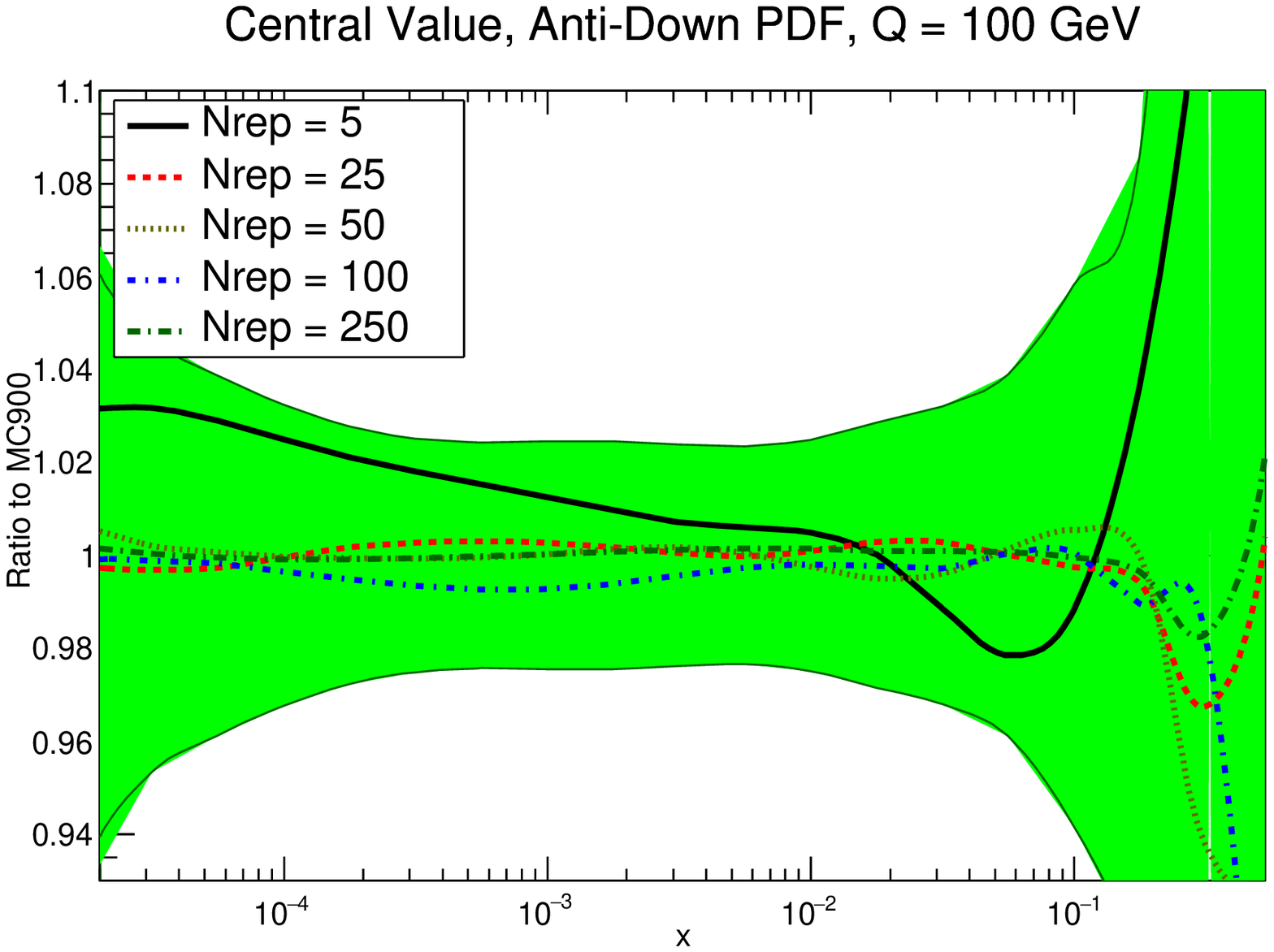}
  \includegraphics[scale=0.33]{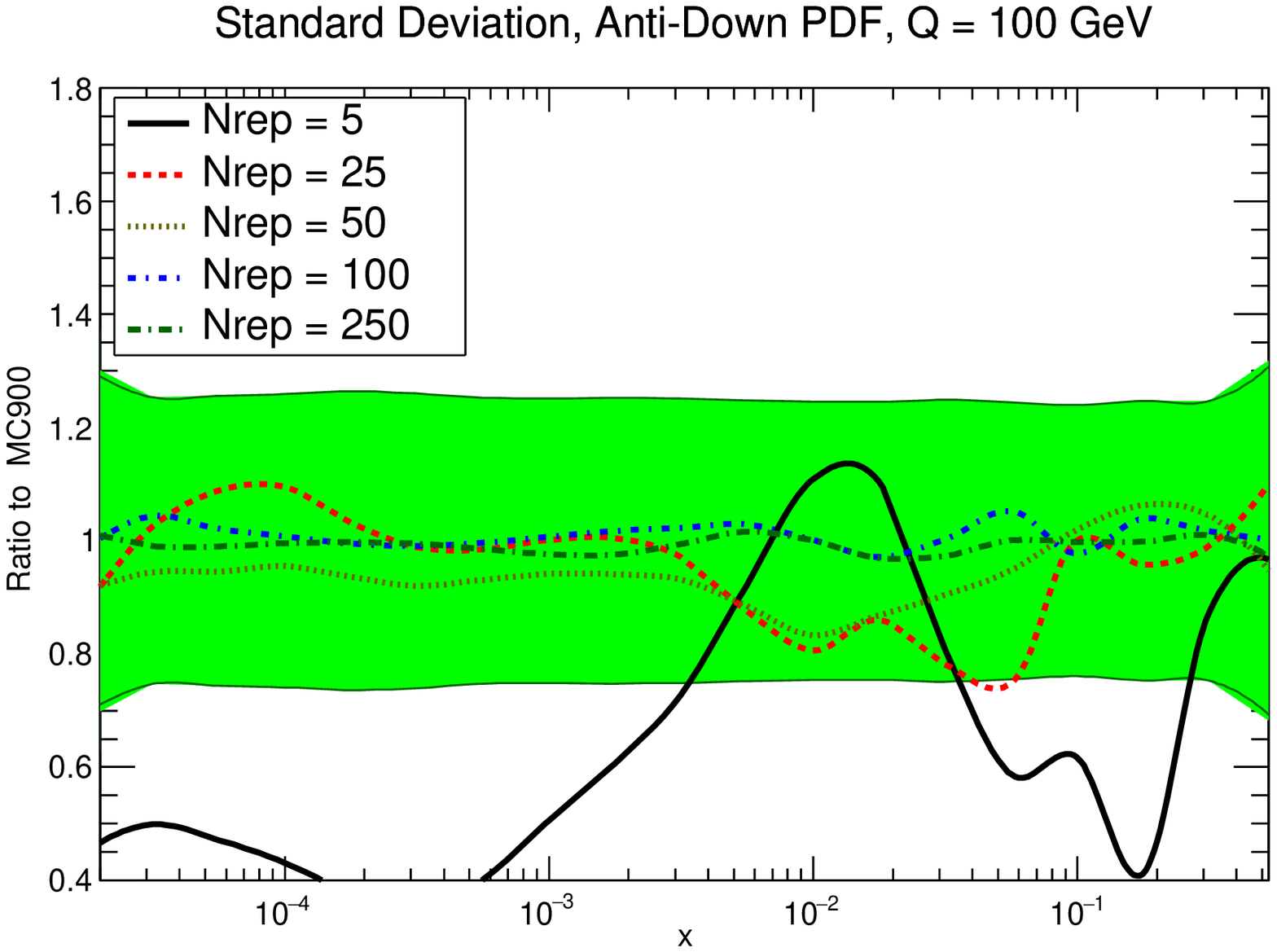}
  \caption{\small Comparison of the 
    central values (left plots) and one-sigma
    intervals (right plots) for the CMC-PDFs with different
    values of 
    $N_{\rm rep}$ (5, 25, 50, 100 and 250 respectively),
    for the gluon (upper plots) and the down anti-quark
    (lower plots).
    Results are shown normalized to the central value and the
    standard deviation of the MC900 prior combined set, respectively.
    We also show the one-sigma PDF band (left plots) and the
    variance of the variance (right plots) as a full green
    band.
  }
  \label{fig:cmcpdf-validation2}
\end{figure}

To gauge the dependence of the agreement between the prior and
the compressed Monte Carlo sets, it is illustrative to compare
central values and variances for the different values of
$N_{\rm rep}$ in the compression.
This comparison is shown for the gluon and the down anti-quark in
Fig.~\ref{fig:cmcpdf-validation2}.
In the left plots, we compare the central value of the PDF
for different values of $N_{\rm rep}$, normalized to the prior
result.
We also show the one-sigma PDF band, which is useful to compare
the deviations found in the compressed set with the typical
statistical fluctuations.
We see that starting from $N_{\rm rep}\simeq 25$ replicas, the central
values of the compressed sets fluctuate much less than the size of
the PDF uncertainties.
In the right plot of Fig.~\ref{fig:cmcpdf-validation2} we show
the corresponding comparison at the level of standard deviations,
again normalized to the standard deviation of the prior set.
Here for reference the green band shows
the variance of the variance itself, which is typically
of the order of 20\%-30\% in a Monte Carlo PDF set~\cite{Demartin:2010er}.
Here we see that with $N_{\rm rep}\simeq 100$ replicas or more, the variance of
the compressed set varies by a few percent at most, much less than
the statistical fluctuations of the PDF uncertainty
itself.

\begin{figure}[h]
  \centering 
  \includegraphics[scale=0.33]{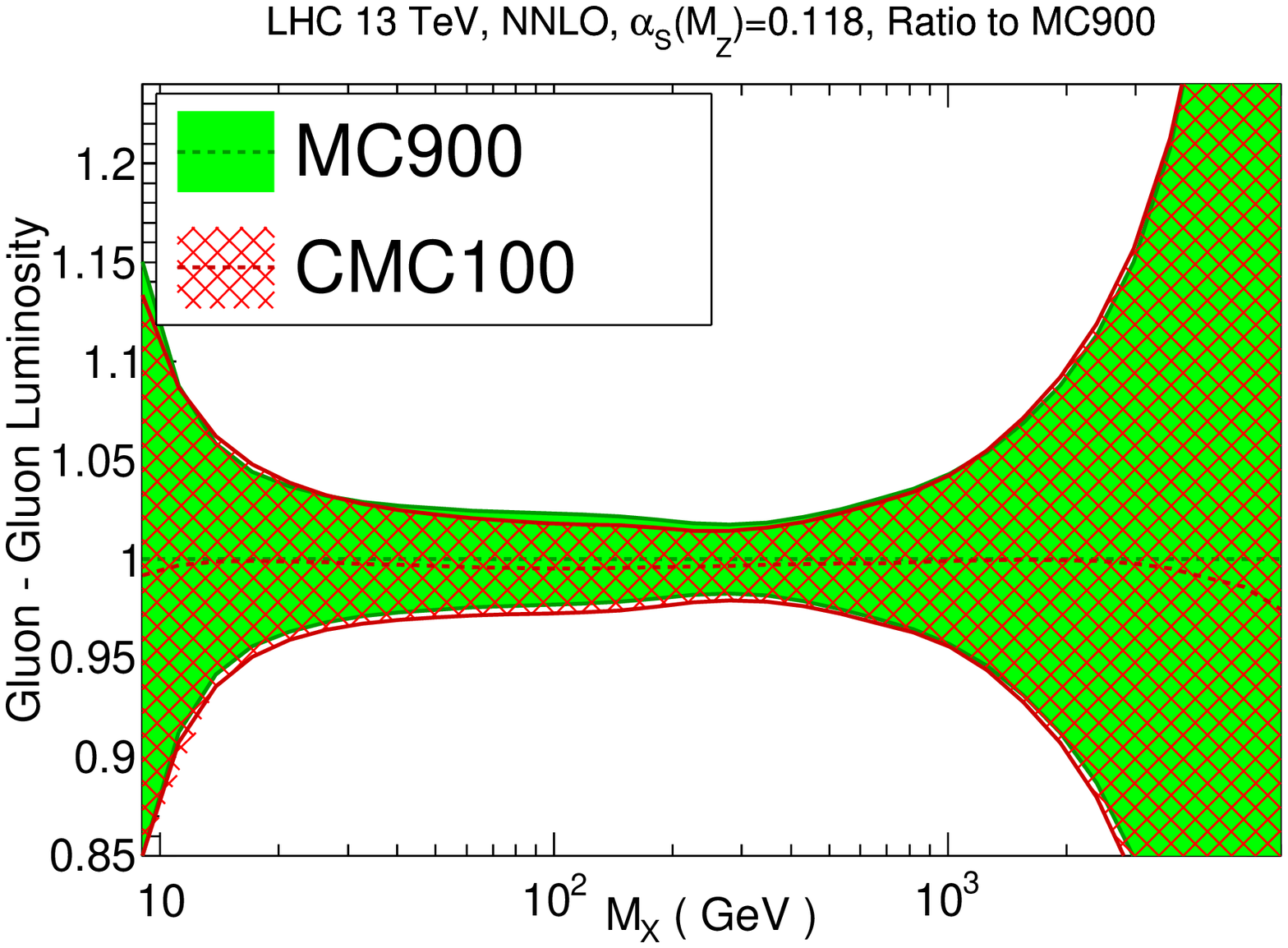}
  \includegraphics[scale=0.33]{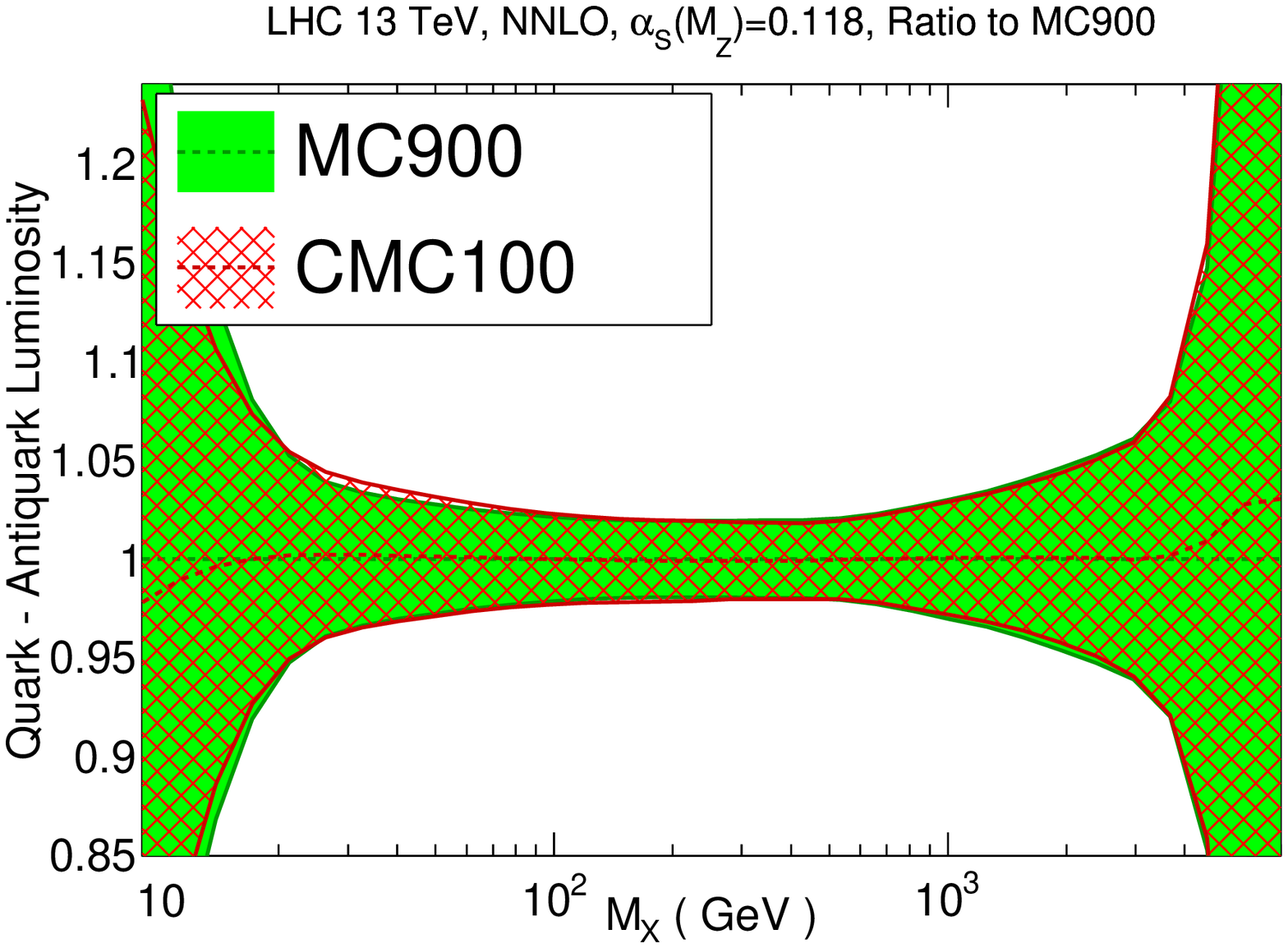}
  \includegraphics[scale=0.33]{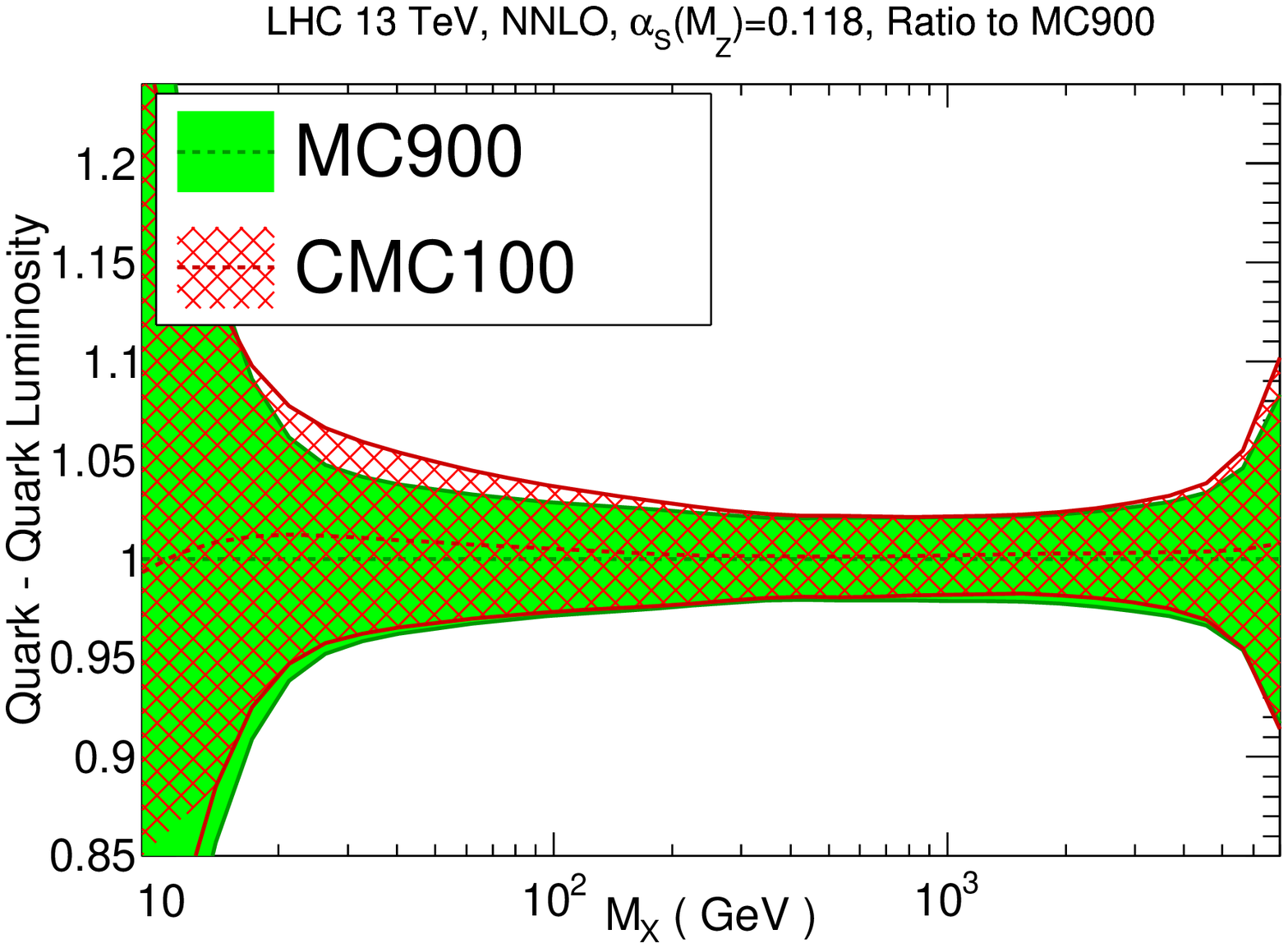}
  \includegraphics[scale=0.33]{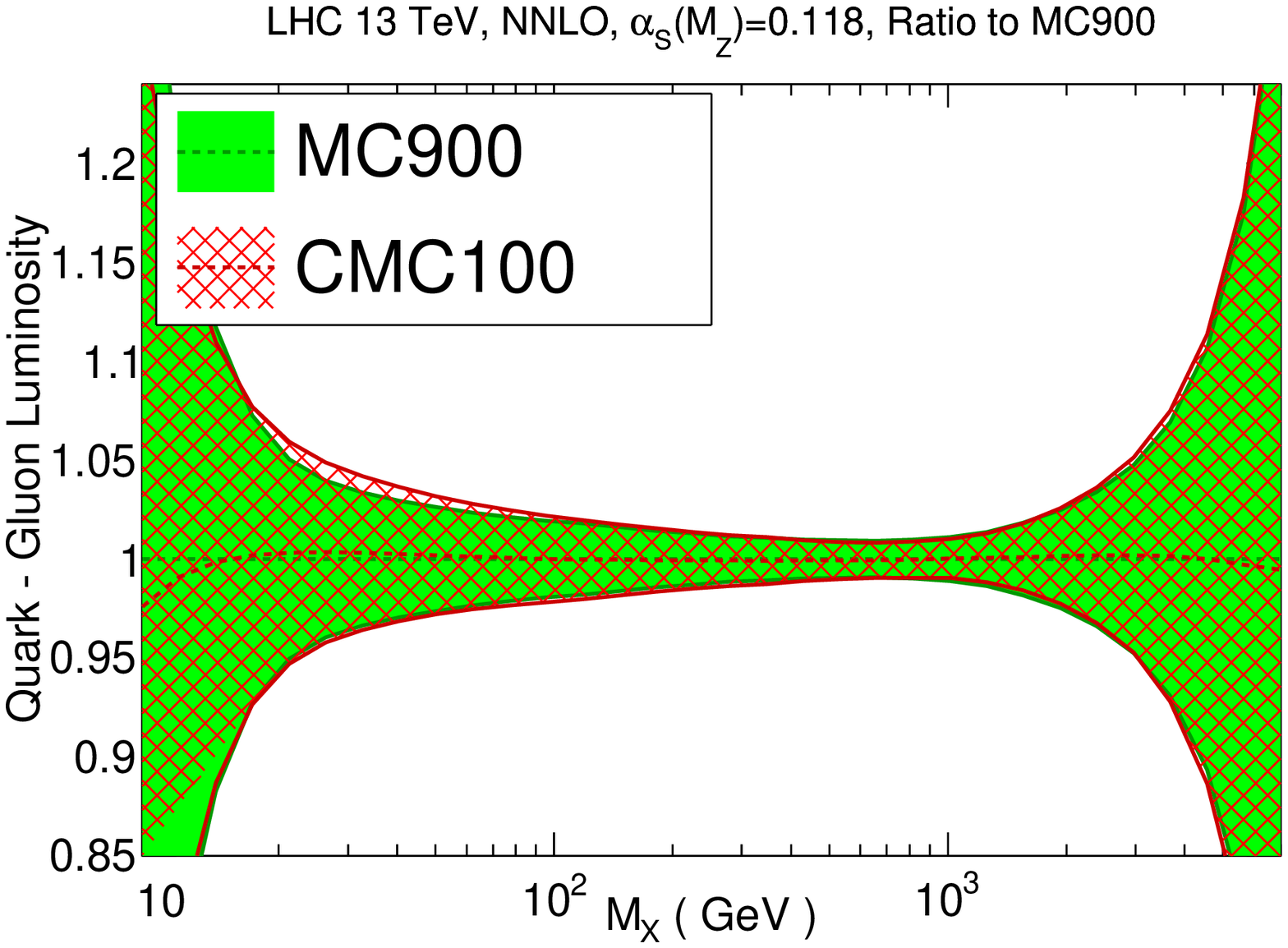}
  \caption{\small
    Same as Fig.~\ref{valnnpdf2} for the comparison
    between the prior set MC900 and the compressed set CMC100
   }
  \label{fig:cmcpdf-lumis}
\end{figure}

As in the case of the native Monte Carlo sets, it is also useful
here for the CMC-PDFs to compare the parton luminosities
between the original and the compressed sets.
This comparison is shown in Fig.~\ref{fig:cmcpdf-lumis},
which is the analog of Fig.~\ref{valnnpdf2} in the case of
CMC-PDFs.
As in the case of the native sets, we find also here good
agreement at the level of PDF luminosities.
As we will see in the next section, this agreement will also
translate to all LHC cross-sections and differential distributions
that we have explored.

\begin{figure}[h]
  \centering 
  \includegraphics[scale=0.52]{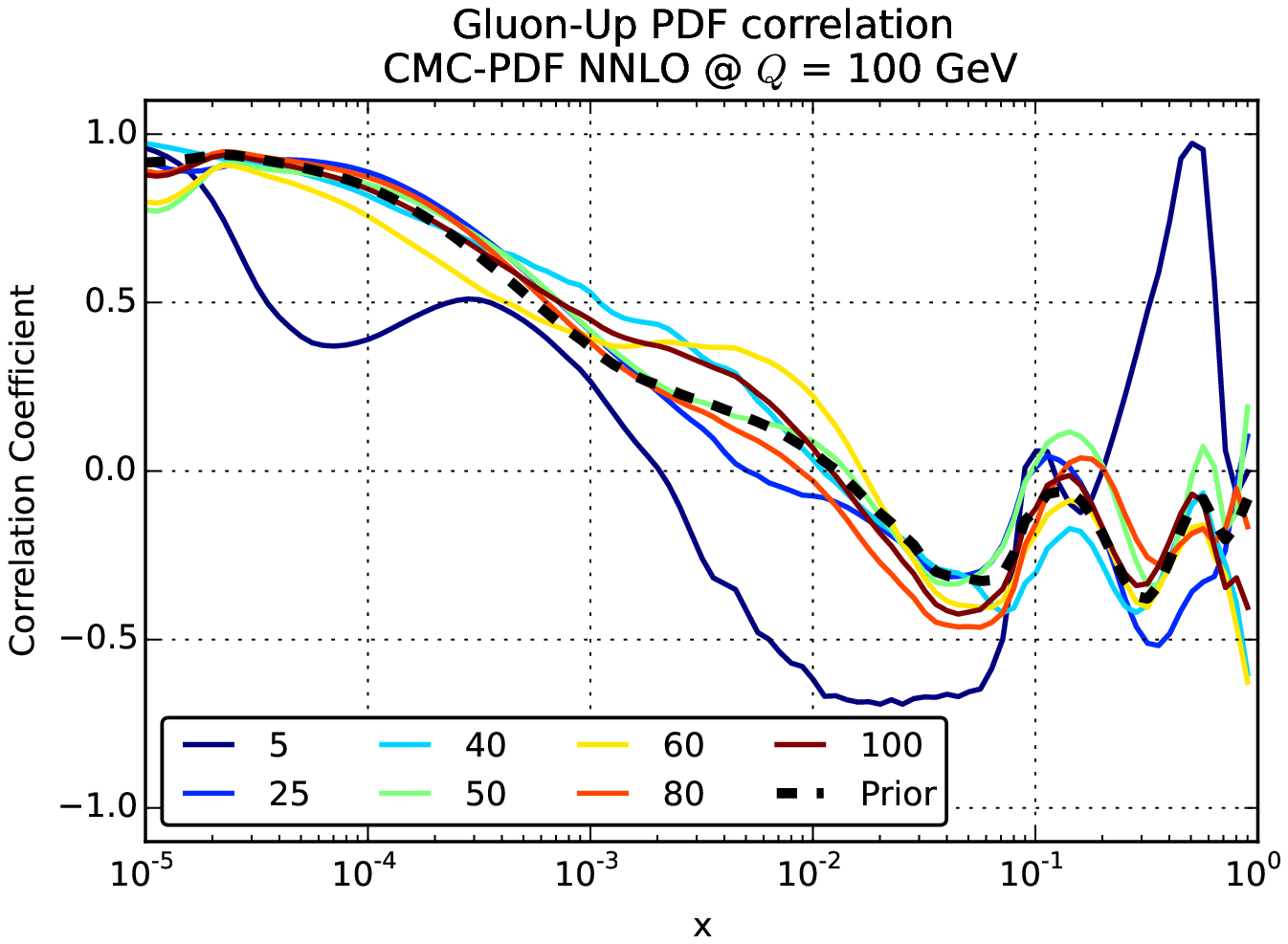}
  \includegraphics[scale=0.52]{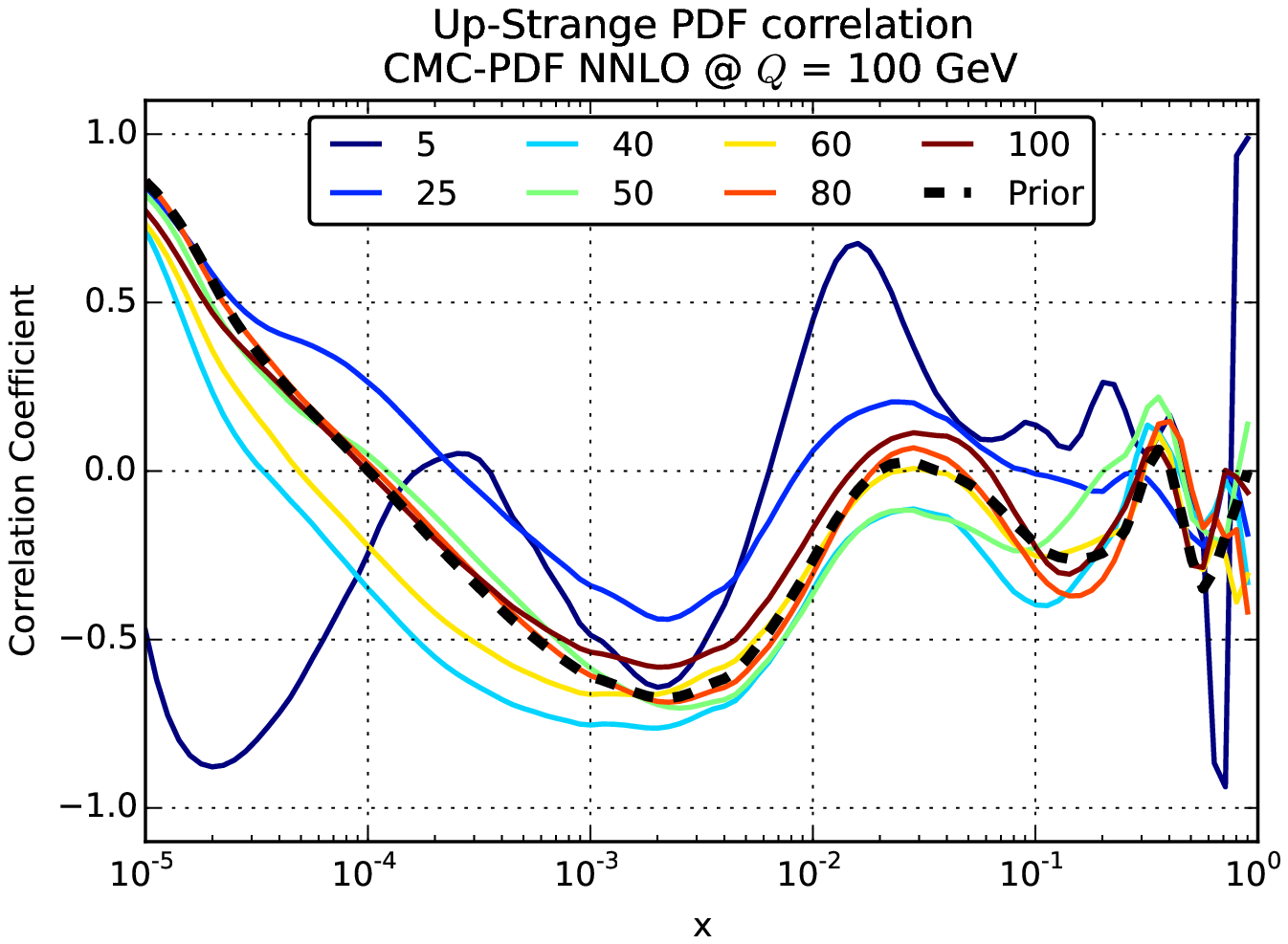}\\
  \includegraphics[scale=0.52]{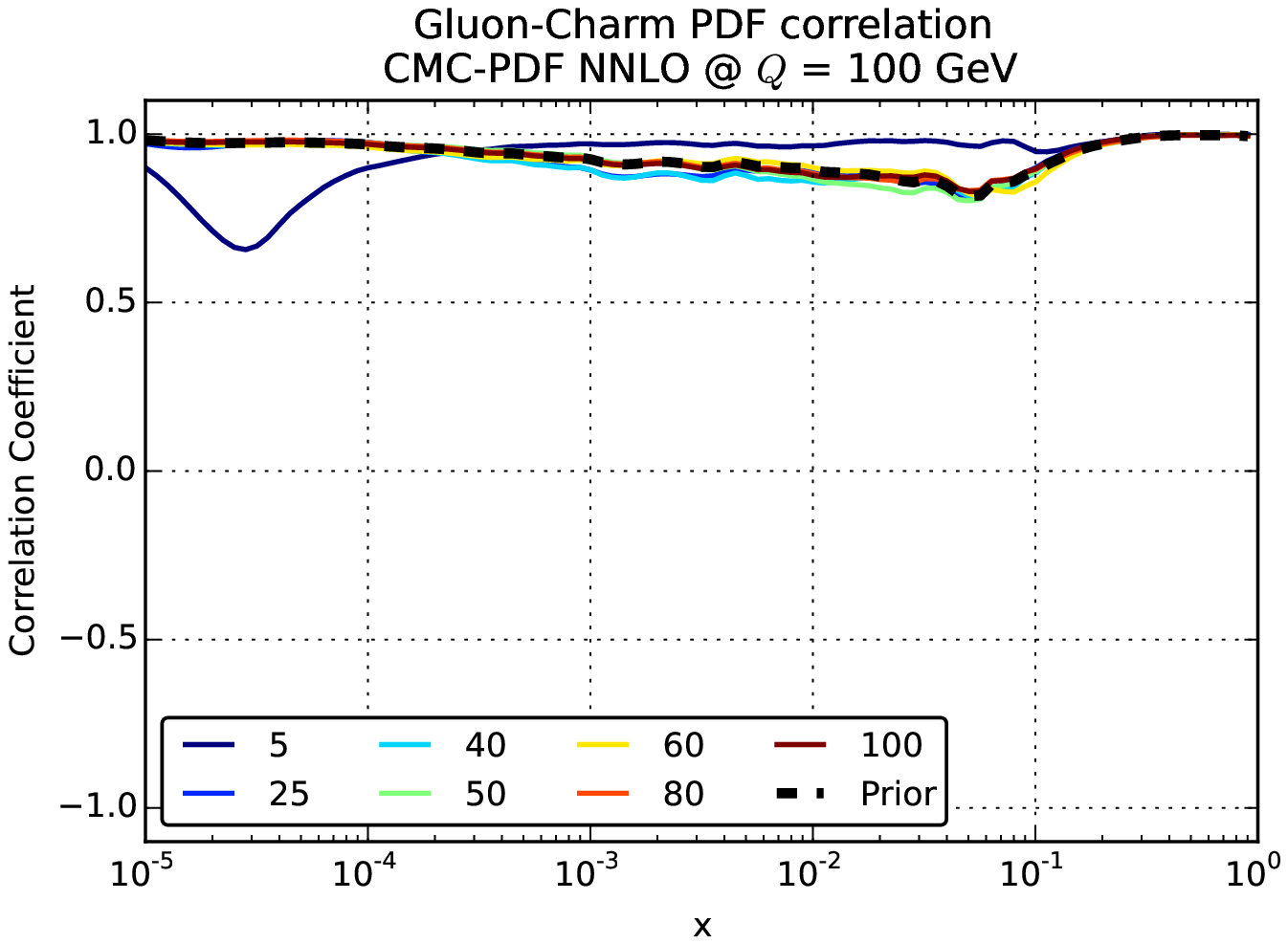}
  \includegraphics[scale=0.52]{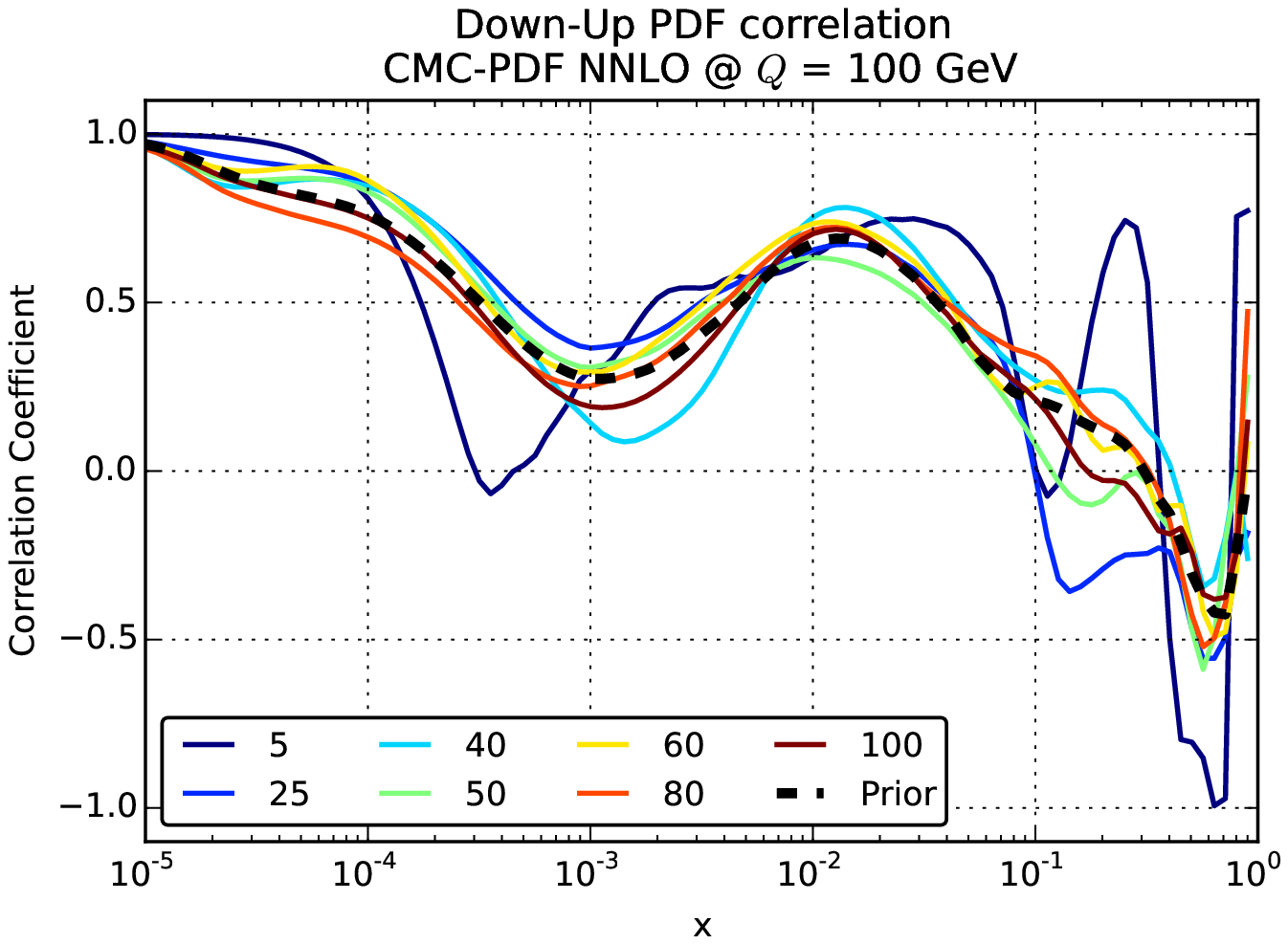}
  \caption{\small Same as Fig.~\ref{valnnpdf4}
    for correlation coefficients of the CMC-PDFs,
    evaluated at  $Q=$100 GeV, for a range of values of
    $N_{\rm rep}$ in the compressed set, from 5 to 100 replicas,
    compared with the prior MC900 result.
    From left to right and from top to bottom
    we show the correlation between gluon and up quark, between
    up and strange quarks, between gluon and charm quark, and between
    the down and up quarks.
  }
  \label{fig:cmcpdf-correlations2}
\end{figure}

Having verified in a number of ways that central values
and variances of the PDFs are successfully preserved by the
compression, we turn to study PDF correlations.
We have verified that a similar
level of agreement as in the case of the native MC sets,
Fig.~\ref{valnnpdf4}, is achieved also here.
To illustrate this point, in Fig.~\ref{fig:cmcpdf-correlations2} we
show a comparison of the correlation coefficients as a function of
$x$, for $Q=100$ GeV, for different PDF combinations, between the original
CMC-PDF set with
$\widetilde{N}_{\rm rep}=900$ replicas and
the compressed sets for different values of $N_{\rm rep}$.
From left to right and from top to bottom
we show the correlation between gluon and up quark, between
up and strange quarks, between gluon and charm quark, and between
the down and up quarks.
We see that already with $N_{\rm rep}=100$ replicas the result
for the correlation is close enough to the prior with
$\widetilde{N}_{\rm rep}=900$ replicas.
%

\begin{figure}[h]
  \centering 
  \includegraphics[scale=0.7]{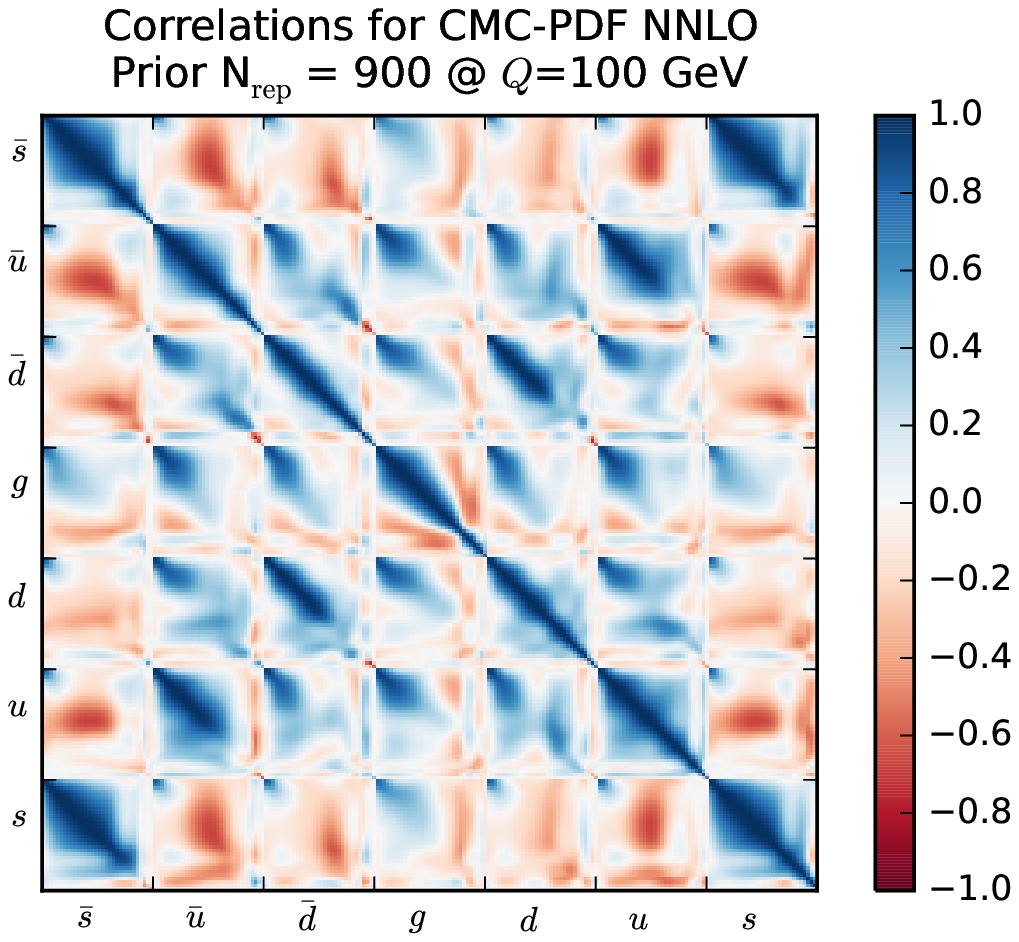}
  \includegraphics[scale=0.7]{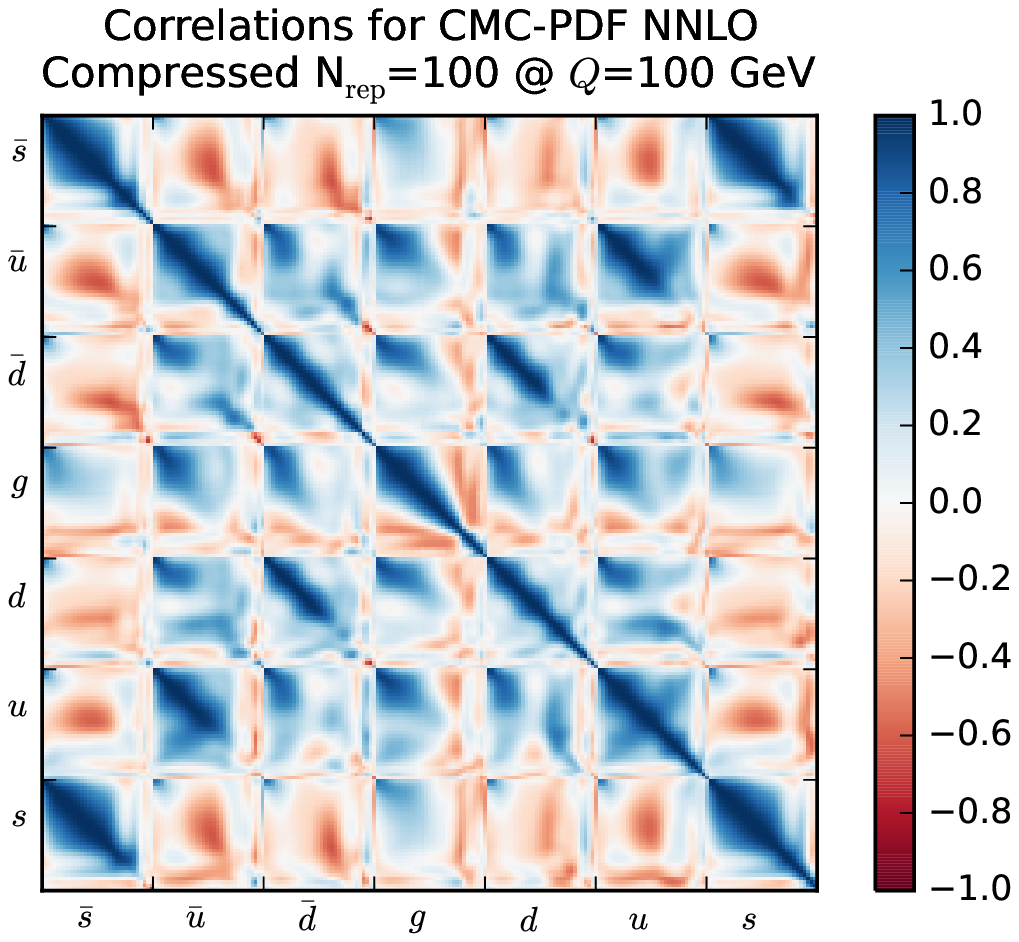}
  \includegraphics[scale=0.7]{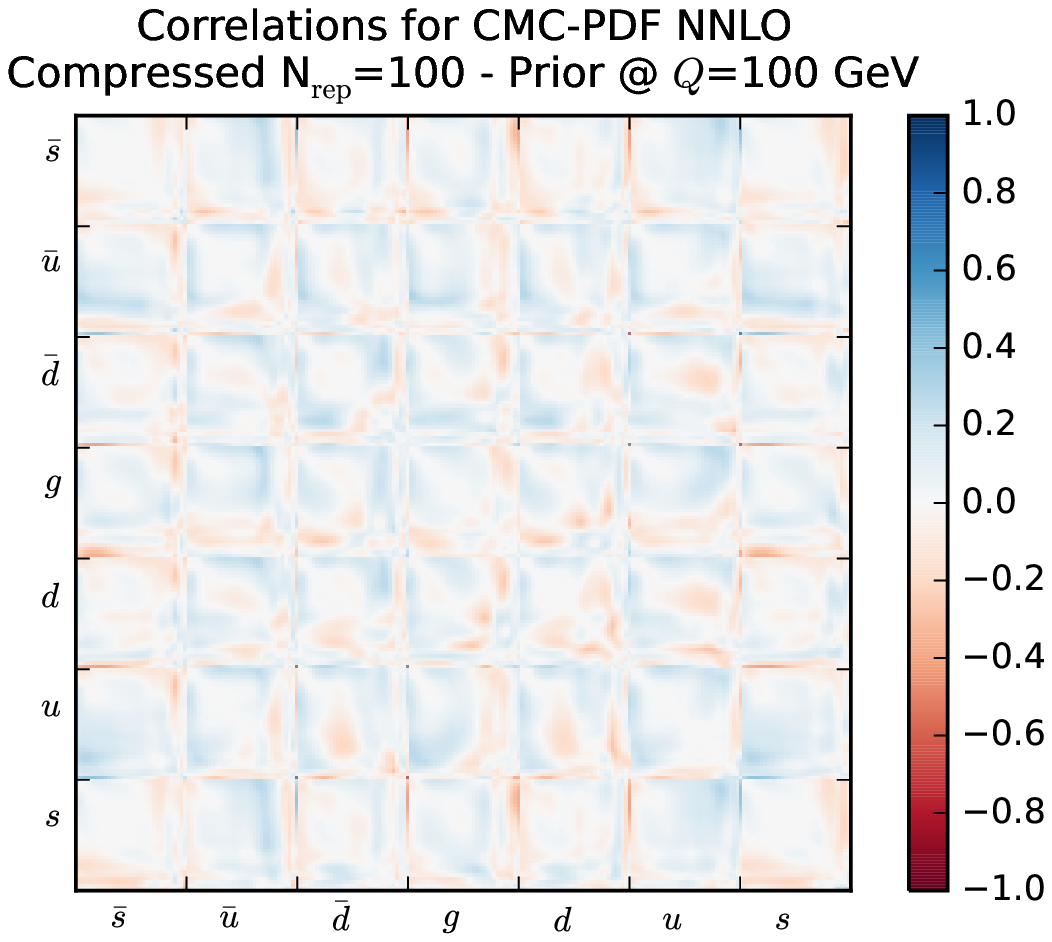}
  \caption{\small Same as Fig.~\ref{fig:pycorrelation} for the correlation
    matrix of the
    CMC-PDFs at $Q=100$ GeV, comparing the prior combination
    MC900  (left plot)
    and the CMC-PDF100 set (right plot).
    In the bottom plot we show the difference between the correlation
    coefficients in the two cases.
  }
  \label{fig:cmcpdf-correlationsmatrix}
  \label{fig:pycorrelationCMC}
\end{figure}

The analogous version of Fig.~\ref{fig:pycorrelation} for the correlation
matrix of the CMC-PDFs is shown in Fig.~\ref{fig:pycorrelationCMC}.
As in the case of the native MC sets, also for the CMC-PDFs
the broad pattern of the correlation matrix of the
original combination
with $\widetilde{N}_{\rm rep}=900$ replicas is maintained
by the compression to $N_{\rm rep}=100$ replicas,
as is quantified by the bottom plot, representing the differences
between the correlation coefficients in the two cases.

\clearpage

\section{CMC-PDFs and LHC phenomenology}
\label{sec:lhcpheno}

Now we present the validation of the compression algorithm 
applied to the combination of Monte Carlo PDF sets for a
variety of LHC cross-sections.
We will compare the results of the original combined
Monte Carlo set MC900
with those
of the CMC-PDFs with $N_{\rm rep}=100$ replicas (CMC-PDF100).
This validation has been performed both at the level
of inclusive cross-sections and of differential
distributions with realistic kinematical cuts.
All cross-sections will be computed with the NNLO sets,
even when the hard cross-sections are computed at NLO,
which is suitable for the present illustration purposes.

First of all, we  compare the MC900 prior and the CMC-PDFs for
benchmark inclusive LHC cross-sections, and then we perform
the validation for LHC differential distributions including
realistic kinematical cuts.
In the latter case we use fast NLO interfaces for
the calculation of these LHC observables: this allows us to straightforwardly
repeat the validation when different PDF sets are used for
the compression without the need to repeat any calculation.
Finally, we verify that the correlations between physical observables
are also maintained by the compression algorithm, both
for inclusive cross-sections and for differential
distributions.

\subsection{LHC cross-sections and differential distributions}

\begin{figure}[t]
\centering 
\includegraphics[scale=0.70]{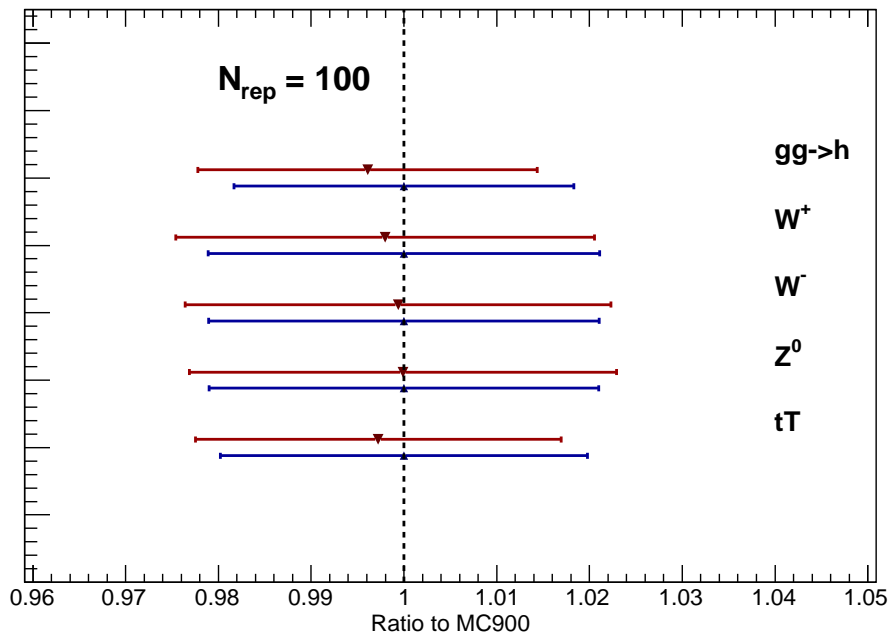}
\caption{\small Comparison of the predictions of the
  Monte Carlo combined prior MC900
   with those of the CMC-PDFs
  with  $N_{\rm rep}=100$ replicas, normalized to
  the central value of the former, for a number
  of benchmark inclusive  NNLO cross-sections at the
  LHC with $\sqrt{s}=13$ TeV.
  The error bands correspond to the PDF uncertainty bands for each of the sets.
  See text for more details.
}
\label{fig:inclnnlo}
\end{figure}

We begin with the validation of the CMC-PDF predictions at the level
of inclusive cross-sections.
The following results have been computed for the LHC at a centre-of-mass
energy of 13 TeV.
In Fig.~\ref{fig:inclnnlo} we compare
the results obtained with the prior Monte Carlo combined set
and with the CMC-PDFs with $N_{\rm rep}=100$
replicas, everything normalized to the central value
of the prior set.
The processes that have been included in Fig.~\ref{fig:inclnnlo}
are the same as those considered in the benchmark comparisons of Sect.~\ref{sec:bench}.
As we can see from Fig.~\ref{fig:inclnnlo}, in all cases the agreement
at the central value level is always at the permille level, and
also the size of the PDF uncertainties is very similar between the original
and compressed set.
Taking into account the fluctuations of the PDF uncertainty itself,
shown in Fig.~\ref{fig:cmcpdf-validation2}, it is clear that the predictions from
the original and the compressed sets are statistically equivalent.

Having established that the compression works for total cross-sections,
one might question if perhaps
the accuracy degrades when we move to differential distributions,
especially if one considers extreme regions of the phase space and
the effects of 
realistic final state kinematical cuts.
To verify that this is not the case, now we consider a number of
differential processes computed using
{\sc\small MCFM}~\cite{Campbell:2000bg} and
{\sc\small NLOjet++}~\cite{Nagy:2003tz}
interfaced to {\sc\small APPLgrid}~\cite{Carli:2010rw} as
well as {\sc\small MadGraph5\_aMC@NLO}~\cite{Alwall:2014hca}
interfaced to {\sc\small aMCfast}~\cite{amcfast} and
{\sc\small APPLgrid}.
All processes are computed for $\sqrt{s}=7$ TeV,
and the matrix-element
calculations have been performed
at fixed NLO perturbative order.
The advantage of using fast NLO grids is that it is straightforward to repeat
the validation without having to redo the full NLO computation
when a  different set
of input PDFs is used for the combination.
Note that while for simplicity we only show the results for
selected bins, we have verified that the 
agreement also holds for the complete differential distribution.
%

\begin{figure}[t]
\centering 
\includegraphics[scale=0.70]{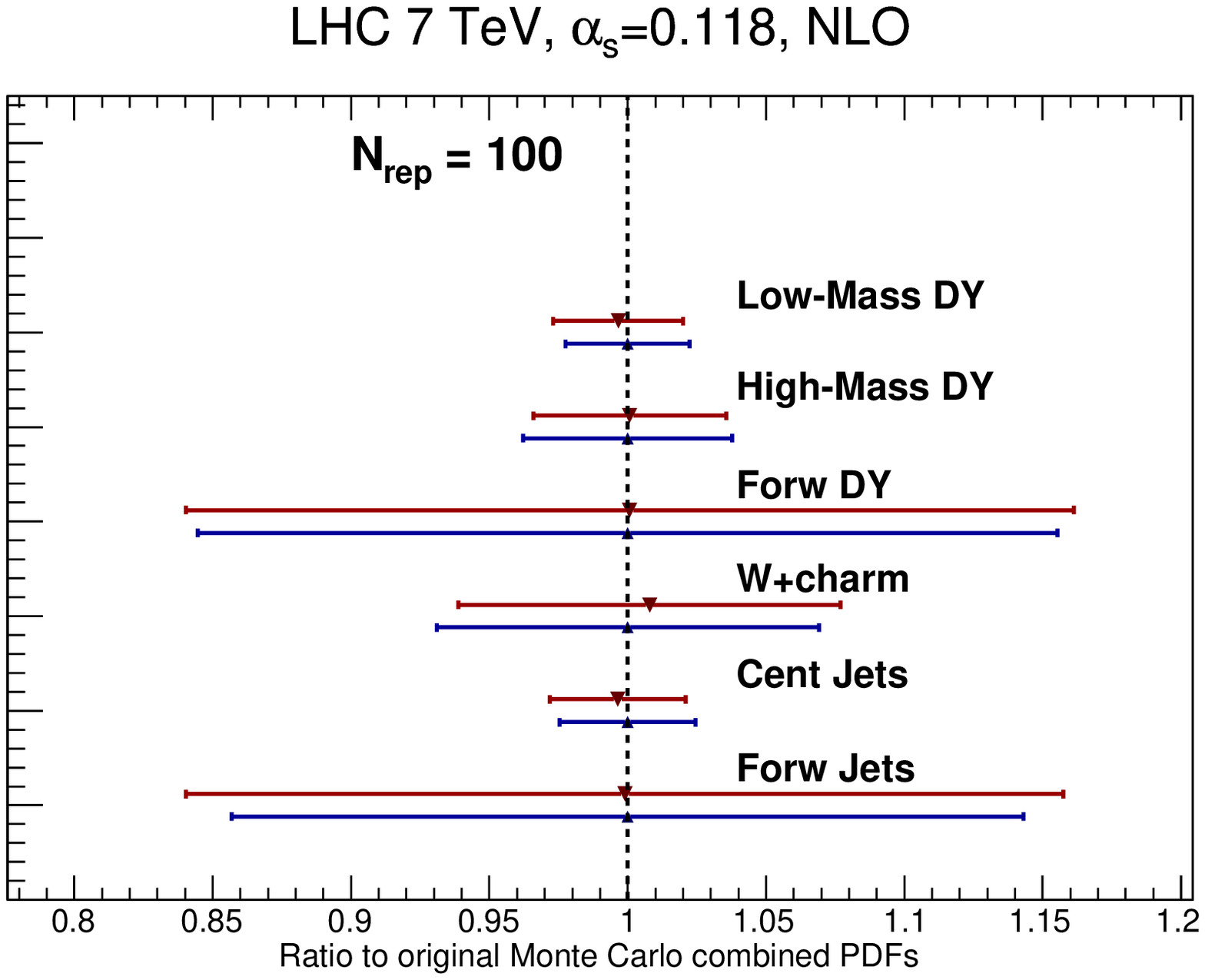}
\caption{\small Same as Fig.~\ref{fig:inclnnlo},
 for a variety of
NLO differential distributions computed with {\sc\small MCFM}
and {\sc\small NLOjet++} interfaced to {\sc\small APPLgrid}
for the LHC with $\sqrt{s}=7$ TeV.
See text for the details of the choice of binning in each process.}
\label{fig:diffnlo}
\end{figure}

The corresponding version of Fig.~\ref{fig:inclnnlo} for the case of LHC 7 TeV
differential distributions is shown in Fig.~\ref{fig:diffnlo}.
The theoretical calculations are provided for the following
processes:
\begin{itemize}
\item The ATLAS high-mass Drell-Yan measurement~\cite{Aad:2013iua},
  integrated over rapidity $|y_{ll}|\le 2.1$, and binned as a function
  of the di-lepton invariant mass pair $M_{ll}$.
  Here we show the prediction for the highest mass bin, $M_{ll}\in \lc 1.0,1.5\rc$ TeV.
\item The CMS double differential Drell-Yan measurement~\cite{CMSDY}
  in the low-mass region, $20~{\rm GeV} \le M_{ll} \le 30$ GeV, as a function of
  the di-lepton rapidity $y_{ll}$. The prediction is shown for the lowest rapidity bin,
  $y_{ll}\in \lc 0.0,0.1\rc$.
\item The CMS $W^+$ lepton rapidity
  distribution~\cite{Chatrchyan:2013mza}.
  The prediction is shown for the the  lowest rapidity bin,
  $y_{l}\in \lc 0.0,0.1\rc$.
\item The CMS measurement of $W^+$ production in association with charm
  quarks~\cite{Chatrchyan:2013uja}, as a function of the lepton
  rapidity $y_l$.
  The prediction is shown for the lowest rapidity bin,
  $y_{l}\in \lc 0.0,0.3\rc$.
\item The ATLAS inclusive jet production measurement~\cite{Aad:2011fc}
  in the central rapidity region, $|y_{\rm jet}|\le 0.3$, as a
  function of the jet $p_T$.
   The prediction is shown for the lowest $p_T$  bin,
  $p_{T}\in \lc 20,30\rc$ GeV.
\item The same ATLAS inclusive jet production
  measurement~\cite{Aad:2011fc} now in the forward rapidity region,
  $3.6 \le |y_{\rm jet}|\le 4.4$, as a function of the jet $p_T$.
 The prediction is shown for the highest $p_T$  bin,
  $p_{T}\in \lc 110,160\rc$ GeV.
\end{itemize}
More details about the selection cuts applied to these
processes can be found in the original
references and in the NNPDF3.0 paper~\cite{Ball:2014uwa}, though note
that here no comparison with experimental data is attempted.
The various observables of Fig.~\ref{fig:diffnlo}
probe a wide range of PDF combinations, from light
quarks and anti-quarks (low and high-mass Drell-Yan) and strangeness
($W$+charm) to the gluon (central and forward jets) in a wide range of
Bjorken-$x$ and momentum transfers $Q^2$.

As we can see from Fig.~\ref{fig:diffnlo}, the level of
the agreement between the MC900 prior
and the CMC-PDFs with $N_{\rm rep}=100$ is similar to that of the inclusive
cross-sections.
This is also true for other related processes that we have also
studied, but that are not shown explicitly here.
This agreement is of course understood from the fact that the compression is performed
at the level of parton distributions, as shown in Sect.~\ref{sec:results}.
Note also that the agreement found for the processes in Fig.~\ref{fig:diffnlo}
is particularly remarkable since in some cases, like forward Drell-Yan or
forward jet production, the underlying PDFs are probed at large-$x$, where
deviations from the Gaussian behavior are sizable:
 even in this case, the compression algorithm is successful in reproducing the
 mean and variance of the prior probability distribution.
 
Another illustrative way of checking that the compression algorithm really preserves
the non-Gaussian features of the prior is provided by the probability distribution
of specific LHC cross-sections in which such features are clearly observed.
To better visualize the probability density $P(\sigma)$ estimated from the Monte Carlo
sample we use the \emph{Kernel Density Estimation} (KDE) method.
In this technique,
the probability distribution is obtained by averaging a kernel
function $K$ centered at the predictions $\{\sigma_{i}\}$
obtained for each individual PDF replica:
\begin{equation}
P(\sigma)=\frac{1}{N_{{\rm rep}}}\sum_{i=1}^{N_{{\rm rep}}}K\left(\sigma-\sigma_{i}\right)\, .\label{eq:KDE}
\end{equation}
Here we choose the function $K$ to be a normal distribution, that is
\begin{equation}
K(\sigma-\sigma_{i})=\frac{1}{h\sqrt{2\pi}}e^{\frac{-(\sigma-\sigma_{i})^{2}}{h}}\,,\label{eq:kdenormal}
\end{equation}
where we set the parameter $h$, known as bandwidth, so that it is the
optimal choice if the underlying data was Gaussian.
This choice is known as the
Silverman rule.\footnote{It can be shown that this choice amounts
  to using a bandwidth of
  \begin{equation}
h=\left(\frac{4s^{5}}{3N_{{\rm rep}}}\right)^{\frac{1}{5}}\,,\label{eq:kdebw}
  \end{equation}
  where $s$ is the standard deviation of the sample.
  }

\begin{figure}[t]
\centering 
\includegraphics[scale=0.82]{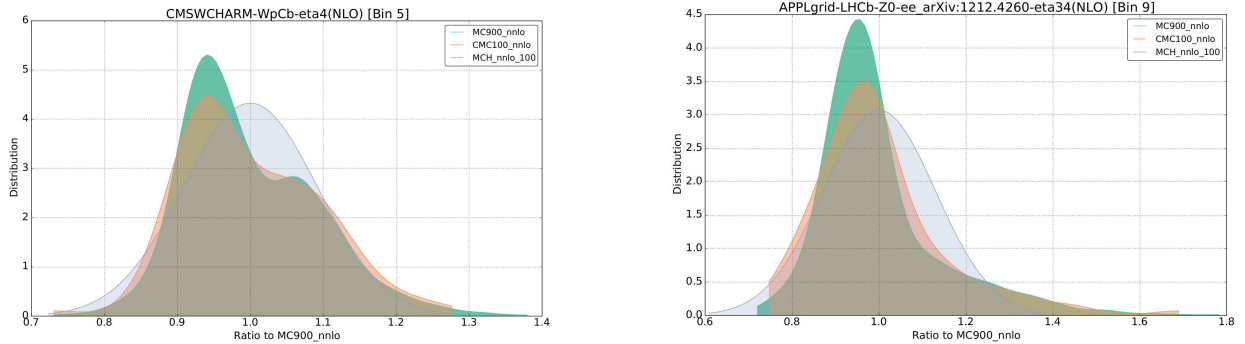}
\caption{\small The probability distribution for two LHC cross-sections: the
  CMS $W$+charm production in the most forward bin (left plot) and the
  LHCb $Z\to e^+e^-$ rapidity distribution for $\eta_{Z}= 4$ (right plot).
  We compare the original prior MC900 with the results
  from the CMC-PDF100 and MCH100 reduced sets.
}
\label{fig:nongaussian}
\end{figure}

In Fig.~\ref{fig:nongaussian} we compare the
probability distributions, obtained using the KDE method,
for two LHC cross-sections: the
  CMS $W$+charm production in the most forward bin (left plot) and the
  LHCb $Z\to e^+e^-$ rapidity distribution for $\eta_{Z}= 4$ (right plot).
  We compare the original prior MC900 with the CMC-PDF100 and MCH100 reduced sets.
  In the case of the $W$+charm cross-section, which is directly sensitive to the poorly-known
  strange PDF, the prior shows a double-hump structure, which is reasonably well
  reproduced by the CMC-PDF100 set, but that disappears if a Gaussian reduction,
  in this case MCH100, is used.
  For the LHCb forward $Z$ production, both the prior and CMC-PDF100 are significantly skewed,
  a feature which is lost in the gaussian reduction of MCH100.

\subsection{Correlations between LHC cross-sections}
\label{sec-corrxsec}

Any reasonable algorithm for the combination of PDF sets
should reproduce not only the central values and the variances of the
prior distribution,
but also the correlations between physical observables.
This is relevant for phenomenological applications at the
LHC, where PDF-induced correlations are used for instance to determine the
degree of correlation of the systematic uncertainties between
different processes.
Using the PDF4LHC recommendations, the
PDF-induced correlations between different Higgs production channels
were estimated in Ref.~\cite{Dittmaier:2012vm},
and this information is now extensively used
in the Higgs analyses of ATLAS and CMS.

\begin{figure}[t]
\centering 
\includegraphics[scale=0.38]{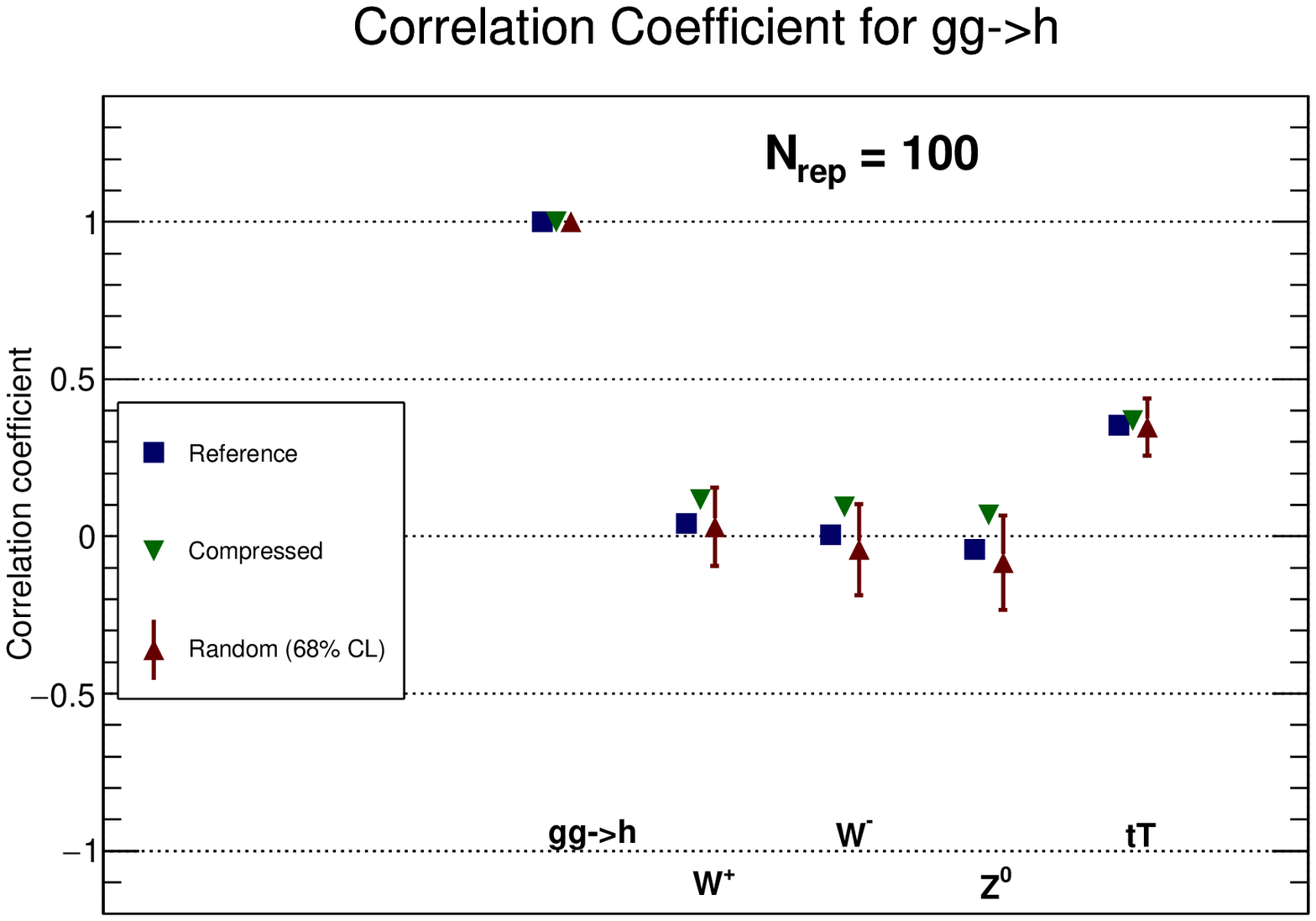}
\includegraphics[scale=0.38]{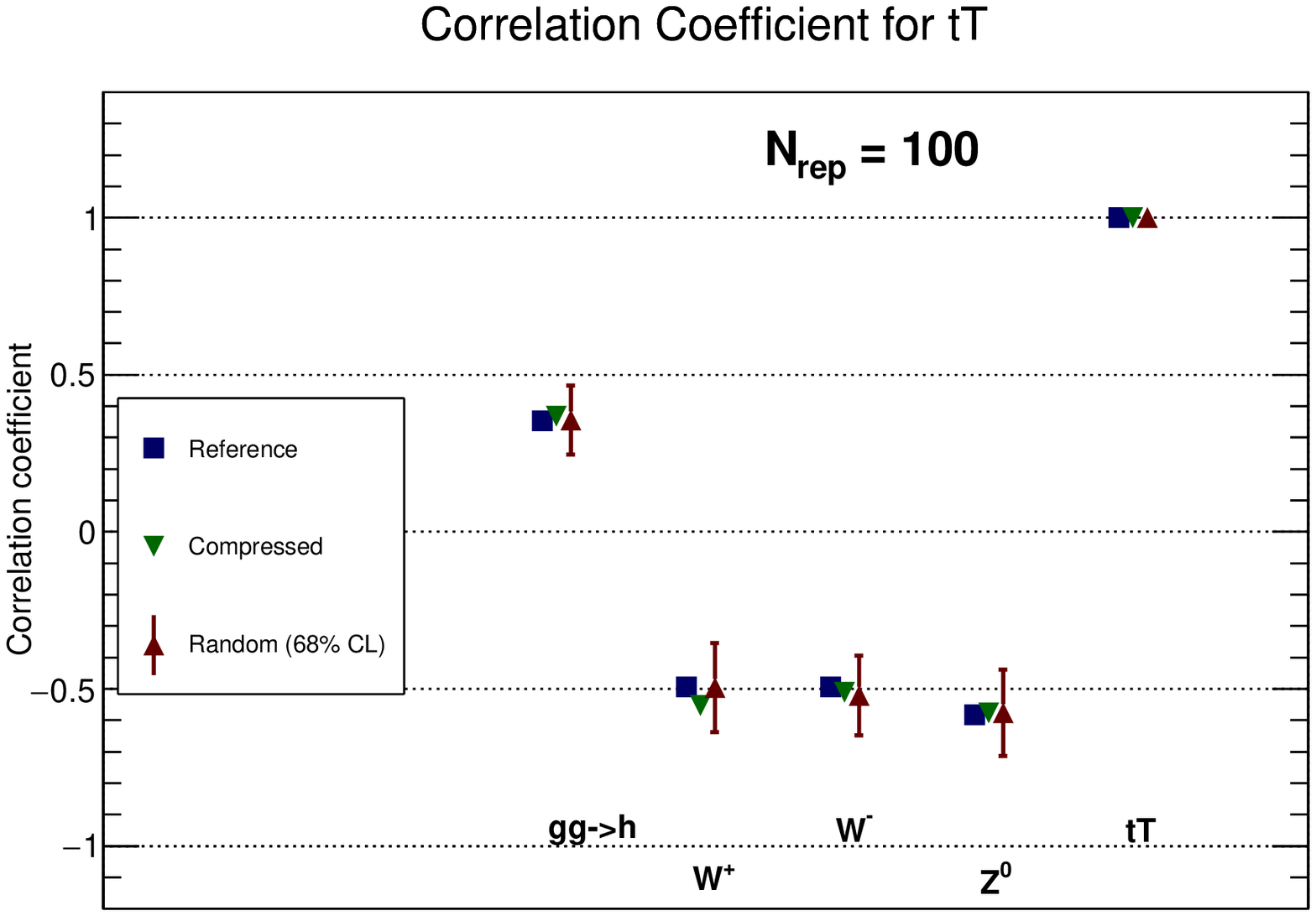}
\includegraphics[scale=0.38]{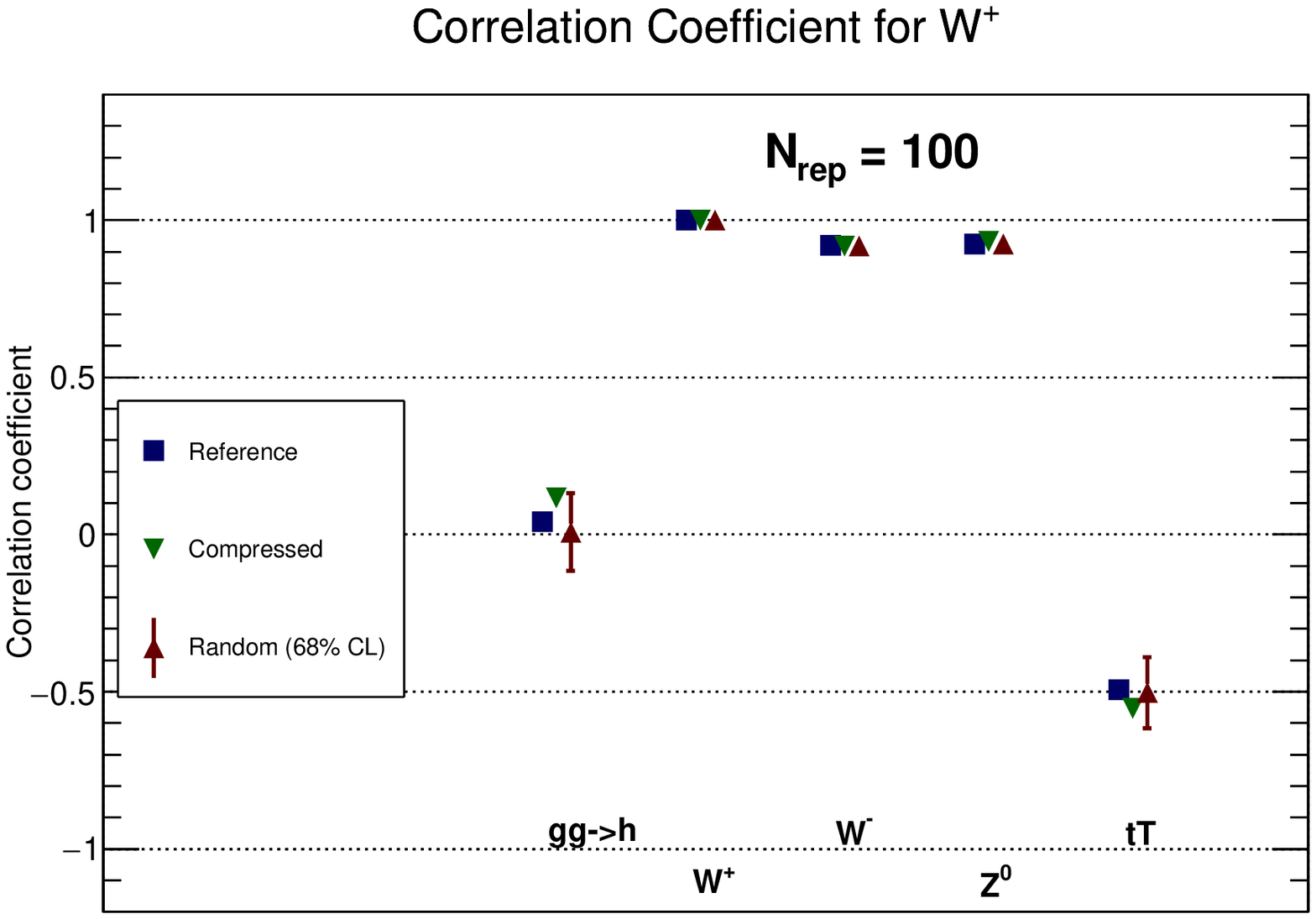}
\includegraphics[scale=0.38]{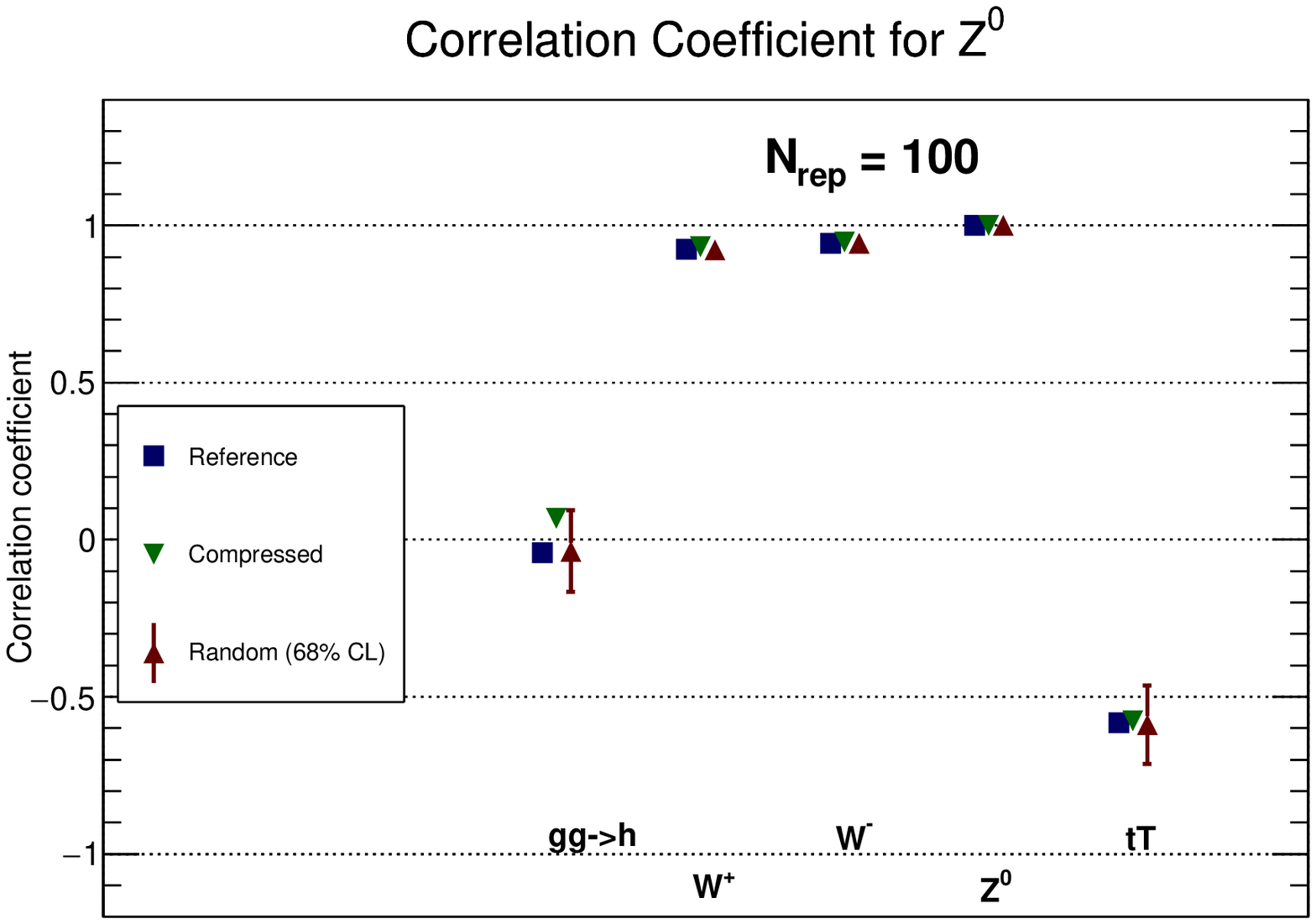}
\caption{\small Comparison of the correlation coefficients computed
  from the reference Monte Carlo combined set and from the
  CMC-PDFs with $N_{\rm rep}=100$ replicas.
  We show here the results for the correlations
  between the inclusive LHC cross-sections, using the settings
  described in the text.
  Each plot contains the correlation coefficient of a given
  cross-section
  with respect to all the other inclusive cross-sections considered
  here.
  To gauge the effectiveness of the
  compression algorithm, we also show the 68\% confidence-level
  interval for the correlation coefficients computed
  from $N_{\rm rand}=1000$ random partitions of $N_{\rm rep}=100$
  replicas each.}
\label{fig:correlations1}
\end{figure}

\begin{figure}[t]
  \centering
\includegraphics[scale=0.38]{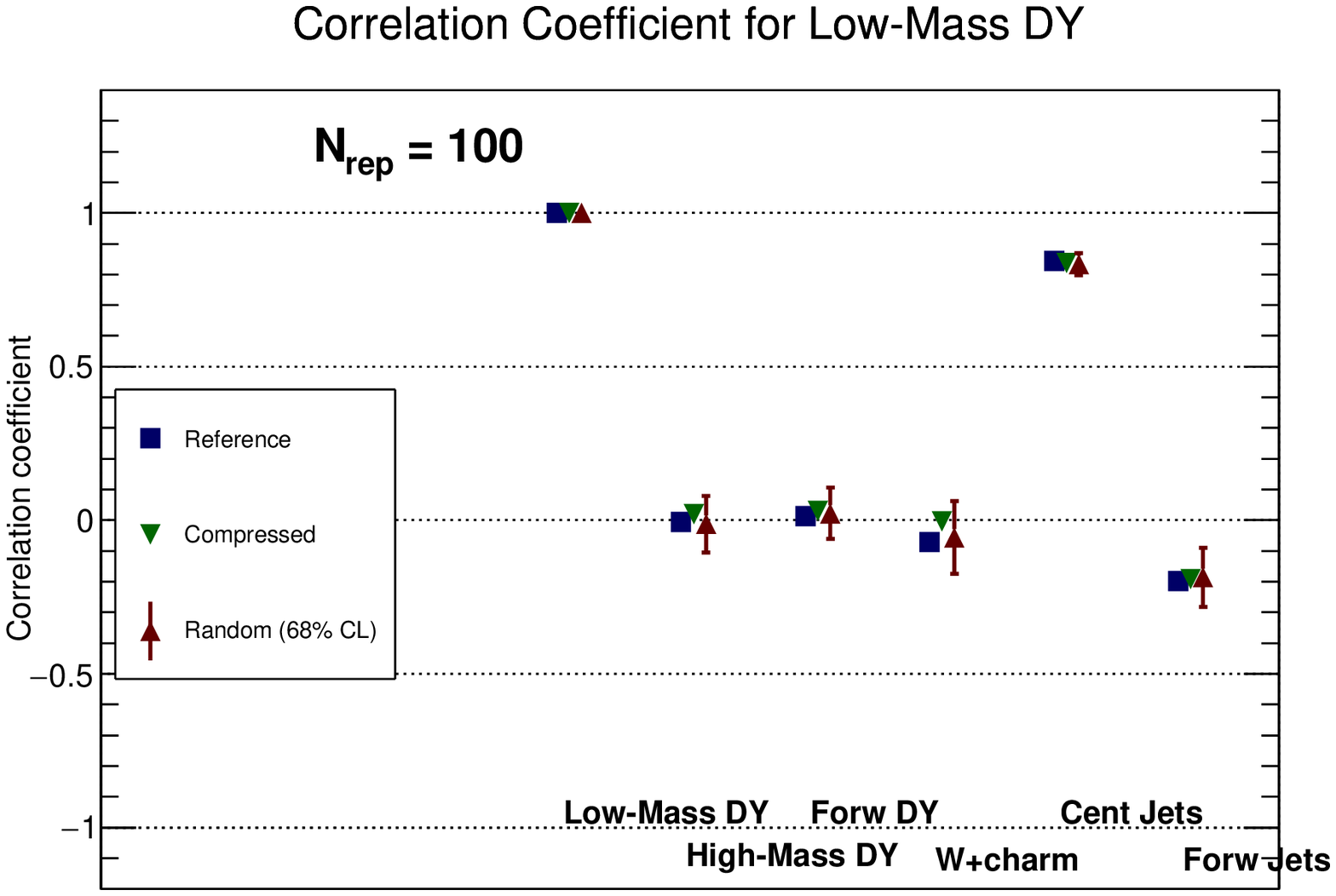}
\includegraphics[scale=0.38]{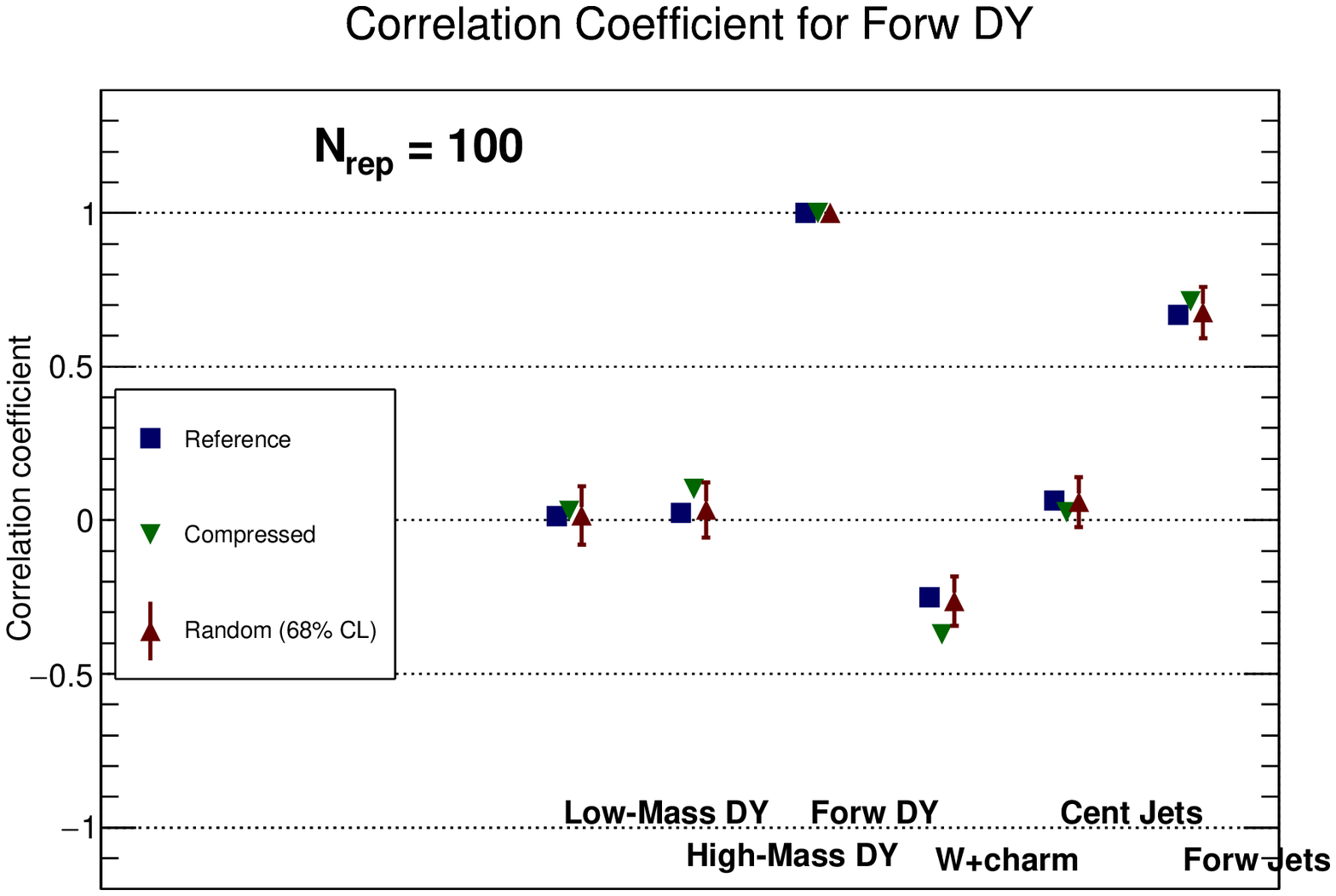}
\includegraphics[scale=0.38]{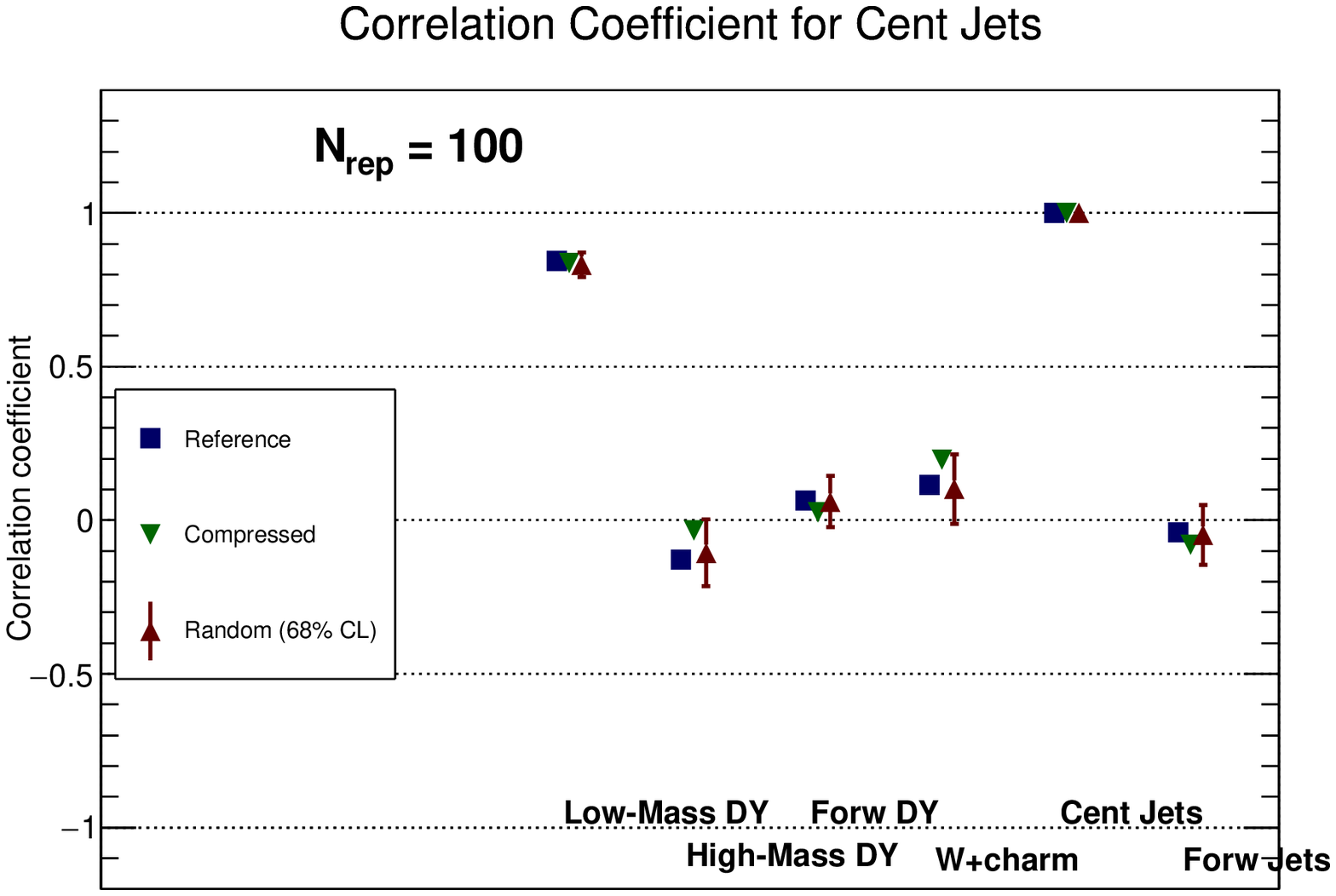}
\includegraphics[scale=0.38]{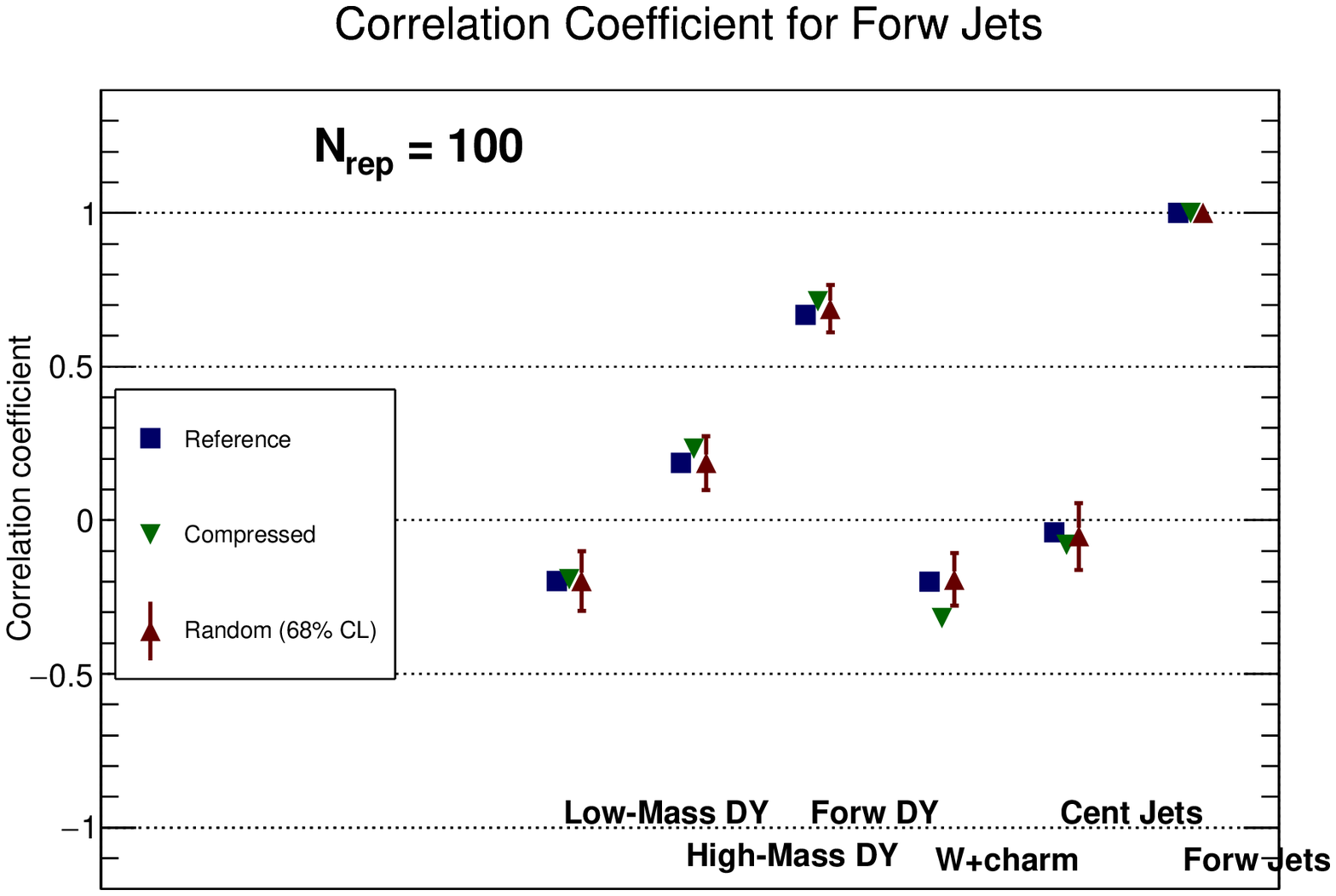}
\caption{\small Same as Fig.~\ref{fig:correlations1} for
  some of the various LHC
  NLO differential cross-sections discussed in the text.
  From top to bottom and from left to right we show the correlations
  for low-mass Drell-Yan, forward Drell-Yan, and central 
  and forward jets.
}
\label{fig:correlations2}
\end{figure}

To validate that the compression algorithm presented here
also maintains the correlations of the original set,
we have computed the correlations between all processes
used in the previous section, both for the MC900 prior
and for the CMC-PDF100 set.
The results are shown in Fig.~\ref{fig:correlations1}, for
the NLO and NNLO inclusive cross-sections shown in Fig.~\ref{fig:inclnnlo}, and in
Fig.~\ref{fig:correlations2}, for the case of
differential distributions shown in Fig.~\ref{fig:diffnlo}.
We have also verified that from $N_{\rm rep}\simeq 50$
replicas onwards the correlations are very well reproduced
by the compressed set.

To gauge the effectiveness of the compression algorithm,
in Figs.~\ref{fig:correlations1} and~\ref{fig:correlations2}
we also show the 68\% confidence-level
  interval for the correlation coefficients computed
  from $N_{\rm rand}=1000$ random partitions of $N_{\rm rep}=100$
  replicas: we see the compression in general outperforms the results
  from a random selection of a $N_{\rm rep}=100$ replica set.
  The agreement of the correlations at the level of LHC observables
  is a direct consequence of course that correlations are maintained by
  the compression at the PDF level, as discussed in detail in
  Sect.~\ref{sec:resultsCMCPDFs}.
  Only for very few cases the correlation coefficient of the CMC-PDF set
  is outside the 68\% confidence-level range of the random selections, and this
  happens only when correlations are very small to begin with,
  so this fact is not relevant for phenomenology.

To summarize, the results of this section show that at the level of
LHC phenomenology, CMC-PDFs with $N_{\rm rep}=100$ replicas can be
reliably used instead of the original Monte Carlo
combination of PDF sets, thereby allowing a substantial reduction of the
CPU-time burden associated with the calculation of the theory
predictions for the original $\widetilde{N}_{\rm rep}=900$ replicas by almost
a full order of magnitude.


\section{Summary and delivery}
\label{sec:conclusions}

In this work we have presented a novel strategy for the
combination of individual PDF sets, based on the Monte Carlo method
followed by a compression algorithm.
The resulting Compressed Monte Carlo PDFs, or CMC-PDFs for short,
are suitable to be used to estimate
PDF uncertainties in theoretical predictions of generic LHC
processes.
As compared to the original PDF4LHC
recommendation, the new approach
we advocate here is both more straightforward
to use, based on a single combined PDF set,
and less computationally expensive:
$N_{\rm rep}\simeq 100$ replicas are enough to preserve the statistical features
of the prior combination with sufficient accuracy for
most relevant applications. 
Using as an illustration the combination of the recent NNPDF3.0, CT14 and
MMHT14 NNLO sets,
we have verified that the compression algorithm successfully
reproduces the predictions of the prior combined MC set for a wide
variety of LHC processes and their correlations.

The compressed PDF sets at NLO and NNLO, with $N_{\rm rep}=100$ replicas
each, and $\alpha_s(M_Z)=0.118$,
will be made available in {\sc\small LHAPDF6}~\cite{Buckley:2014ana}
as part of the upcoming PDF4LHC 2015 recommendations.
Additional members to estimate the combined PDF+$\alpha_s$ uncertainty
will also be included in the same grid files,
 and
 new functions will be provided in {\sc\small LHAPDF} 6.1.6 to facilitate
 the computation of this combined PDF+$\alpha_s$ uncertainty.
In addition, we have also made publicly available the compression
algorithm used in this work:
\begin{center}
\url{https://github.com/scarrazza/compressor}
\end{center}
This {\tt compressor} code~\cite{compressor} includes a script to
combine Monte Carlo sets from different groups into a single
MC set,
the compression algorithm and the validation suite.
A concise user manual for this code can be found in Appendix~\ref{sec:compressioncode}: the  code produces CMC-PDF sets
directly in the {\sc\small LHAPDF6} format ready to be used
for
phenomenological applications.

We would like to emphasize that it is beyond the scope of this paper
to determine which specific PDF sets should be used in the present
or future PDF4LHC
combination: this is an issue in which only the PDF4LHC Steering Committee
has the mandate to decide.
We have used the most updated NNLO sets from NNPDF, CT and MMHT for consistency
with the current prescription, but using the publicly available code
it is possible to construct CMC-PDFs from any other choice
of sets.
We note however that for the combination of PDF sets that are based
on very different input datasets or theory assumptions
as compared to the three global sets, the determination of the
number of replicas from each set that should be included in the combination
is a complex problem which still requires to be understood.

Examples of applications where combined PDF sets,
as implemented by the CMC-PDFs, should be used
include the computation of PDF uncertainties for
acceptances and efficiencies, due to extrapolations or
interpolations, to estimate the PDF uncertainties
in the extraction of
Higgs couplings or other fundamental SM parameters such as
$M_W$ from LHC data, and to obtain limits
in  searches for BSM physics.
Even in these cases, whenever possible, 
providing results obtained using individual
PDF sets should be encouraged, since such comparisons shed
light on the origin of the total PDF uncertainties for each particular
application, and provide guidance about how they might reduce this
PDF uncertainty.
Needless to say, in all PDF-sensitive
Standard Model comparisons between experimental
data and theory models,
only the individual PDF sets should be used, rather than only a combined
PDF set.
The latter might be suitable only if PDF uncertainties are much
smaller than all other theoretical and experimental uncertainties.

It is also important to emphasize that the CMC-PDFs,
as well as any other method
for the 
combination of PDF sets, do not replace
the individual PDF sets: CMC-PDFs are simply a user-convenient
method to easily obtain the results of the combination of the individual
PDF sets.
For this reason, it should be clear that
whenever the CMC-PDF sets are used,
not only the present publication should be cited, but also the original
publications corresponding to the individual PDF sets
used as input to the combination.

Let us conclude by stating the obvious fact that
the availability of a method for the combination
of different sets does not reduce, but if anything  strengthens, the need
to keep working in reducing the PDF uncertainties in the individual
sets, both in terms of improved theory, more constraining data
and refined methodology, 
as well as to continue the benchmarking exercises between groups
that have been performed in the
past~\cite{Butterworth:2014efa,Ball:2012wy,LHhq} and that are instrumental to
understand (and
eventually reduce) the 
differences between different groups.

\subsection*{Acknowledgments}
We are grateful to Joey Huston, Jun Gao and Pavel Nadolsky for many illuminating discussions
about the comparison between the CMC-PDF and the Meta-PDF approaches.
We also are grateful to all the members of the NNPDF Collaboration for
fruitful discussions during the development of this project.
This work of  S.~C. is partially supported by an Italian
PRIN2010 grant  and by a European Investment Bank EIBURS
grant.
The work of J.~I.~L. is funded by a FIS2010-16185 grant
from the Spanish MINNECO.
The work of J.~R. is supported by an
STFC Rutherford Fellowship ST/K005227/1 and by the European Research Council Starting Grant ``PDF4BSM''.

\appendix
\section{The compression code}
\label{sec:compressioncode}

The numerical implementation of the compression algorithm used in this
work is available in the {\tt compressor v1.0.0} package. Here we
provide instructions about how to download, install and run the
code. The code was designed for Unix systems.

\subsection*{Download}

Open a terminal and download the latest release available from
\begin{center}
  {\bf \url{https://github.com/scarrazza/compressor/releases}~}
\end{center}
or clone the master development branch from the {\tt GitHub}
repository:
\begin{lstlisting}
  $ git clone https://github.com/scarrazza/compressor.git
\end{lstlisting}

\subsection*{Installation}
Compressor requires three external public libraries in order to work
properly:
{\tt LHAPDF6}\footnote{\url{http://lhapdf.hepforge.org/}}~\cite{Buckley:2014ana},
{\tt ROOT}\footnote{\url{http://root.cern.ch/}} and
{\tt GSL}\footnote{\url{http://www.gnu.org/software/gsl/}}. In order to
install the package compile with the configure script:
\begin{lstlisting}
  $ cd compressor
  $ ./configure --prefix=/path/to/install/location
  $ make && make install
\end{lstlisting}
This operation will copy the {\tt bin/compressor} binary to {\tt /usr/local/bin} (or the location given by {\tt --prefix}).

\subsection*{Running the code}

After installing this package, the {\tt compressor} program is
available for the user.

\begin{lstlisting}
  $ compressor --help
  usage: ./compressor [REP] [PDF prior name] [energy Q=1] [seed=0] [compress=1]
\end{lstlisting}
The first two arguments are required:
\begin{itemize}
\item {\tt REP}: the number of required compressed replicas
\item {\tt PDF prior name}: the name of the prior {\tt LHAPDF6} grid
\item {\tt energy} $Q$: the input energy scale used by the compression algorithm ({\tt default} = 1 GeV)
\item {\tt seed}: the random number seed ({\tt default} = 0)
\item {\tt compress}: switches on/off the minimization step ({\tt default} = true)
\end{itemize}

\subsection*{Output}

After running {\tt compressor} a folder with the prior set name is created.

\begin{lstlisting}
$ compressor 100 MyPriorSet
  ...
$ ls MyPriorSet/
  erf_compression.dat         # contains the erf. values for the compressed set
  erf_random.dat              # contains the erf. values for the random set
  replica_compression_100.dat # list of compressed replicas from the prior
\end{lstlisting}

The script {\tt /bin/compressor\_buildgrid} creates the compressed
{\tt LHAPDF6} grid:
\begin{lstlisting}
 $ ./compressor_buildgrid --help
   usage: ./compressor_buildgrid [prior set name] [number of compressed replicas]
\end{lstlisting}

Finally, in order to generate the ERF plots place the
{\tt /bin/compressor\_validate.C} script in the output folder and run:

\begin{lstlisting}
 $ root -l compressor_validate.C
\end{lstlisting}


\begin{thebibliography}{10}

\bibitem{Forte:2010dt}
S.~Forte, {\it {Parton distributions at the dawn of the LHC}},  {\em Acta
  Phys.Polon.} {\bf B41} (2010) 2859--2920,
  [\href{http://arxiv.org/abs/1011.5247}{{\tt arXiv:1011.5247}}].

\bibitem{Forte:2013wc}
S.~Forte and G.~Watt, {\it {Progress in the Determination of the Partonic
  Structure of the Proton}},  {\em Ann.Rev.Nucl.Part.Sci.} {\bf 63} (2013)
  291--328, [\href{http://arxiv.org/abs/1301.6754}{{\tt arXiv:1301.6754}}].

\bibitem{Perez:2012um}
E.~Perez and E.~Rizvi, {\it {The Quark and Gluon Structure of the Proton}},
  {\em Rep.Prog.Phys.} {\bf 76} (2013) 046201,
  [\href{http://arxiv.org/abs/1208.1178}{{\tt arXiv:1208.1178}}].

\bibitem{Ball:2012wy}
R.~D. Ball, S.~Carrazza, L.~Del~Debbio, S.~Forte, J.~Gao, et~al., {\it {Parton
  Distribution Benchmarking with LHC Data}},  {\em JHEP} {\bf 1304} (2013) 125,
  [\href{http://arxiv.org/abs/1211.5142}{{\tt arXiv:1211.5142}}].

\bibitem{Watt:2011kp}
G.~Watt, {\it {Parton distribution function dependence of benchmark Standard
  Model total cross sections at the 7 TeV LHC}},  {\em JHEP} {\bf 1109} (2011)
  069, [\href{http://arxiv.org/abs/1106.5788}{{\tt arXiv:1106.5788}}].

\bibitem{DeRoeck:2011na}
A.~De~Roeck and R.~S. Thorne, {\it {Structure Functions}},  {\em
  Prog.Part.Nucl.Phys.} {\bf 66} (2011) 727--781,
  [\href{http://arxiv.org/abs/1103.0555}{{\tt arXiv:1103.0555}}].

\bibitem{Rojo:2015acz}
J.~Rojo et~al., {\it {The PDF4LHC report on PDFs and LHC data: Results from Run
  I and preparation for Run II}},  \href{http://arxiv.org/abs/1507.00556}{{\tt
  arXiv:1507.00556}}.

\bibitem{Heinemeyer:2013tqa}
{LHC Higgs Cross Section Working Group}, S.~Heinemeyer, C.~Mariotti,
  G.~Passarino, and R.~Tanaka~(Eds.), {\it {Handbook of LHC Higgs Cross
  Sections: 3. Higgs Properties}},  {\em CERN-2013-004} (CERN, Geneva, 2013)
  [\href{http://arxiv.org/abs/1307.1347}{{\tt arXiv:1307.1347}}].

\bibitem{AbelleiraFernandez:2012cc}
{\bf LHeC Study Group} Collaboration, J.~Abelleira~Fernandez et~al., {\it {A
  Large Hadron Electron Collider at CERN: Report on the Physics and Design
  Concepts for Machine and Detector}},  {\em J.Phys.} {\bf G39} (2012) 075001,
  [\href{http://arxiv.org/abs/1206.2913}{{\tt arXiv:1206.2913}}].

\bibitem{Borschensky:2014cia}
C.~Borschensky, M.~Krämer, A.~Kulesza, M.~Mangano, S.~Padhi, et~al., {\it
  {Squark and gluino production cross sections in pp collisions at $\sqrt{s}$ =
    13, 14, 33 and 100 TeV}},
{\em Eur.\ Phys.\ J.\ C} {\bf 74}, no. 12, 3174 (2014) [\href{http://arxiv.org/abs/1407.5066}{{\tt
  arXiv:1407.5066}}].

\bibitem{Kramer:2012bx}
M.~Kramer, A.~Kulesza, R.~van~der Leeuw, M.~Mangano, S.~Padhi, et~al., {\it
  {Supersymmetry production cross sections in $pp$ collisions at $\sqrt{s}=7$
  TeV}},  \href{http://arxiv.org/abs/1206.2892}{{\tt arXiv:1206.2892}}.

\bibitem{Bozzi:2011ww}
G.~Bozzi, J.~Rojo, and A.~Vicini, {\it {The Impact of PDF uncertainties on the
  measurement of the W boson mass at the Tevatron and the LHC}},  {\em
  Phys.Rev.} {\bf D83} (2011) 113008,
  [\href{http://arxiv.org/abs/1104.2056}{{\tt arXiv:1104.2056}}].

\bibitem{ATL-PHYS-PUB-2014-015}
{\bf ATLAS} Collaboration, G.~Aad et~al., {\it {Studies of theoretical
  uncertainties on the measurement of the mass of the $W$ boson at the LHC}},
  Tech. Rep. ATL-PHYS-PUB-2014-015, CERN, Geneva, Oct, 2014.

\bibitem{Bozzi:2015hha}
G.~Bozzi, L.~Citelli, and A.~Vicini, {\it {PDF uncertainties on the W boson
    mass measurement from the lepton transverse momentum distribution}},
Phys.\ Rev.\ D {\bf 91}, no. 11, 113005 (2015),
 [\href{http://arxiv.org/abs/1501.05587}{{\tt arXiv:1501.05587}}].

\bibitem{CooperSarkar:2011aa}
{\bf H1 and ZEUS} Collaboration, A.~Cooper-Sarkar, {\it {PDF Fits at HERA}},
  {\em PoS} {\bf EPS-HEP2011} (2011) 320,
  [\href{http://arxiv.org/abs/1112.2107}{{\tt arXiv:1112.2107}}].

\bibitem{Ball:2014uwa}
{\bf NNPDF} Collaboration, R.~D. Ball et~al., {\it {Parton distributions for
  the LHC Run II}}, JHEP {\bf 1504}, 040 (2015) [\href{http://arxiv.org/abs/1410.8849}{{\tt
  arXiv:1410.8849}}].

\bibitem{Alekhin:2013nda}
S.~Alekhin, J.~Bluemlein, and S.~Moch, {\it {The ABM parton distributions tuned
  to LHC data}},  {\em Phys.Rev.} {\bf D89} (2014) 054028,
  [\href{http://arxiv.org/abs/1310.3059}{{\tt arXiv:1310.3059}}].

\bibitem{Gao:2013xoa}
J.~Gao, M.~Guzzi, J.~Huston, H.-L. Lai, Z.~Li, et~al., {\it {CT10
  next-to-next-to-leading order global analysis of QCD}},  {\em Phys.Rev.} {\bf
  D89} (2014), no.~3 033009, [\href{http://arxiv.org/abs/1302.6246}{{\tt
  arXiv:1302.6246}}].

\bibitem{Harland-Lang:2014zoa}
L.~A. Harland-Lang, A.~D. Martin, P.~Motylinski, and R.~S. Thorne, {\it {Parton
    distributions in the LHC era: MMHT 2014 PDFs}},
 Eur.\ Phys.\ J.\ C {\bf 75}, no. 5, 204 (2015)
  [\href{http://arxiv.org/abs/1412.3989}{{\tt arXiv:1412.3989}}].

\bibitem{Accardi:2011fa}
A.~Accardi, W.~Melnitchouk, J.~Owens, M.~Christy, C.~Keppel, et~al., {\it
  {Uncertainties in determining parton distributions at large x}},  {\em
  Phys.Rev.} {\bf D84} (2011) 014008,
  [\href{http://arxiv.org/abs/1102.3686}{{\tt arXiv:1102.3686}}].

\bibitem{Jimenez-Delgado:2014twa}
P.~Jimenez-Delgado and E.~Reya, {\it {Delineating parton distributions and the
  strong coupling}},  {\em Phys.Rev.} {\bf D89} (2014), no.~7 074049,
  [\href{http://arxiv.org/abs/1403.1852}{{\tt arXiv:1403.1852}}].

\bibitem{Dulat:2015mca}
S.~Dulat, T.~J. Hou, J.~Gao, M.~Guzzi, J.~Huston, P.~Nadolsky, J.~Pumplin,
  C.~Schmidt, D.~Stump, and C.~P. Yuan, {\it {The CT14 Global Analysis of
  Quantum Chromodynamics}},  \href{http://arxiv.org/abs/1506.07443}{{\tt
  arXiv:1506.07443}}.

\bibitem{Alekhin:2010dd}
S.~Alekhin, J.~Bl{\"u}mlein, P.~Jimenez-Delgado, S.~Moch, and E.~Reya, {\it
  {NNLO Benchmarks for Gauge and Higgs Boson Production at TeV Hadron
  Colliders}},  {\em Phys. Lett.} {\bf B697} (2011) 127--135,
  [\href{http://arxiv.org/abs/1011.6259}{{\tt arXiv:1011.6259}}].

\bibitem{Botje:2011sn}
M.~Botje et~al., {\it {The PDF4LHC Working Group Interim Recommendations}},
  \href{http://arxiv.org/abs/1101.0538}{{\tt arXiv:1101.0538}}.

\bibitem{Alekhin:2011sk}
S.~Alekhin, S.~Alioli, R.~D. Ball, V.~Bertone, J.~Bl{\"u}mlein, et~al., {\it
  {The PDF4LHC Working Group Interim Report}},
  \href{http://arxiv.org/abs/1101.0536}{{\tt arXiv:1101.0536}}.

\bibitem{Nadolsky:2008zw}
P.~M. Nadolsky et~al., {\it {Implications of CTEQ global analysis for collider
  observables}},  {\em Phys. Rev.} {\bf D78} (2008) 013004,
  [\href{http://arxiv.org/abs/0802.0007}{{\tt arXiv:0802.0007}}].

\bibitem{Martin:2009iq}
A.~D. Martin, W.~J. Stirling, R.~S. Thorne, and G.~Watt, {\it {Parton
  distributions for the LHC}},  {\em Eur. Phys. J.} {\bf C63} (2009) 189--285,
  [\href{http://arxiv.org/abs/0901.0002}{{\tt arXiv:0901.0002}}].

\bibitem{Ball:2010de}
{\bf {The NNPDF }} Collaboration, R.~D. Ball et~al., {\it {A first unbiased
  global NLO determination of parton distributions and their uncertainties}},
  {\em Nucl. Phys.} {\bf B838} (2010) 136--206,
  [\href{http://arxiv.org/abs/1002.4407}{{\tt arXiv:1002.4407}}].

\bibitem{Lai:2010nw}
H.-L. Lai et~al., {\it {Uncertainty induced by QCD coupling in the CTEQ global
  analysis of parton distributions}},  {\em Phys. Rev.} {\bf D82} (2010)
  054021, [\href{http://arxiv.org/abs/1004.4624}{{\tt arXiv:1004.4624}}].

\bibitem{Martin:2009bu}
A.~D. Martin, W.~J. Stirling, R.~S. Thorne, and G.~Watt, {\it {Uncertainties on
  $\alpha_S$ in global PDF analyses}},  {\em Eur. Phys. J.} {\bf C64} (2009)
  653--680, [\href{http://arxiv.org/abs/0905.3531}{{\tt arXiv:0905.3531}}].

\bibitem{Demartin:2010er}
F.~Demartin, S.~Forte, E.~Mariani, J.~Rojo, and A.~Vicini, {\it {The impact of
  PDF and $\alpha_s$ uncertainties on Higgs Production in gluon fusion at
  hadron colliders}},  {\em Phys. Rev.} {\bf D82} (2010) 014002,
  [\href{http://arxiv.org/abs/1004.0962}{{\tt arXiv:1004.0962}}].

\bibitem{PDF4LHCrecom}
{PDF4LHC Steering Committee}, ``{PDF4LHC Recommendations}.''
  {\url{http://www.hep.ucl.ac.uk/pdf4lhc/PDF4LHCrecom.pdf}}.

\bibitem{alphasprop}
{PDF4LHC Steering Committee}, ``{Procedure for Adding PDF and $\alpha_s$
  Uncertainties}.'' {\url{http://www.hep.ucl.ac.uk/pdf4lhc/alphasprop.pdf}}.

\bibitem{Agashe:2014kda}
{\bf Particle Data Group} Collaboration, K.~Olive et~al., {\it {Review of
  Particle Physics}},  {\em Chin.Phys.} {\bf C38} (2014) 090001.

\bibitem{Watt:2012tq}
G.~Watt and R.~S. Thorne, {\it {Study of Monte Carlo approach to experimental
  uncertainty propagation with MSTW 2008 PDFs}},  {\em JHEP} {\bf 1208} (2012)
052, [\href{http://arxiv.org/abs/1205.4024}{{\tt arXiv:1205.4024}}],
   Supplementary material available from \url{http://mstwpdf.hepforge.org/random/}.


\bibitem{Watt:2013}
{G.~Watt, talk presented at PDF4LHC meeting, 17th April 2013, CERN, Geneva}.
\newblock {\url{http://indico.cern.ch/event/244768/contribution/5}}.

\bibitem{Alwall:2014hca}
J.~Alwall, R.~Frederix, S.~Frixione, V.~Hirschi, F.~Maltoni, et~al., {\it {The
  automated computation of tree-level and next-to-leading order differential
  cross sections, and their matching to parton shower simulations}},  {\em
  JHEP} {\bf 1407} (2014) 079, [\href{http://arxiv.org/abs/1405.0301}{{\tt
  arXiv:1405.0301}}].

\bibitem{Frederix:2011ss}
R.~Frederix, S.~Frixione, V.~Hirschi, F.~Maltoni, R.~Pittau, et~al., {\it
  {Four-lepton production at hadron colliders: aMC@NLO predictions with
  theoretical uncertainties}},  {\em JHEP} {\bf 1202} (2012) 099,
  [\href{http://arxiv.org/abs/1110.4738}{{\tt arXiv:1110.4738}}].

\bibitem{Alioli:2010xd}
S.~Alioli, P.~Nason, C.~Oleari, and E.~Re, {\it {A general framework for
  implementing NLO calculations in shower Monte Carlo programs: the POWHEG
  BOX}},  {\em JHEP} {\bf 1006} (2010) 043,
  [\href{http://arxiv.org/abs/1002.2581}{{\tt arXiv:1002.2581}}].

\bibitem{Gavin:2012sy}
R.~Gavin, Y.~Li, F.~Petriello, and S.~Quackenbush, {\it {W Physics at the LHC
  with FEWZ 2.1}},  {\em Comput.Phys.Commun.} {\bf 184} (2013) 208--214,
  [\href{http://arxiv.org/abs/1201.5896}{{\tt arXiv:1201.5896}}].

\bibitem{Gao:2013bia}
J.~Gao and P.~Nadolsky, {\it {A meta-analysis of parton distribution
  functions}},  {\em JHEP} {\bf 1407} (2014) 035,
  [\href{http://arxiv.org/abs/1401.0013}{{\tt arXiv:1401.0013}}].

\bibitem{Ball:2012cx}
{\bf NNPDF} Collaboration, R.~D. Ball, V.~Bertone, S.~Carrazza, C.~S. Deans,
  L.~Del~Debbio, et~al., {\it {Parton distributions with LHC data}},  {\em
  Nucl.Phys.} {\bf B867} (2013) 244--289,
  [\href{http://arxiv.org/abs/1207.1303}{{\tt arXiv:1207.1303}}].

\bibitem{Pumplin:2009bb}
J.~Pumplin, {\it {Parametrization dependence and $\Delta \chi^2$ in parton
  distribution fitting}},  {\em Phys.Rev.} {\bf D82} (2010) 114020,
  [\href{http://arxiv.org/abs/0909.5176}{{\tt arXiv:0909.5176}}].

\bibitem{Carrazza:2015aoa}
S.~Carrazza, S.~Forte, Z.~Kassabov, J.~I. Latorre, and J.~Rojo, {\it {An
  Unbiased Hessian Representation for Monte Carlo PDFs}},  {\em Eur. Phys. J.}
  {\bf C75} (2015), no.~8 369, [\href{http://arxiv.org/abs/1505.06736}{{\tt
  arXiv:1505.06736}}].

\bibitem{tau}
J.~Rojo and J.~I. Latorre, {\it Neural network parametrization of spectral
  functions from hadronic tau decays and determination of qcd vacuum
  condensates},  {\em JHEP} {\bf 01} (2004) 055,
  [\href{http://arxiv.org/abs/hep-ph/0401047}{{\tt hep-ph/0401047}}].

\bibitem{DelDebbio:2004qj}
{\bf The NNPDF} Collaboration, L.~Del~Debbio, S.~Forte, J.~I. Latorre,
  A.~Piccione, and J.~Rojo, {\it Unbiased determination of the proton structure
  function f2(p) with estimation},  {\em JHEP} {\bf 03} (2005) 080,
  [\href{http://arxiv.org/abs/hep-ph/0501067}{{\tt hep-ph/0501067}}].

\bibitem{Ball:2010gb}
{\bf The NNPDF} Collaboration, R.~D. Ball et~al., {\it {Reweighting NNPDFs: the
  W lepton asymmetry}},  {\em Nucl. Phys.} {\bf B849} (2011) 112--143,
  [\href{http://arxiv.org/abs/1012.0836}{{\tt arXiv:1012.0836}}].

\bibitem{Ball:2011gg}
R.~D. Ball, V.~Bertone, F.~Cerutti, L.~Del~Debbio, S.~Forte, et~al., {\it
  {Reweighting and Unweighting of Parton Distributions and the LHC W lepton
  asymmetry data}},  {\em Nucl.Phys.} {\bf B855} (2012) 608--638,
  [\href{http://arxiv.org/abs/1108.1758}{{\tt arXiv:1108.1758}}].

\bibitem{Buckley:2014ana}
A.~Buckley, J.~Ferrando, S.~Lloyd, K.~Nordström, B.~Page, et~al., {\it
  {LHAPDF6: parton density access in the LHC precision era}},  {\em
  Eur.Phys.J.} {\bf C75} (2015), no.~3 132,
  [\href{http://arxiv.org/abs/1412.7420}{{\tt arXiv:1412.7420}}].

\bibitem{compressor}
S.~Carrazza, ``{A compression tool for Monte Carlo PDF sets}.''
  {\url{https://github.com/scarrazza/compressor}}.

\bibitem{Salam:2008qg}
G.~P. Salam and J.~Rojo, {\it {A Higher Order Perturbative Parton Evolution
  Toolkit (HOPPET)}},  {\em Comput. Phys. Commun.} {\bf 180} (2009) 120--156,
  [\href{http://arxiv.org/abs/0804.3755}{{\tt arXiv:0804.3755}}].

\bibitem{Ball:2013bra}
R.~D. Ball, M.~Bonvini, S.~Forte, S.~Marzani, and G.~Ridolfi, {\it {Higgs
  production in gluon fusion beyond NNLO}},  {\em Nucl.Phys.} {\bf B874} (2013)
  746--772, [\href{http://arxiv.org/abs/1303.3590}{{\tt arXiv:1303.3590}}].

\bibitem{Czakon:2011xx}
M.~Czakon and A.~Mitov, {\it {Top++: a program for the calculation of the
    top-pair cross-section at hadron colliders}},
Comput.\ Phys.\ Commun.\  {\bf 185}, 2930 (2014)
  [\href{http://arxiv.org/abs/1112.5675}{{\tt arXiv:1112.5675}}].

\bibitem{Anastasiou:2003ds}
C.~Anastasiou, L.~J. Dixon, K.~Melnikov, and F.~Petriello, {\it {High precision
  QCD at hadron colliders: Electroweak gauge boson rapidity distributions at
  NNLO}},  {\em Phys. Rev.} {\bf D69} (2004) 094008,
  [\href{http://arxiv.org/abs/hep-ph/0312266}{{\tt hep-ph/0312266}}].

\bibitem{Dittmaier:2012vm}
{LHC Higgs Cross Section Working Group}, S.~Dittmaier, C.~Mariotti,
  G.~Passarino, and R.~Tanaka~(Eds.), {\it {Handbook of LHC Higgs Cross
  Sections: 2. Differential Distributions}},  {\em CERN-2012-002} (CERN,
  Geneva, 2012) [\href{http://arxiv.org/abs/1201.3084}{{\tt arXiv:1201.3084}}].

\bibitem{Hardy:1978lp}
G.~H. Hardy, J.~E. Littlewood, and G.~P\'olya, {\em {Inequalities}}.
\newblock {Cambridge University Press}, 1978.

\bibitem{Cover:2006}
T.~M. Cover, {\em {Elements of information theory}}.
\newblock {John Wiley and sons}, 2006.

\bibitem{Bertone:2013vaa}
V.~Bertone, S.~Carrazza, and J.~Rojo, {\it {APFEL: A PDF Evolution Library with
  QED corrections}},  {\em Comput.Phys.Commun.} {\bf 185} (2014) 1647--1668,
  [\href{http://arxiv.org/abs/1310.1394}{{\tt arXiv:1310.1394}}].

\bibitem{Carrazza:2014gfa}
S.~Carrazza, A.~Ferrara, D.~Palazzo, and J.~Rojo, {\it {APFEL Web: a web-based
  application for the graphical visualization of parton distribution
  functions}},  {\em J.Phys.} {\bf G42} (2015), no.~5 057001,
  [\href{http://arxiv.org/abs/1410.5456}{{\tt arXiv:1410.5456}}].

\bibitem{Campbell:2000bg}
J.~M. Campbell and R.~K. Ellis, {\it {Radiative corrections to Z b anti-b
  production}},  {\em Phys. Rev.} {\bf D62} (2000) 114012,
  [\href{http://arxiv.org/abs/hep-ph/0006304}{{\tt hep-ph/0006304}}].

\bibitem{Nagy:2003tz}
Z.~Nagy, {\it {Next-to-leading order calculation of three-jet observables in
  hadron hadron collision}},  {\em Phys. Rev.} {\bf D68} (2003) 094002,
  [\href{http://arxiv.org/abs/hep-ph/0307268}{{\tt hep-ph/0307268}}].

\bibitem{Carli:2010rw}
T.~Carli, D.~Clements, A.~Cooper-Sarkar, C.~Gwenlan, G.~P. Salam, et~al., {\it
  {A posteriori inclusion of parton density functions in NLO QCD final-state
  calculations at hadron colliders: The APPLGRID Project}},  {\em Eur.Phys.J.}
  {\bf C66} (2010) 503--524, [\href{http://arxiv.org/abs/0911.2985}{{\tt
  arXiv:0911.2985}}].

\bibitem{amcfast}
V.~Bertone, R.~Frederix, S.~Frixione, J.~Rojo, and M.~Sutton, {\it {aMCfast:
  automation of fast NLO computations for PDF fits}},  {\em JHEP} {\bf 1408}
  (2014) 166, [\href{http://arxiv.org/abs/1406.7693}{{\tt arXiv:1406.7693}}].

\bibitem{Aad:2013iua}
{\bf ATLAS} Collaboration, G.~Aad et~al., {\it {Measurement of the high-mass
  Drell--Yan differential cross-section in pp collisions at $\sqrt{s}$=7 TeV
  with the ATLAS detector}},  {\em Phys.Lett.} {\bf B725} (2013) 223--242,
  [\href{http://arxiv.org/abs/1305.4192}{{\tt arXiv:1305.4192}}].

\bibitem{CMSDY}
{\bf CMS Collaboration} Collaboration, S.~Chatrchyan et~al., {\it {Measurement
  of the differential and double-differential Drell-Yan cross sections in
  proton-proton collisions at $\sqrt{s} =$ 7 TeV}},  {\em JHEP} {\bf 1312}
  (2013) 030, [\href{http://arxiv.org/abs/1310.7291}{{\tt arXiv:1310.7291}}].

\bibitem{Chatrchyan:2013mza}
{\bf CMS} Collaboration, S.~Chatrchyan et~al., {\it {Measurement of the muon
  charge asymmetry in inclusive pp to WX production at $\sqrt{s}$ = 7 TeV and
  an improved determination of light parton distribution functions}},  {\em
  Phys.Rev.} {\bf D90} (2014) 032004,
  [\href{http://arxiv.org/abs/1312.6283}{{\tt arXiv:1312.6283}}].

\bibitem{Chatrchyan:2013uja}
{\bf CMS} Collaboration, S.~Chatrchyan et~al., {\it {Measurement of associated
    W + charm production in pp collisions at $\sqrt{s}$ = 7 TeV}},
JHEP {\bf 1402}, 013 (2014)
  [\href{http://arxiv.org/abs/1310.1138}{{\tt arXiv:1310.1138}}].

\bibitem{Aad:2011fc}
{\bf ATLAS} Collaboration, G.~Aad et~al., {\it {Measurement of inclusive jet
  and dijet production in pp collisions at $\sqrt{s}$ = 7 TeV using the ATLAS
  detector}},  {\em Phys. Rev.} {\bf D86} (2012) 014022,
  [\href{http://arxiv.org/abs/1112.6297}{{\tt arXiv:1112.6297}}].

\bibitem{Butterworth:2014efa}
J.~Butterworth, G.~Dissertori, S.~Dittmaier, D.~de~Florian, N.~Glover, et~al.,
  {\it {Les Houches 2013: Physics at TeV Colliders: Standard Model Working
  Group Report}},  \href{http://arxiv.org/abs/1405.1067}{{\tt
  arXiv:1405.1067}}.

\bibitem{LHhq}
J.~Rojo et~al., ``{Chapter 22 in: J.~R.~Andersen et al., "The SM and NLO
  multileg working group: Summary report"}.'' arXiv:1003.1241, 2010.

\end{thebibliography}
\providecommand{\href}[2]{#2}\begingroup\raggedright\endgroup

\end{document}